\documentclass[a4paper,aps,pra,superscriptaddress,nofootinbib,reprint]{revtex4-2}

\usepackage[utf8]{inputenc}
\usepackage[T1]{fontenc}
\usepackage{bm}
\usepackage{bbm}
\usepackage{physics}
\usepackage{siunitx}
\usepackage[colorlinks]{hyperref}

\DeclareMathAlphabet{\mathcal}{OMS}{cmsy}{m}{n}
\DeclareSymbolFont{largeymbols}{OMX}{cmex}{m}{n}

\usepackage{amsmath}
\usepackage{amssymb}
\usepackage[symbol]{footmisc}

\usepackage{tcolorbox}
\tcbuselibrary{skins}
\NewTotalTColorBox[auto counter]{\ABox}{ +m +m }{ 
    notitle,
    colback=blue!5!white,
    colbacklower=white,
    frame hidden,
    boxrule=0pt,
    bicolor,
    sharp corners,
    borderline west={4pt}{0pt}{blue!50!black},
    fontupper=\sffamily
}{
    \textcolor{blue!50!black}{
        \sffamily
        \textbf{Box~\thetcbcounter.}%
    }%
    #1
}

\newcommand{\tc}[2]{\textcolor{#1}{#2}}
\newcommand{\teo}[1]{\tc{teal}{#1}}

\begin{document}

\title{Experimental Aspects of Indefinite Causal Order in Quantum Mechanics}

\author{Lee A. Rozema}
\thanks{These authors contributed equally to this work.}
\affiliation{University of Vienna, Faculty of Physics, Vienna Center for Quantum Science and Technology (VCQ), Austria.}
\affiliation{Research Platform for Testing the Quantum Gravity Interface (TURIS), University of Vienna, Austria.}

\author{Teodor Str\"{o}mberg}
\thanks{These authors contributed equally to this work.}
\affiliation{University of Vienna, Faculty of Physics, Vienna Center for Quantum Science and Technology (VCQ), Austria.}
\affiliation{Research Platform for Testing the Quantum Gravity Interface (TURIS), University of Vienna, Austria.}

\author{Huan Cao}
\thanks{These authors contributed equally to this work.}
\affiliation{University of Vienna, Faculty of Physics, Vienna Center for Quantum Science and Technology (VCQ), Austria.}
\affiliation{Research Platform for Testing the Quantum Gravity Interface (TURIS), University of Vienna, Austria.}

\author{Yu Guo}
\thanks{These authors contributed equally to this work.}
\affiliation{CAS Key Laboratory of Quantum Information, University of Science and Technology of China, Hefei, 230026, China;}
\affiliation{CAS Center For Excellence in Quantum Information and Quantum Physics, University of Science and Technology of China, Hefei, 230026, China}
\affiliation{Hefei National Laboratory, University of Science and Technology of China, Hefei 230088, China}

\author{Bi-Heng Liu}
\affiliation{CAS Key Laboratory of Quantum Information, University of Science and Technology of China, Hefei, 230026, China;}
\affiliation{CAS Center For Excellence in Quantum Information and Quantum Physics, University of Science and Technology of China, Hefei, 230026, China}
\affiliation{Hefei National Laboratory, University of Science and Technology of China, Hefei 230088, China}

\author{Philip Walther}
\email{lee.rozema@univie.ac.at\\philip.walther@univie.ac.at}
\affiliation{University of Vienna, Faculty of Physics, Vienna Center for Quantum Science and Technology (VCQ), Austria.}
\affiliation{Research Platform for Testing the Quantum Gravity Interface (TURIS), University of Vienna, Austria.}
\affiliation{Institute for Quantum Optics and Quantum Information (IQOQI) Vienna, Austrian Academy of Sciences, Austria.}
\affiliation{Christian Doppler Laboratory for Photonic Quantum Computer, University of Vienna, Austria.}

\begin{abstract}
In the past decade, the toolkit of quantum information has been expanded to include processes in which the basic operations do not have definite causal relations. Originally considered in the context of the unification of quantum mechanics and general relativity, these causally indefinite processes have been shown to offer advantages in a wide variety of quantum information processing tasks, ranging from quantum computation to quantum metrology. Here we overview these advantages and the experimental efforts to realise them. We survey both the different experimental techniques employed, as well as theoretical methods developed in support of the experiments, before discussing the interpretations of current experimental results and giving an outlook on the future of the field.
\end{abstract}

\flushbottom
\maketitle

\thispagestyle{empty}

\section{Introduction to Indefinite Causal Order}
Quantum information processing tasks are commonly described using the quantum circuit model, in which quantum states evolve through a series of gates in a fixed order. If one associates the application of the gates in such a circuit with events in spacetime, then the classically controlled order of these gates gives rise to a causal structure in which two events always have a definite causal relation, such as ``A happened and then B”. By departing from the quantum circuit model, however, and instead using a quantum system to control the order of the gates, one arrives at quantum processes in which the constituent events do not have a definite causal structure. {Events in such processes are said to have an indefinite causal order (ICO)}. This notion is distinct from the superficially similar sounding field of causal inference\cite{Ried2015inferringCausal,Maclean2017mixturesOfCausalRelations,Carvacho2017Carvacho}.

{While ICO processes were originally proposed by Hardy\cite{hardy2007quantumGravity,hardy2009quantumGravComp} in the context of quantum superpositions of gravitational fields, Chiribella et al.\cite{Chiribella2013quantum} were the first to study them from a pure quantum-information perspective, introducing} the quantum switch: a process in which two gates U and V act in a superposition of causal orders, controlled by an auxiliary quantum system. This process was shown to strictly outperform any causally-ordered circuit at the computational problem of determining whether a pair of gates commute or anti-commute, an advantage that was also demonstrated by Procopio et al.~\cite{Procopio2015Experimental} Following the introduction of the quantum switch, Colnaghi et al.\cite{Colnaghi2012quantumComputation} generalized the process to scenarios involving more gates, while Oreshkov et al.\cite{oreshkov2012quantum} developed a framework to analyse processes with more general causal structures. This framework, the process matrix formalism, was leveraged by Araujo et al.\cite{Araujo2015Witnessing} who, in analogy to entanglement witnesses, introduced the notion of causal witnesses. Such a witness is an observable of a quantum process whose expectation value satisfies an inequality for all causally separable processes; that is, processes that can be described as probabilistic mixtures of processes with a definite causal structure. Armed with these tools, Rubino et al.\cite{rubino2017ExperimentalVerification}, and later Goswami et al.\cite{goswami2018Indefinite}, conclusively verified the indefinite causal structure of an experimental quantum process.

Following these initial studies, the potential of ICO processes was explored in a variety of contexts. Guerin et al.\cite{Guerin2016communicationComplexity} showed how an ICO process can achieve an exponential advantage in communication complexity for a tailored task, later realised by Wei et al.\cite{wei2019experimentalCommunication}; a quantum switch was used by Schiansky et al.\cite{Schiansky22TimeReversal} to show a universal protocol for time reversal; Yin et al. \cite{yin2023experimentalSuperHeisenberg} applied ICO to quantum metrology, and demonstrated super-Heisenberg sensitivity; and Cao et al.\cite{cao2022quantumSimulation} simulated an ICO-enhanced quantum refrigeration cycle, in which thermalising channels bring a quantum system out of thermal equilibrium. This broad yet non-exhaustive set of experimental applications is a reason why ICO processes continue to attract significant attention within the field of quantum information processing. 

In this Review, we will give an overview of the current state of applications of ICO processes, and survey the range of experimental methods used to realise them. We will also discuss methods used in the experimental characterisation of ICO processes, as well as explore loopholes in and criticisms of the experimental demonstrations performed thus far. While early experimental results in the field were previously reviewed  by Goswami and Romero\cite{goswami2020review}, this article aims to provide a broader and more holistic view of the experimental aspects of indefinite causality in quantum mechanics. Finally, since a comprehensive {theoretical} introduction to indefinite causality is beyond the scope of this article, we instead refer interested readers to a piece by Brukner\cite{Brukner2014quantumCausality}.

\section{Experimental Realisations of Indefinite Causal Orders}

\begin{figure*}
\centering
\includegraphics[width=\linewidth]{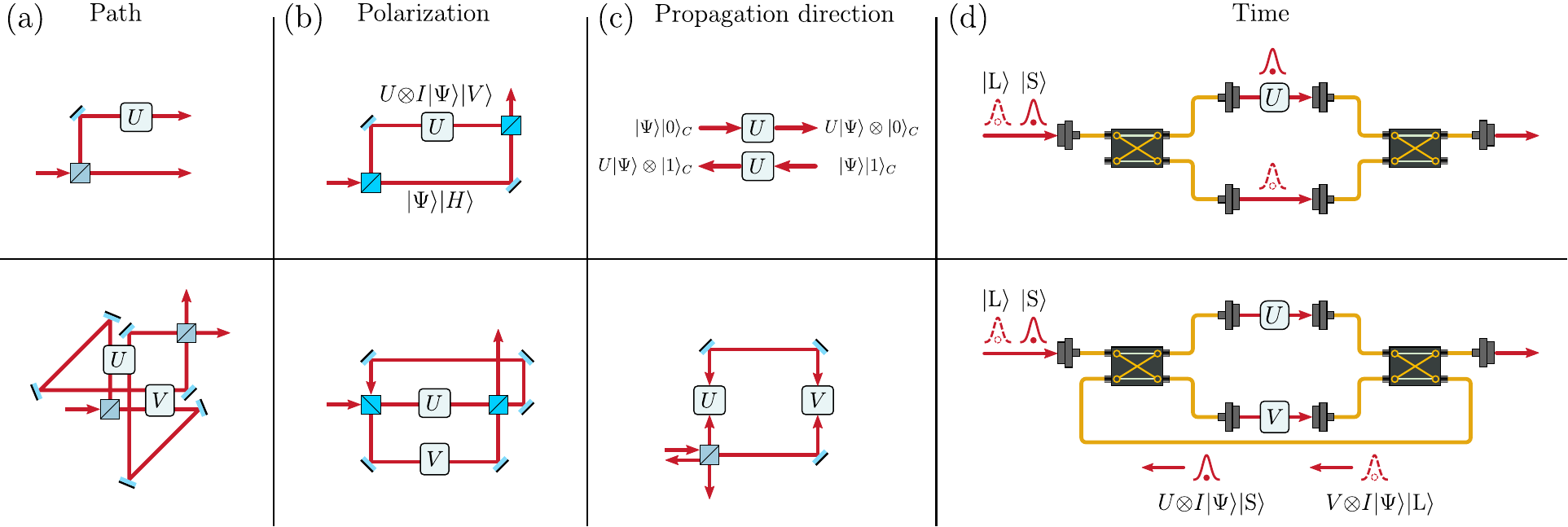}
\caption{\textbf{Different control state encodings to realize the quantum switch.} Four different photonic target- and control-state that have been used to demonstrate indefinite causal processes. (a) \textit{Top:} A balanced beam splitter can prepare a superposition of path-encoded control states. \textit{Bottom:} These outputs of the beam splitter can be aligned to propagate through two unitary channels in different orders, thereby realising a quantum switch. (b) \textit{Top:} A polarizing beam splitter (PBS) maps a polarization-encoded control qubit state to a path state, enabling coherent unitary control. The second PBS recombines the path degrees of freedom. \textit{Bottom:} By connecting the output of the second PBS to the input of the first, a quantum switch can be realised. (c) \textit{Top:} Connecting the two outputs of a beamsplitter forms a closed path with two propagation directions. \textit{Bottom:} Placing two unitary channels in this path naturally correlates the propagation direction with the gate order, thereby generating a quantum switch. (d) \textit{Top:} In a time encoding, the control state is prepared in a superposition of time bins by passing through a short ($\ket{S}$) or long $\ket{L}$ delay line. Ultrafast optical switches are used to perform controlled unitaries, by only routing the $\ket{S}$ state through the unitary channel. \textit{Bottom:} A quantum switch can be constructed by routing the output of the second switch to the first.
} 
\label{fig:exp}
\end{figure*}

Most experimental work on ICO has focused on the two-party quantum switch. This is a two-qubit process consisting of a control qubit $C$ and a target qubit $T$. The quantum switch then takes two gates $U$ and $V$, applying them in a superposition of both orders to the target qubit, where the precise superposition is dependent on the state of the control qubit as:
\begin{equation}
\label{eq:qswitch}
    (U,V)\mapsto U_T V_T\otimes\ketbra{0}{0}_C + V_T U_T\otimes \ketbra{1}{1}_C.
\end{equation}
To date, all physical realisations of this process have encoded both the target and control systems in the same particle. {Specifically, using two different degrees of freedom (DoFs) of a single photon}.
Since every localized optical device can be thought of as performing a controlled operation conditioned on the path of a particle, passing through the device or not, controlled operations between any two internal DoFs can be realised by first mapping one of these to the photon path (Fig. \ref{fig:exp}a, top). This insight, which underlies all the experiments discussed here, was first hinted at in Ref.~\cite{Andersson2005comparison}, {where the authors proposed a scheme for probabilistically determining whether two unitaries commute}. {The coherent unitary control needed for ICO processes} was first realised in Ref.~\cite{Zhou2011AddingControl} where the authors experimentally transformed a black-box unitary into a controlled version of itself {(Fig. \ref{fig:exp}b, top)}, {a task later shown to be forbidden in the quantum circuit model~\cite{Araujo2014controlUnknown}}. Building upon these two works, it was {subsequently} shown how to control the order of two gates in a photonic setting\cite{Araujo2014ComputationalAdvantage,Friis2014ImplementingQuantumControl}.

The first experiments {on ICO} used a modified version of the scheme from Ref.~\cite{Andersson2005comparison} to deterministically superpose gate orders using a path control qubit and a polarisation target system~\cite{Procopio2015Experimental,rubino2017ExperimentalVerification}. Follow-up experiments, however, have made use of a multitude of encodings, and ICO processes have been demonstrated using a polarization control and transverse-electric mode (TEM) target system~\cite{goswami2018Indefinite, goswami2020IncreasingCommunication}, propagation direction control and time-bin target~\cite{wei2019experimentalCommunication}, propagation direction control and polarization target~\cite{stromberg23demonstration,Schiansky22TimeReversal,liu2023experimentally}, time-bin control and polarization target~\cite{Antesberger2023tomography}, and even path control and a continuous variable target~\cite{yin2023experimentalSuperHeisenberg}. These different encodings will be examined below. 

\subsection{Path Control}
To date, the majority of quantum switch realisations have used the path DoF of single photons as the control system. This is a natural choice, since as discussed any optical device can be interpreted as a controlled operation between the position of the device and the photonic DoF that it acts on. A coherently controlled operation with a qubit control system can therefore be realised by a superposition of two optical paths, one passing through the optical device and one bypassing it, as illustrated in {the top panel of Fig.~\ref{fig:exp}a}. Similarly, the order of two quantum operations can be controlled by aligning two paths of a photon such that they pass through the optical devices in opposite orders {(bottom panel of Fig. \ref{fig:exp}a)}. More complex {superpositions of} orders of operations can be realised by increasing the number of paths over which the photon is superposed.

\subsubsection{Polarization Target}
The first realisation of a quantum switch used the polarisation state of a single photon to encode the target system\cite{Procopio2015Experimental}. A balanced beam splitter was used to prepare a superposition of two photon paths, which were aligned to propagate through two sets of three wave plates in two different orders. The experiment demonstrated the ability of the quantum switch to determine if two gates $U$ and $V$, realised by the two sets of wave plates, commute or anti-commute with unity success probability, thereby indirectly serving as a witness for an indefinite causal order. The key difference with respect to the probabilistic approach of Ref.~\cite{Andersson2005comparison} was the use of a folded Mach-Zehnder interferometer (MZI) geometry, in which both arms of the interferometer were made to propagate through both unitary transformations, but in different orders. 
This alignment necessitates two different paths through the polarization optics black boxes implementing $U$ and $V$, which requires these elements to be spatially uniform, and has additionally been a source of discussion about the interpretation of the experiment {(see Sec. \ref{sec:loopholes} for more details)}. 
The commutator was measured by projecting the control system in Eq. \eqref{eq:qswitch} on the $\sigma_x$ basis using the second beam splitter of the MZI, followed by non-polarisation-resolving detection of the photon path. Pairs of gates that anti-commute have the effect of introducing a relative $\pi$ phase between the two interferometer arms, thereby performing a SWAP operation on the output ports when the interferometer is balanced. However, since the information about the commutativity of the two gates is encoded in the interference condition such an interferometer cannot be stabilised by classical light co-propagating through the polarisation optics. The experiment therefore instead utilised passive phase stabilisation in concert with a lock-and-hold technique. Several subsequent experiments adopted the methods of Ref.~\cite{Procopio2015Experimental} to other realisations of ICO processes~\cite{rubino2017ExperimentalVerification,Rubino2021Communication,Rubino2022experimentalEntanglement,guo2020experimental,cao2022quantumSimulation,cao2022Semideviceindependent,zhu2023prl,Min23}. In Refs.~\cite{guo2020experimental,cao2022Semideviceindependent,cao2022quantumSimulation} the problem of phase stability was addressed by actively locking the interferometer using a carefully aligned laser beam so that it co-propagated with the single photons without being subjected to the polarization rotations inside the quantum switch.

\subsection{Polarization Control}
There have been several demonstrations of ICO processes using the polarization DoF for the control system, and these experiments are based on the idea of Refs.~\cite{Zhou2011AddingControl,Araujo2014controlUnknown} to use a birefringent element to entangle the photon polarization with the photon path, performing {the target operations} on an additional DoF conditioned on the path, and finally disentangling the polarization and path, {as shown in Fig~\ref{fig:exp}b}.
{In these experiments, only a single path passes through each optical element, in contrast to situation with a path-control qubit.  Nevertheless, the target operations are still applied on two orthogonal optical modes (polarization in this case).}

\subsubsection{TEM Target}
The first experiment to make use of a polarization-encoded control system was the measurement of a causal witness~\cite{goswami2018Indefinite} {(discussed in detail in Sec. \ref{Sec:Characterizing}). In this experiment,} the target system was encoded in the two Hermite-Guassian modes $\mathrm{HG}_{01}$ and $\mathrm{HG}_{10}$ of the transverse electric field, which form a two-dimensional subspace of the space of transverse electric modes. The setup consisted of a PBS-based MZI that was traversed twice, cancelling any phases due to path-length fluctuations and removing the need for stabilization of the setup. The transformations of the target-state TEM modes were realised using a set of inverting prisms and cylindrical lenses, and in contrast to the experiments using a path-polarization encoding the two control qubit states could be made to fully overlap inside the devices realising the transformation. {These advantages come at the cost of optically more complex target-state transformations, making high-fidelity operation challenging.} {This, however, did not stand in the way of} follow-up work using the same methods to demonstrate communication through a depolarizing channel~\cite{goswami2020IncreasingCommunication}.

\subsubsection{CV target}
The polarization control system was also used together with a continuous-variable (CV) target to demonstrate super-Heisenberg quantum metrology \cite{yin2023experimentalSuperHeisenberg}. For this task, the different evolutions of the target had to satisfy the Weyl relation \cite{reed1980methods}, which is not possible in a discrete variable system. Thus, the target system was taken to be the transverse coordinate of the photon, and the evolutions were taken to be position and momentum displacements. The position displacement was realised using \SI{2}{\milli\meter} thick birefringent MgF$_2$ and the momentum displacement was realised using an optical wedge pair. A challenge in this experiment was to precisely calibrate the direction of the beams on each path to ensure that they strictly propagated along a predefined axis. The directions of the beams are aligned with a deflection angle that is two orders of magnitude lower than that introduced by the displacements. This was achieved by controlling the distance of the beams fluctuating within less than \SI{5}{\micro\meter} along a \SI{4}{\meter} calibration optical path.
{The ability to use essentially the same apparatus to encode both discrete and CV systems highlights the flexibility of photonic quantum switches.}

\subsection{Propagation Direction Control}
Instead of encoding the control system in two distinct photon paths, there have also been experiments that encoded this system in two different propagation directions along the same optical path. For the two-party quantum switch, {shown in Fig.~\ref{fig:exp}c}, such a configuration automatically generates the superposition of gate orders and furthermore bestows an insensitivity to path-length fluctuations since the two control system states traverse the same path. A challenge associated with this approach is that the black boxes acting on the target state have to be invariant under the reversal of propagation direction.

\subsubsection{Time-bin Target}
The first experiment to make use of a propagation-direction encoding of the control system was a demonstration of an exponential communication advantage using an ICO\cite{wei2019experimentalCommunication}. This task required a bit string to be encoded in the target system, which therefore had to be high dimensional. This made time bins a suitable choice, since the encoding can in principle be extended indefinitely. The operations implemented in the experiment consisted of shift operations on the computational (time-bin) basis states, as well as time-bin-dependent phase shifts. The former was realised using high-precision fiber patch cables, and the latter used two fast phase modulators. Through the use of fiber delays, these modulators acted independently on photon wave packets in the two different propagation directions, and these decoupled operations could therefore be chosen to be identical. 

\subsubsection{Polarization Target}
The transformations generated by standard polarization optics, such as waveplates, are not fully invariant under reversal of the propagation direction. 
However, there exists a subset of polarization rotations in SU(2) that is invariant for such devices. 
{Thus, this encoding is especially suitable for} applications of the quantum switch that do not strictly require a universal qubit-gate set, such as a time-rewinding protocol\cite{Schiansky22TimeReversal} and an ICO version of Deutsch's algorithm\cite{liu2023experimentally}; {both of which} have been demonstrated. At the cost of more complex polarization black boxes it is possible to realise fully invariant transformations, and therefore a quantum switch with propagation direction control with a universal gate set\cite{stromberg23demonstration}. 
{However, the passive phase stability comes at a cost; in particular,} the fixed interference condition ensured by the Sagnac geometry restricts the range of possible control-qubit measurements when only using passive optical elements.

\subsection{Time-bin Control}
The temporal DoF of photons has the attractive property of being largely decoherence free, since co-propagating temporally offset single-photon wave packets do not acquire relative phase, while also offering a large Hilbert space. This property was exploited in Ref.~\cite{wei2019experimentalCommunication} to encode a large target system, but could also be leveraged for a high-dimensional control system to realise superpositions of gate orders beyond the two-party quantum switch.
{This was proposed in Ref.~\cite{rambo2016functional}.}
To date there has only been a single demonstration of a quantum switch using a time-bin control system\cite{Antesberger2023tomography}, likely due to the complexities associated with manipulating and measuring time-bin qubits. When measuring a time-bin state on a superposition basis, care must be taken that the measurement apparatus shares a phase reference with the state-creation stage. The experiment presented in Ref.~\cite{Antesberger2023tomography} tackled this challenge by reusing the state-generation stage for the measurement of the control qubit, thereby ensuring a vanishing phase difference. However, similarly to the case of propagation-direction control, without the use of fast phase modulators this limited the measurements of the control system to the $\sigma_z$ and $\sigma_y$ bases. The target state was encoded in the polarization DoF and controlled operations between the control and the target were realised using ultra-fast optical switches that actively routed the light through the polarization optics in two different orders. {This is illustrated in Fig.~\ref{fig:exp}d}. The experimental apparatus was used to perform tomography on the quantum switch, and reconstruct the process matrix. This protocol measured {9216} different probabilities, demonstrating the utility of passively stable platforms.

\subsection{Variations on the quantum switch}
The experiments described above have all been realisations of the two-party quantum switch {with qubit control systems}  and were restricted to unitary operations on the target system. While the quantum switch is uniquely defined by its action on unitary transformations\cite{dong2023quantum}, it remains a valid supermap for nonunitary transformations as well, and there have been several experiments demonstrating this\cite{rubino2017ExperimentalVerification,cao2022Semideviceindependent,Antesberger2023tomography,goswami2020IncreasingCommunication,guo2020experimental,Rubino2021Communication}. Additionally, ICO processes with higher-dimensional control systems have been studied extensively from a theoretical perspective \cite{Abbott2016MultipartiteCausalCorrelations,Abbott2017GenuinelyMultipartite,Wechs2019definition}, and one such process has been the subject of an experiment\cite{taddei2021computational}.

\subsubsection{High Dimensional Control Systems}
\label{Sec:realization:highD}
For pairs of gates the quantum switch can efficiently answer the computational question of whether or not the gates commute, but when considering more than two gates this is no longer a well-posed question. However, there are analogous problems for $2^k =N$ permutations of $N$ qubit gates, called Hadamard promise problems. An instance of such a problem for $N=4$ was solved using a {simplified version of the} four-party quantum switch in Ref.~\cite{taddei2021computational}, {which superposed only four of the possible 24 orders of the gates}.
The experiment used a path (polarization) control (target) encoding, which lends itself well to higher-dimensional control systems since the number of photon paths can in principle be extended arbitrarily. However, a direct extension of the approach pioneered in Ref.~\cite{procopio2020sending} leads to {prohibitively} complex optical geometries for higher-dimensional control systems. The experiment therefore instead used multi-core fibers to multiplex several modes through a single fiber, and the polarization optics acting on the target state were placed in the Fourier plane of the fiber launchers to ensure spatial overlap of the four modes. The initialization and measurement of the quqart control state was performed using four-core fiber beamsplitters realising Hadamard transformations\cite{carine2020multi}.
\subsubsection{Non-Unitary Operations in the Quantum Switch}
\label{Sec:realization:nonU}
Indefinite causal processes involving non-unitary operations have been demonstrated for a variety of different quantum channels. The first of these to be realised were measure-repreparation channels, which can be performed as a von Neumann measurement where detection is delayed until after exiting the quantum switch in order to maintain coherence. For polarization target encodings, such measurements were realised using a PBS\cite{rubino2017ExperimentalVerification} and in Ref.~\cite{cao2022Semideviceindependent} two measurements were performed using beam displacers. In both cases, the measurement outcome is encoded in the photon path, thereby doubling the number of paths with each measurement. For experiments that do not encode the control system in the photon path this presents a problem, and the time-bin-control experiment in Ref.~\cite{Antesberger2023tomography} solved this problem by only recording one measurement outcome at a time.

A second type of nonunitary quantum channel that has been realised inside a quantum switch is a decohering or depolarizing channel. In two experiments using polarization\cite{guo2020experimental} and TEM\cite{goswami2020IncreasingCommunication} target-state encodings such channels were realised by exploiting the fact that they can be decomposed as mixtures of unitary Pauli channels. The decoherence then emerges in the data analysis through an appropriate averaging of measurement outcomes for the different Pauli channels. In a subsequent experiment\cite{Rubino2021Communication} polarization-encoded target states were decohered by rapidly modulating the unitary channel using liquid crystal retarders. The time scale of this modulation was not fast enough to randomize the transformation on a shot-to-shot basis, but was nevertheless fast enough to attribute the bulk of the decoherence to the channel rather than the post-measurement averaging.

{Yet another example of a} nonunitary channel implemented in a quantum switch is amplitude damping\cite{cao2022quantumSimulation}, which was studied in the context of ICO in quantum thermodynamics, in a realisation of a quantum refrigeration protocol proposed in Ref.~\cite{felce2020QuantumRefrigeration}. Similar to the realisations of decohering channels, the generalized amplitude damping channel can be decomposed into pairs of Kraus operators, and classical averaging over these channels produces the desired map. The Kraus operators in the decomposition are themselves nonunitary maps and were realised on the polarization target system through two MZIs embedded inside the quantum switch. The non-trace-preserving nature of the Kraus operators manifests as loss when the photon exists in an output of the MZI that is not collected. {Similar methods, using beam-displacer polarization MZIs, were applied in Ref.~\cite{zhu2023prl} to realise a non-unitary quantum-battery charging process in terms of its Kraus operators.}

\subsubsection{The quantum switch in non-photonic platforms}
There have been two experiments on ICO with matter-based qubits, specifically superconducting qubits\cite{Felce2021IBMswitch} and an NMR system\cite{nei2022NMRswitch}. These platforms differ from the photonic ones previously discussed in that they only have a single experimentally accessible DoF, and consequently the gate order cannot be coherently controlled. While future experiments using matter-based qubits may make use of coherent unitary control, as it has been shown to be possible for ion-based systems\cite{Friis2014ImplementingQuantumControl}, the two experiments discussed here instead implemented circuit-model simulations of the ICO processes in which each quantum channel is realised twice. In the superconducting experiment\cite{Felce2021IBMswitch} the quantum refrigeration cycle proposed in Ref.~\cite{felce2020QuantumRefrigeration}, which was also the subject of Ref.~\cite{cao2022quantumSimulation}, was used to bring a qubit state out of equilibrium with two reservoirs using only thermalizing channels. These channels were implemented unitarily as SWAP operations with randomly drawn qubit states in the reservoir, and the thermal ensembles were realised using classical averaging of measurement outcomes. A modified version of the refrigeration protocol was demonstrated in an NMR platform\cite{nei2022NMRswitch}. In contrast to the pure states used in Ref.~\cite{Felce2021IBMswitch} the qubits representing the thermal reservoirs were decohered using magnetic field gradients as part of the experimental state initialization.

\subsection{Processes aside from the quantum switch}
The experiments discussed so far all fit within the framework of indefinite causality and can be accurately described using the process matrix formalism {(described in the Appendix A)}.
Recently, however, processes that go beyond this framework have been proposed. The first such process is a map that acts on a quantum channel \cite{chiribella2022indefiniteTimeDirection} or a thermodynamic process \cite{Rubino2021superpositionThermodynamicEvolutions} and transforms it into a coherent superposition of itself and its transpose or its adjoint. Since the transpose or the adjoint of a channel can be understood as its time-reversal this process has been dubbed the quantum time flip \cite{chiribella2022indefiniteTimeDirection}.
Unlike the quantum switch, this is only valid for a subset of quantum channels. These channels are said to be bidirectional. A composition of two time flips can outperform any causally ordered process, and even any process captured by the process matrix formalism, in a specific channel discrimination task. This advantage has been witnessed experimentally\cite{stromberg2022timeDirections,guo2022inputOutput} using polarization encoded target states and methods similar to those used in realisations of the quantum switch. In Ref.~\cite{stromberg2022timeDirections} the time flip was realised on unitary transformations by exploiting the basis-dependent nature of the transposition operation, and in Ref.~\cite{guo2022inputOutput} the time flip was applied to measure-reprepare channels with a single outcome, analogous to\cite{Antesberger2023tomography}. A theoretical study of the applicability of the time flip to communication tasks has also been carried out\cite{Zixuan2023indefiniteIO}. {It was also shown that the quantum time flip acting on a thermodynamic process can be {implemented} with a photonic system, and used to estimate the work distribution and the dissipative work \cite{Rubino2022InferringWork}.}

\begin{figure*}
\centering
\includegraphics[width=0.95\linewidth]{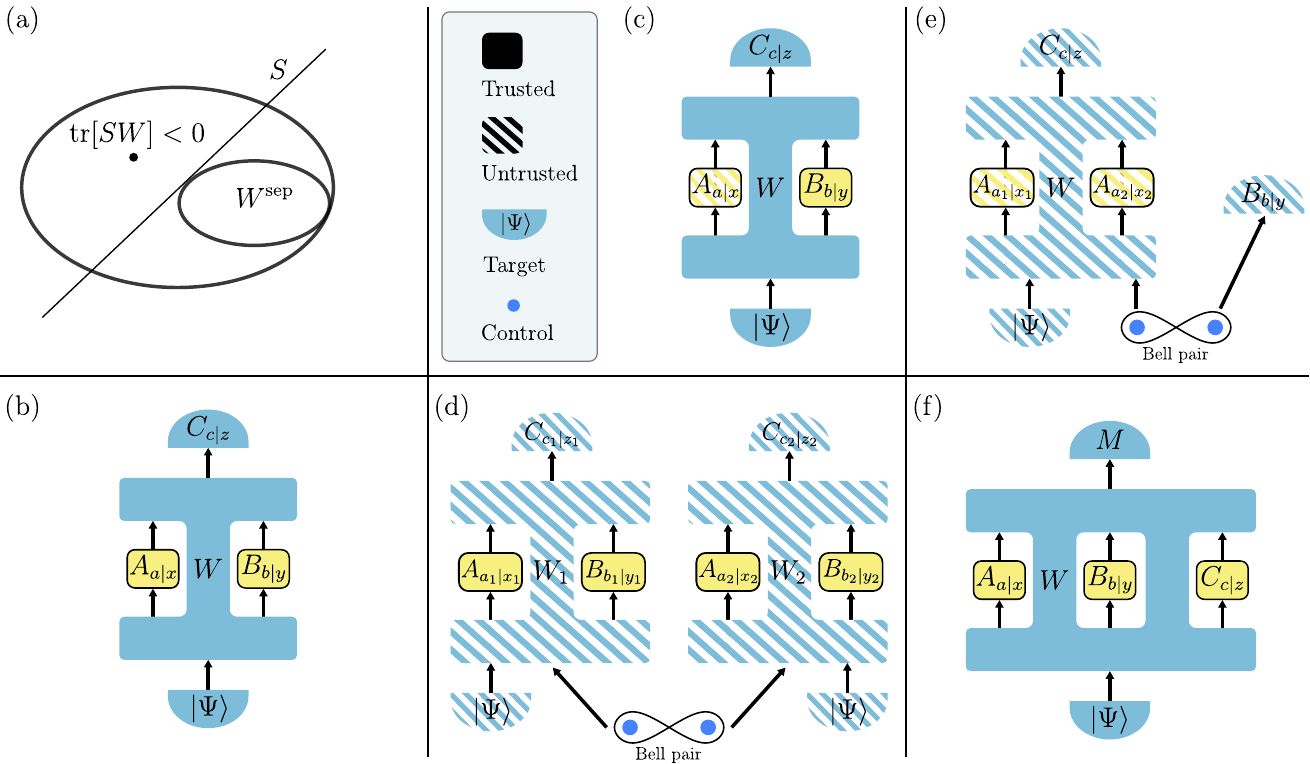}
\caption{Certification of indefinite causal order. {In panels b-f, the solid filled areas represent trusted devices, while dashed areas are untrusted.
Process matrices are represented as the areas labeled with W. Quantum instruments, represented by square boxes filled with yellow, are considered as inputs for process matrices.
The semicircles represent the measurement (upper half) or preparation (lower half) of quantum states. (a) The causal correlations for causally separable processes $W^{\mathrm{sep}}$ are depicted by a convex polytope. A hyperplane $S$ separates causally separable processes. Negative values of $\mathrm{tr}\left[S \ W\right]<0$ signify that $W$ is causal nonseparable. (b)(c) Evaluation of causal nonseparability in different scenarios. The distinction between full device-dependence and other scenarios is given by which instruments within the quantum switch $W$ are trusted and which are untrusted. (d) Theory-independent protocol to verify an indefinite causal order, wherein the control qubits of two switches are entangled. 
For this panel only, the dashed areas represents components that are not assumed to be described by quantum theory.
This approach is not device independent, as it requires specific properties of  the initial states and laboratory operations to be verified. 
(e) Device-independent approach utilizing ancillary distributed entanglement and spacelike-separated observers. The control qubit of quantum switch is entangled with its ancilla. (f) High-dimensional switch allowing the order of additional parties to be superposed.} 
} 
\label{fig:ICOdiagram}
\end{figure*}

\section{Characterizing Indefinite Causal Order}
\label{Sec:Characterizing}

So far we have discussed ICO in a rather qualitative manner, i.e., a process with an ICO is one wherein it cannot be said if $A$ occurs before or after $B$.
This notion has been rigorously formalized, based on the  \textit{process matrix formalism}, introduced in Appendix A.
In brief, one can use the Choi–Jamiołkowski (CJ) isomorphism to consider quantum processes as quantum states. 
Based on this, one can then define causally separable processes as those with a well-defined causal order and causally non-separable processes as those with an ICO.
The mathematics behind this approach are analogous to those describing separable and entangled states. This allows for the adaptation of a host of tools, originally designed for entanglement certification, to the problem of characterizing ICO.
Similar to entanglement certification, ICO certification can be performed with device-dependent (DD), semi-device independent (semi-DI), or fully device-independent (DI) methods. 
Historically, the first proposal for ICO verification was a DI technique known as a causal inequality \cite{oreshkov2012quantum}.
However, an experimental DI verification has not yet been performed.  Thus, we will first discuss DD ICO characterization techniques, and then semi-DI techniques, and conclude this section with an outlook discussing DI proposals, including the aforementioned causal inequality.

\subsection{Device Dependent Techniques}
\label{Sec:devDep}
\subsubsection{Causal Witnesses}

Causal witnesses were developed to verify an ICO \cite{Araujo2015Witnessing}, and are somewhat analogous to entanglement witnesses. 
They are a DD ICO characterization method, and were one of the first techniques for verifying an ICO to be experimentally implemented \cite{rubino2017ExperimentalVerification,goswami2018Indefinite}.
In the case of a process with two parties, A and B, the goal is to determine whether there is a well-defined causal order between them. In other words: the target system is sent first to $A$ and then to $B$, first to $B$ and then to $A$, or the process is a convex mixture of those two scenarios. 
Using the CJ isomorphism to describe a quantum process matrix $W$ as a quantum state, two fixed order processes $W_{A\rightarrow B}$ and $W_{B\rightarrow A}$ can be defined.
One can employ the separating hyperplane theorem {(see Fig. \ref{fig:ICOdiagram}a)} to distinguish causally non-separable processes ($W^{\mathrm{non-sep}}$) from causally separable processes: the aforementioned convex mixtures: $W^{\mathrm{sep}} = p W_{A\rightarrow B} + (1-p) W_{B\rightarrow A}$.
A causal witness, then, is a Hermitian operator $S$ whose expectation value is negative $\mathrm{tr}\left[S \ W^{\mathrm{non-sep}}\right]< 0$ only if the process matrix is causally non-separable.
The interpretation of the actual value of this expectation value is somewhat subtle, and depends on the precise normalization of $S$. Some refer to it as the ``causal non-separability''
$\mathrm{tr}\left[S \ W^{\mathrm{non-sep}}\right]:=\mathrm{\mathrm{CNS}}$ \cite{rubino2017ExperimentalVerification}.
The CNS is related to how much noise a given witness can tolerate and still detect the presence of an ICO, and is sometimes also called ``robustness''\cite{Antesberger2023tomography}.

Above we said that measuring a causal witness requires one to measure the expectation value of a quantum process.
{Although one does not typically discuss the expectation values of processes, this concept is mathematically well-defined since the CJ isomorphism allows us to consider quantum processes as states.
Experimentally measuring a causal witness requires one to to decompose the witness in terms of operations that are accessible in the laboratory.  This is similar to how one measures an entanglement witness by decomposing it terms of local measurements.}
{The causal witness procedure is represented in Fig. \ref{fig:ICOdiagram}b, where the solid blue area labeled $W$ represents the unknown process matrix we wish to characterize.
Then to assess the ICO of $W$ we must vary the input states (labeled $\ket{\psi}$), the post-process measurements (labeled $C_{c|z}$), as well as Alice and Bob's channels (labeled $A_{a|x}$ and $B_{b|y}$, respectively. Here $x(y)$ refers to the specific channel chosen and $a(b)$ is the output (if any) of the given channel.}). 
Alice and Bob's operations can be unitary gates, quantum channels, positive operator-valued measures (POVMs), or measure-and-prepare instruments \cite{Branciard2016witnesses}. The precise choice of channels will affect the maximum obtainable value of the CNS, and not all sets of operations can be used to detect an ICO.

The first experimental proof of a causally non-separable process using a causal witness was conducted by Rubino et al. \cite{rubino2017ExperimentalVerification}. 
In this work, to achieve a larger CNS one of the parties in the switch was equipped with a measure and repreparation channel, as discussed in Sec. \ref{Sec:realization:nonU}. The work also represents the first time a non-unitary operation was placed in an ICO process.
Shortly after, Goswami et al. also demonstrated the violation of a causal witness using only unitary operations \cite{goswami2018Indefinite}. The violation of a similar unitary-only witness, formulated using the methods developed in Ref.~\cite{Bavaresco2021StrictHierarchy}, was later shown by Strömberg et al.~\cite{stromberg23demonstration}.
{In this context, it is also worth pointing to Ref.~\cite{Antesberger2023tomography}. Although this work focused on experimentally performing process tomography on the quantum switch, it also discusses the measurement of different causal witnesses on the switch; in particular, witnesses constructed using different sets of channels and normalized differently to represent different notions of robustness.}


\subsubsection{Multipartite Causal Witnesses}

Although most investigations of ICO have focused on bipartite cases, one can also define ICO for multipartite cases (i.e., those involving more than two parties).
This situation is similar to that of multipartite entanglement; wherein, multipartite states have a much richer state-space structure, leading to different classes of entanglement \cite{svetlichny1987distinguishing, seevinck2002bell}. 
For ICO, consider, for example, the case of three parties $A$, $B$ and $C$ {(see Fig. \ref{fig:ICOdiagram}f)}. 
If one can group specific subsets of parties together {and ascribe a well-defined order of the subsets to other parties}, {then this would not constitute a genuine multipartite noncausal correlation.}
For example, {if $B$ and $C$ are grouped together as a new party, and then $A$ occurs in their causal past, then this is not multipartite ICO}
{even if $B$ and $C$ are causally nonseparable}.
Thus, genuine multipartite ICO should exclude the existence of partial causal orderings, and specify a given correlation for which no subset of parties admitted can be grouped as above. 
This idea has been formalized in several theoretical studies 
\cite{Abbott2016MultipartiteCausalCorrelations, Abbott2017GenuinelyMultipartite,Wechs2019definition}. 
Experimentally studying this problem with currently accessible methods requires one to incorporate N parties in quantum switch, i.e., to build a  quantum N-switch. 
As discussed in Sec. \ref{Sec:realization:highD}, a process similar to the four-switch, wherein a subset of all possible orders were superposed, has been experimentally realized \cite{taddei2021computational}. 
This work demonstrated the computational advantages of Hadamard promise problem involving four unknown unitary gates.
However, the number of settings required to implement the four-partite witness was prohibitively high, and prevented the authors from verifying a multipartite ICO. Nevertheless, the work provides a recipe to create a causal witnesses for their process.

\subsubsection{Process Matrix Tomography }

Process matrices represent so-called higher-order quantum processes.
This naming reflects the fact that they can take as input both quantum states and quantum operations, and can return states, operations or both.
In particular, for the quantum switch, Alice and Bob's channels are said to be the inputs to the process.
Given that to fully characterise a higher-order process, one must input a complete set of operations for each input channel, the scaling of higher-order process tomography is even worse than that of standard process tomography.
To practically combat this challenge, Antesberger et al.\cite{Antesberger2023tomography} recently constructed a passively-stable fiber-based quantum switch, to facilitate the required large number of measurements.
{Complete higher-order process tomography would have required the measurement of $13824$ different probabilities to reconstruct the process matrix of the two-party quantum switch. However, due to experimental limitations in measuring the time-bin control system, the experiment used additional constraints and only measured $9216$ different probabilities \cite{Antesberger2023tomography}. 
This work represents the first experimental technique to completely characterize a process with an ICO.
We point out that in addition to describing ICO, higher-order processes can be used to describe non-Markovian processes, and higher-order process tomography has been experimentally implemented in this context \cite{Giarmatzi2023MultiTimeTomo,Kavan_experimental,Kavan_experimental2,guo21Markovorder}.\textbf{}

\subsection{Semi-Device Independent Approaches}
 
Although the DD techniques described in Sec. \ref{Sec:devDep} can detect the presence of an ICO, they require two main assumptions to made about the experimental apparatus.
First, they require one to assume that quantum theory applies to the experiment, and, second, these techniques require one to trust the experimental apparatus.
A protocol that removes both of these assumptions completely is said to be DI, while partially removing them results in a semi-DI protocol.
%
In 2019, in a step towards device independence, Zych and Brukner conceived of a ``theory independent'' protocol by leveraging Bell's inequality \cite{zych2019bell}. 
They considered a scenario wherein two states initially cannot violate a Bell inequality. They then considered sending these two states into two different quantum switches, with entangled control qubits, as the respective target state {(see Fig. \ref{fig:ICOdiagram}d)}.
They argued, without using quantum theory, that unless the order of operations inside the quantum switches is indefinite, the target systems cannot be used to violate a Bell inequality.
This proposal was later adapted to a photonic setting by Rubino et al. \cite{Rubino2022experimentalEntanglement}, who performed an experiment using two quantum switches sharing entangled path control qubits. 
In the experiment, the causally non-separable process was validated by the violation of a Bell inequality with a Bell parameter of $S_{\mathrm{target}}=2.55\pm0.08$, when from an initially separable state. 

In a different semi-DI approach, one can use quantum mechanical descriptions of parts of the experimental devices while making no assumptions about the other parts {(Fig. \ref{fig:ICOdiagram}c)}.
This is what is typically meant by semi-device independence in quantum information scenarios such as EPR-steering \cite{wiseman2007steering,uola2014joint,quintino2014joint} and quantum cryptography \cite{branciard2012one}.
Semi-DI approaches for causal witness were developed in Refs.~\cite{bavaresco2019Semideviceindependent,Dourdent2022SemiDeviceIndependent}. 
For example, Bavaresco et al. \cite{bavaresco2019Semideviceindependent} considered the quantum switch as a tripartite process where Alice and Bob are the parties in the switch and Charlie is a third party who performs measurements after the switch.
They considered various scenarios wherein not all three parties are completely trusted, and showed how to derive semi-DI causal witnesses for each case.
Building on this proposal, Cao et al. \cite{cao2022Semideviceindependent} performed an experiment using a photonic quantum switch in which both parties inside the quantum switch performed measurement and reprepare channels.
This experiment allowed for two parties to be untrusted, but one of the trusted parties should be in the switch (i.e. Alice or Bob).
This experiment was also notable in that it was the first experiment to equip both parties with such measurement and reprepare channels.

\subsection{Causal Inequality and Device-Independent Techniques}

Bell's inequality is perhaps the most well-known DI protocol \cite{aspect1999bell,clauser1969proposed,bell1964einstein,bell1966problem}.
Its violation has profound implications for our understanding of the nature of nonlocality \cite{weihs1998violation,aspect1981experimental,freedman1972experimental}, telling us, irrespective of quantum theory, that nature allows for nonlocal correlations.
From a more mundane point of view, it can also be used as a tool to certify the presence of entanglement \cite{toth2005detecting,guhne2007toolbox}.
To maintain our analogy between ICO and entanglement certification, ``causal inequalities'' take the place of Bell's inequality for ICO.
These inequalities were first proposed by Oreshkov et al. \cite{oreshkov2012quantum}, and concern the correlations between two parties.
Making three assumptions: (1)  the parties operate in a fixed causal order, (2) they can freely choose their measurement, and \hypertarget{assumption3}{(3)} they operate in closed laboratories, Oreshkov et al. bounded the allowed correlations between the two parties if their order is predefined. By working in a completely DI framework, they showed that these bounds must hold even if a new theory eventually supersedes quantum theory.
Thus a loophole-free violation of such a causal inequality would definitively prove that processes can exist wherein there is no well-defined order between the events. 
They further showed that a quantum process, now known as the Oreshkov-Costa-Brukner (OCB) process, exists that can exceed these correlations.
%
Various follow-up studies have continued to analyze similar causal inequalities \cite{Brukner2015Bounding,Branciard2016simplestCausalInequalities,Abbott2016MultipartiteCausalCorrelations,Oreshkov2016causal,Abbott2017GenuinelyMultipartite,Miklin2017entropicApproach,Wechs2023Existence}.
Although it is possible to write out a process matrix that violates a causal inequality, it is now not yet known how to  experimentally realize such a process.
For example, it is known that the quantum switch cannot be used to violate a causal inequality \cite{Araujo2015Witnessing,Purves2021CannotViolate} and it has been suggested that a physically implementable process should be purifiable \cite{Araujo2017purificationPostulate}, which the OCB and related processes are not.
While the violation of causal inequalities could be simulated using post-selection \cite{Silva2017Connecting,Dimic2020SimulatingICO} this approach would not allow one to make a DI conclusion about the processes' ICO.

The allure of causal inequalities continues to motivate researchers to find alternative approaches to device-independently certify ICO. 
One very recent proposal is based on a strategy in which the control system is entangled with an ancillary spacelike-separated qubit \cite{van2023device,gogioso2023geometry}. This work {(schematically presented in Fig. \ref{fig:ICOdiagram}e)} leverages tools from Bell's theorem and is somewhat inspired by Wigner's friend \cite{bong2020strong}.
It highlights the possibility of transforming broader proofs of Bell nonlocality into the proofs causal nonseparability using quantum switches and potentially other physically implementable processes.
For example, moving beyond the Clauser-Horne-Shimony-Holt
(CHSH) inequality, it has been proposed that Mermin’s proof of nonlocality using GHZ states could be tailored to ICO and provide even stronger DI certification of ICO \cite{van2023possibilistic}.

\section{Applications Exploiting Indefinite Causal Orders}

\begin{figure*}
\centering
\includegraphics[width=\linewidth]{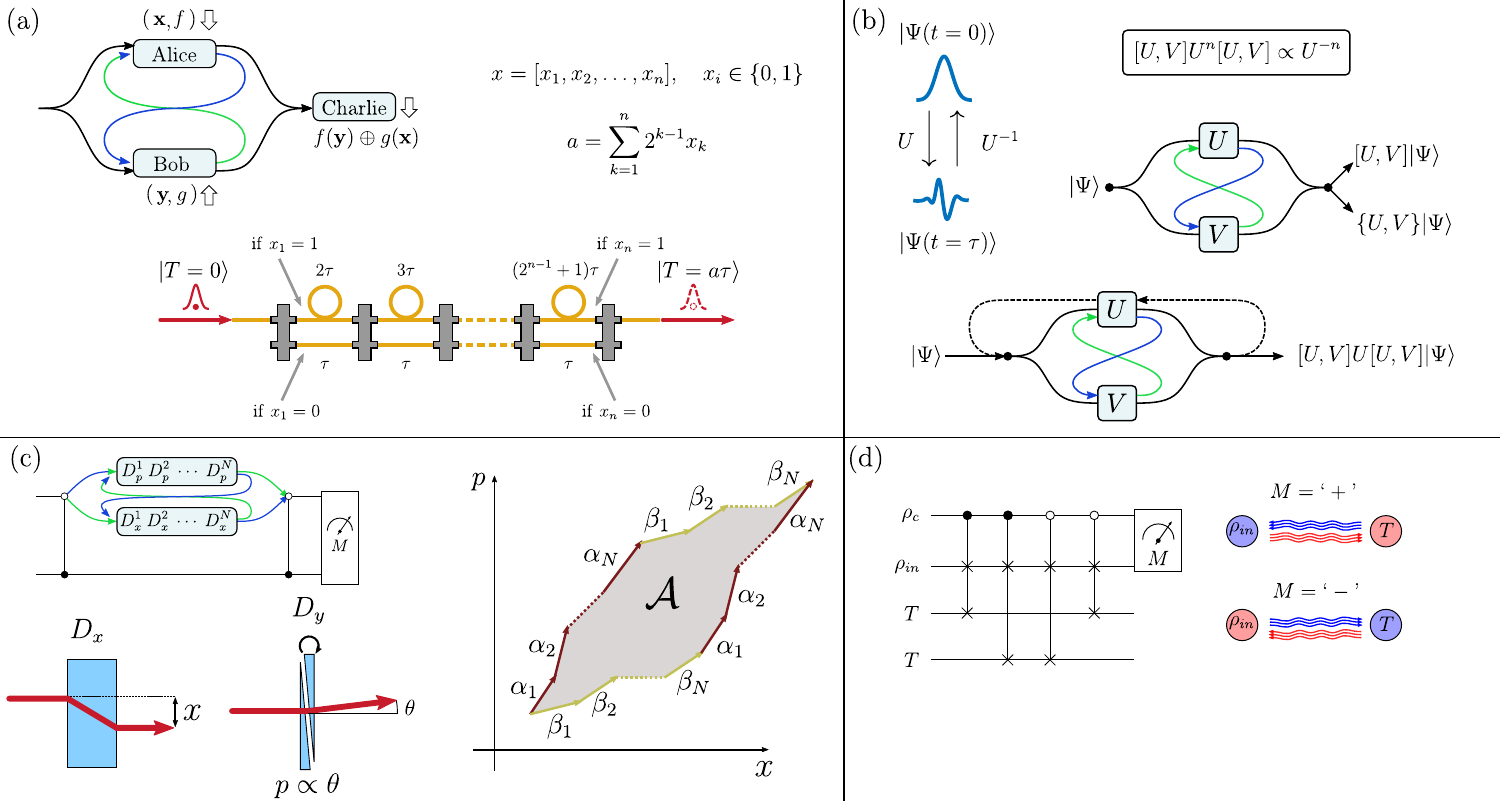}
\caption{\textbf{Experimental applications of ICO processes.} (a) The exponential communication advantage of Ref.~\cite{Guerin2016communicationComplexity} involves operations on a classical bit string, encoded in a quantum system. The realisation of this advantage in Ref.~\cite{wei2019experimentalCommunication} used a high-dimensional time qudit to encode the bit string. Using a sequence of $n$ toggleable fiber delays, $2^n$ different time bins could be realised. (b) In Ref.~\cite{Schiansky22TimeReversal} an ICO process was used to realise the time-reversal of a quantum system. In the experiment, the free evolution of a qubit system was superposed with a perturbed evolution inside a quantum switch. This causes the system to evolve backwards in time, independent of the initial state of the system, as well as its free and perturbed evolution. (c) In Ref.~\cite{yin2023experimentalSuperHeisenberg} an indefinite causal order was used to demonstrate true super-Heisenberg sensing of up to eight independent phase-space displacements. These were realised using the position and momentum displacement of an optical beam, generated by $\mathrm{MgF}_2$ plates and wedge pairs inside the quantum switch. (d) Two thermalizing channels acting in a superposition of orders can bring a system out of thermal equilibrium. This was simulated using a quantum circuit in Refs.~\cite{nei2022NMRswitch,Felce2021IBMswitch}. When post-selecting on the measurement outcome `$+$' (`$-$') at the end of the circuit, heat flows from (to) the input system to (from) a thermal bath.} 
\label{fig:applications}
\end{figure*}

In addition to its fundamental value, an ICO can also lead to advantages in quantum information tasks. One can even establish resource theories for ICO \cite{Taddei2019operationalResource}, just as is done for phenomena such as entanglement \cite{RevModPhys.91.025001}.
Although ICO does not lead to a better computational scaling overall \cite{Araujo2017QuantumComputation}, there are many specific ICO-enhanced applications.

\subsection{Promise Problems}

The first proposal to use an indefinite causal structure as a quantum resource was made by Chiribella in a channel discrimination task \cite{Chiribella2012PerfectDiscrimination}. In this task, two unknown unitary channels are promised to commute or anti-commute and the goal is to determine which relation they obey. 
When the channels are embedded in a causally ordered circuit, perfect discrimination between these two relations requires two queries to at least one of the channels. 
However, it was shown that a quantum superposition of two different causal orders with each channel being used only once could achieve perfect success. In 2015, Lorenzo et al \cite{Procopio2015Experimental}, used a quantum switch, with a path control system and a polarization target system, to distinguish commuting qubit gates from anti-commuting ones. 

This was then generalized to the case involving $N$ unitary gates by Ara\`{u}jo et al. \cite{Araujo2014ComputationalAdvantage}, in a task often called a Fourier promise problem. Ara\`{u}jo et al. showed that an N-switch can distinguish between these $N!$ promises with each gate being queried only once, while the best causal algorithm, through simulation of an N-switch using causally ordered circuits calls the gates $O(N^2)$ times. 
However, this protocol also requires the dimension of the target system to grow as $N$. Furthermore, the advantage was later reassessed by some of the same authors by proposing a more efficient causal algorithm and the results showed that only $O(N\log(N))$ queries were required \cite{Renner2021Reassessing}, reducing the advantage of the N-switch to $\log(N)$. 
Nevertheless, no evidence has so far invalidated the advantage of indefinite causal order in this problem. In any case, the experimental implementation of an N-gate Fourier promise problem is very demanding because it requires the dimensions of both the control and target systems to be at least $N!$. 
In 2021, Taddei et al. \cite{taddei2021computational} addressed both of these challenges by constructing a so-called Hadamard promise problem for four gates. In this game, only a subset of the possible permutations of the four gates are used thus the dimension of the control system could be much less than $N!$. In addition, the black-box unitaries could be chosen as qubit gates, allowing the use of a polarization as the target system. The protocol was demonstrated with a photonic quantum switch in which a 4-dimensional path control and four polarization qubit gates were involved. Shortly after this, Renner and Brukner \cite{renner2022advantage} generalised these Hadamard promise problems to $N$ gates, showing that a best causal quantum algorithm required $O(N\log(N))$ queries to the gates, while using the quantum N-switch only requires each gate to be used once.

\subsection{Communication Complexity}

Communication complexity\cite{RevModPhys.82.665} is an important variant of the standard communication scenario, in which several distributed participants collaborate to calculate a public function that depends on their private input strings. 
It is already well known that quantum resources can yield an exponential advantage in this task \cite{DJproblem,raz1999exponential}, and this is therefore a natural setting to look for ICO advantages. 
In 2014, Baumeler and Wolf \cite{baumeler2014perfectSignaling} found that in a three-party signaling game, the third party can perfectly determine the sum modulo 2 of the other two parties' inputs, which can only succeed with a probability of at most $5/6$ with a causally ordered process. 
Feix et al.\cite{feix2015communicationResource} pointed out that an ICO can provide similar advantages in the $(\log_2 3, 2)$-Hamming game. However, in both of these protocols no general scaling advantage was proposed.

In 2016, Gu\'{e}rin et al.\cite{Guerin2016communicationComplexity} found an exponential scaling advantage of ICO for the two-party exchange evaluation game. 
Here, only one-way communication is permitted and the inputs for each participant contain an $n$-bit string and a private Boolean function.
With the help of ICO, only $n$ qubits of communication are needed to achieve deterministic success, while for causally ordered strategies the required number of qubits of communication grows exponentially as $(2^{n}+n-1)/2$.
The experimental demonstration of this requires a two-party quantum switch acting on an $n$-qubit target system.
A recent experiment by Wei et al.\cite{wei2019experimentalCommunication} has implemented this approach with a target system up to $d=2^{13}$ dimensions (see Fig.\ref{fig:exp}a).
Given the high dimensionality of the target system, this experiment revealed a scaling advantage, reducing the required communication by $(65.2\pm0.3)\%$ and $(30.4\pm0.6)\%$ when compared with any classical strategy and any causally ordered quantum strategy, respectively.

\subsection{Quantum Communication}
\label{sec:app:noise}
In 2018, Ebler et al. realized that it was possible to use the quantum switch to enhance communication rates through noisy channels.
In a series of two papers \cite{Ebler2018EnhancedCommunication,Salek2018QuantumCommunication}, these authors considered communication through two copies of a noisy quantum channel.
In general, if two noisy channels are combined in a fixed causal order, then the net result is a noisier channel; this is formalized in the ``bottleneck inequality'' \cite{wilde2013quantum}.
However, if the channels are combined with an ICO, the overall noise can actually decrease.
This can be achieved for transmitting both classical \cite{Ebler2018EnhancedCommunication} and quantum \cite{Salek2018QuantumCommunication} bits.
For certain noise models even perfect communication can be made possible by the quantum switch \cite{Chiribella2021perfectQuantumCommunication}.
Goswami et al. were the first to experimentally study the transmission of classical information through two completely depolarizing channels\cite{goswami2020IncreasingCommunication}. 
Shortly after this, Yu et al., followed by Rubino et al., experimentally studied ICO enhanced quantum communication \cite{guo2020experimental,Rubino2021Communication}.
These effects seem to contradict standard quantum Shannon theory, which has led to generalizations of the Shannon theory that include ``superposition of trajectories'' \cite{kristjansson2019ResourceTheories,chiribella2019quantumShannon}. 
These results have since been generalized to communication through more than two noisy channels \cite{Procopio2019CommunicationEnhancement,procopio2020sending,Sazim2021ClassicalCommunication}, and even to ICO-enhanced quantum state teleportation in the presence noise \cite{caleffi2020NoiselessCommunications}.

Initial work attributed this effect to ICO, dubbing the effect ``causal activation.''
However, shortly after the initial proposals it was noted that coherence between the target and control system can also lead to noise cancellation \cite{Abbott2020coherentControl, Guerin2019quantumControlledNoise}, which is similar to error filtration \cite{Gisin2005ErrorFiltration}.
In these proposals \cite{Abbott2020coherentControl, Guerin2019quantumControlledNoise}, a photon is placed in a superposition of different trajectories and those trajectories are routed through noisy channels in different manners, but in a fixed order.
Both Pang et al. and Rubino et al. have since studied different noise cancellation protocols experimentally \cite{pang2023experimental,Rubino2021Communication}.
Moreover, similar noise cancellation effects have been proposed in the context of standard quantum computation, wherein correlations with ancillary systems are used to overcome noise in quantum computations \cite{Lee23Error,RamiroSQEM2023,RamiroEnhancing2023}.

While noise reduction may be the most well-known application of ICO to quantum communication,
ICO can also enhance the performance of other quantum communication tasks. Many of which have not yet been experimentally studied. In 2023, Wood proposed an ICO-based QKD protocol\cite{SpencerWood2023qkd}, where the directions of the key shared between Alice and Bob are superposed in a quantum switch. In this QKD protocol, the presence of eavesdroppers can be detected by only measuring the control system, thus
saving a subset of keys that would be consumed during the public comparison phase of standard QKD. 
ICO can also be used to realise the deterministic generation of multipartite entanglement\cite{Koudia2023EntanglementGeneration} and even multipartite higher-dimensional entanglement distribution\cite{Dey2023EntanglementDistribution}. 
The core idea here is to replace multipartite gates with local single-qubit unitary gates applied in an ICO\cite{simonov2023universal}. 
In 2023, Zuo, Hanks, and Kim\cite{Zuo2023EntanglementDistillation} further revealed the advantages of ICO for entanglement distillation. 
In a three-step entanglement distillation protocol, the fidelity of the final entanglement or the success probability can be made higher when two distillation steps are embedded in a quantum switch. 

\subsection{Thermodynamics}

Recent work has shown that ICO can also benefit quantum thermodynamics. In particular, in the construction of quantum refrigeration \cite{felce2020QuantumRefrigeration}, quantum battery charging \cite{chen2021indefinite}, and thermodynamic work extraction \cite{Guha2020Thermodynamic,Guha2022ActivationThermalStates,Simonov2022WorkExtraction}. 
In 2020, Felce and Vedral \cite{felce2020QuantumRefrigeration} discovered that a system can extract or release heat from two identical reservoirs that are placed in a quantum switch. 
An ICO refrigerator was then proposed by combining these ICO-assisted thermalization processes with two classical thermalization processes. The enhancement effect of ICO in work extraction was reported in terms of free energy\cite{Guha2020Thermodynamic} and ergotropy \cite{Guha2022ActivationThermalStates,Simonov2022WorkExtraction}. 
In 2021, Chen and Hasegawa \cite{chen2021indefinite} showed that an ICO-enhanced unitary charging process could achieve full charging.
This has been generalized to nonunitary charging protocols and was experimentally demonstrated using a photonic quantum switch\cite{zhu2023prl}. 
{Therein it was shown that ICO can boost the efficiency of a quantum-battery charging process, as well as give rise to an anomalous effect whereby weaker chargers exhibit higher efficiency.} 
In addition to enhancements based on the quantum switch, quantum thermodynamics with indefinite time directions has also been theoretically \cite{Rubino2021superpositionThermodynamicEvolutions} and experimentally \cite{Rubino2022InferringWork} investigated.

ICO-enhanced thermodynamics studies the heat or workflow between the reservoirs when the reservoirs are placed in the quantum switch. 
In 2022, Cao et al.\cite{cao2022quantumSimulation} reported a photonic implementation of ICO-induced quantum refrigeration, in which a photonic polarization qubit serves to interact with two reservoirs in the ICO. 
The reservoirs were achieved by constructing thermal channels to simulate their interactions with the working system. 
The temperature and energy change were defined as the changes in the population and population of photon in the horizontal and vertical polarization states \cite{Mancino17quantum,Mancino18geometrical}. 
ICO-enhanced quantum refrigeration was also investigated by Nie et al.\cite{nei2022NMRswitch} on a nuclear magnetic resonance (NMR) system (see Fig. \ref{fig:applications}d). In this work, three nuclear spins acted as a qubit and two reservoirs, while a fourth spin was used as a control system. The equivalent four-qubit unitary quantum circuit proposed in Ref.~\cite{felce2020QuantumRefrigeration} was implemented. A similar approach was implemented on an IBM cloud quantum computer~\cite{Felce2021IBMswitch}.  
See also Ref.~\cite{Ball2022fridge} for a summary of ICO-enhanced quantum refrigeration. 

Recently, Capela et al.\cite{Capela2023ReassessingThermodynamic} pointed out that a particular causally ordered non-Markovian process can give an enhancement that is similar to ICO-based proposals \cite{Guha2020Thermodynamic,Simonov2022WorkExtraction}. As we have discussed in more detail in Sec. \ref{sec:app:noise}, the issue is how the control system interacts with the protocol. In some scenarios, coherence between the control and target systems can lead to very similar advantages even if all the operations are applied in a definite causal order. 
Analogous effects have been investigated in ICO-enhanced quantum communication by Liu et al. \cite{Liu23thermodynamics} and the results revealed that an increase in information capacity consumes a thermodynamic resource\teo{:} the free energy of coherence associated with the control system of quantum switch.

\subsection{Quantum Metrology}

ICO can also enhance quantum metrology, which is one of the most recent prominent applications of quantum information. 
In 2019, Frey\cite{Frey2019depolarizingChannelIdentification} showed that ICO can enhance the precision in identifying the properties of a depolarizing channel.
ICO was further used in noisy metrology tasks where other noisy channels are involved in qubit unitary channel estimation games \cite{Ban2023FisherInformation, Chapeau2021NoisyQuantumMetrology}. 
In general, complex measurements are needed to achieve the optimal estimation strategy. 
With ICO, however, one can achieve a near optimal precision with a simple projective measurement on the control qubit. 
This was demonstrated in a photonic experiment where a phase flip channel and a depolarizing channel were used\cite{Min23}. In 2023, multiparameter quantum estimation enhanced by ICO was proposed by Delgado \cite{Delgado2023EstimationPauliChannels}. Despite this progress, Kurdzia\l{}ek et al.\cite{Kurdzialek23} proved that, within discrete-variable systems, ICO could only achieve the Heisenberg scaling, which is the fundamental limit of the precision in quantum metrology.

Achieving scaling beyond the Heisenberg limit has been a long-standing goal of quantum metrology.
An ICO-enhanced metrology task surpassing the Heisenberg limit was proposed in 2020 by Zhao et al.\cite{zhao2019QuantumMetrology}.
The authors proved that quantum switch can help to achieve super-Heisenberg scaling when estimating the product of CV position and momentum displacements of a harmonic oscillator.  
Although the indefinite causal structure for CV systems has been theoretically analyzed\cite{Giacomini_2016}, its experimental implementation requires delicate quantum engineering. 
In 2023, Yin et al.~\cite{yin2023experimentalSuperHeisenberg} realized super-Heisenberg scaling in a photonic system (see Fig. \ref{fig:exp}c). In this experiment, the position $x$ and momentum $p$ displacements were defined in the transverse direction of a single photon and their product was the geometric phase in the $xp$-phase space. Approximately \SI{18.6}{\micro\meter} of position displacement and approximately \SI{2.8e-4}{\radian} of momentum displacement were prepared to ensure a tiny geometric phase ($0.042$). With at most $N=8$ pairs of displacements used in the quantum switch, the experiment observed an RMSE of the geometric phase with a scaling $\propto1/cN^2$ ($c\approx30.65$), showing significant evidence of the super-Heisenberg limit.

\subsection{Other Applications}

ICO can also benefit the way we investigate other phenomena in the foundations of quantum mechanics.
Here we briefly summarize a variety of proposals on this topic.

In 2023, Ban\cite{Ban2023Quantumness} showed that ICO could suppress the decoherence effect of reservoirs on a quantum system, thus reducing the decay of its quantumness, including coherence, temporal steering, entanglement, and Bell nonlocality. Gao et al. \cite{Gao2023MeasuringIncompatibility} showed how to use a quantum switch to efficiently estimate the incompatibility of quantum measurements and further applied this protocol to cluster quantum observables with an unsupervised algorithm. Pan\cite{Pan2023LeggettGarg} showed that the quantum switch can help to close the clumsiness loophole in the Leggett-Garg test, thus providing a loophole-free protocol to examine the boundary between classical worlds and quantum ones. In 2022, Krumm et al.\cite{Krumm2022teleportation} proposed the teleportation of a quantum causal structure by simply teleporting all its inputs and outputs. 
ICO was also proven to be advantageous in channel discrimination tasks where two non-signaling bipartite channels\cite{Chiribella2012PerfectDiscrimination} or even multiple copies of arbitrary unitary channels\cite{Bavaresco2021StrictHierarchy,bavaresco2022UnitaryChannelDiscrimination} were involved. None of the above mentioned protocols have been experimentally implemented.

Another exciting example, that has been experimentally studied, is universal quantum rewinding. The goal of quantum rewinding is to reset a quantum system subjected to an unknown evolution to its state at an earlier time. In 2019, Quintino et al.\cite{quintino19Reversing} showed how to implement the exact inverse of a unitary operation with a failure probability that decays exponentially in the copies of operation used. 
Numerical evidence suggests that ICOs could achieve a higher probability of success than circuits based on parallelized and adaptive strategies. In 2023, a more practical rewinding protocol was presented \cite{Trillo2023universal} and experimentally demonstrated\cite{Schiansky22TimeReversal}. 
The core idea is to utilize the non-commutativity of an unknown target process and an auxiliary process, which makes it well-suited for the quantum switch, which can distinguish commuting from anti-commuting gates. Moreover, the probability of success can be boosted arbitrarily close to one by cascading quantum switches (see Fig.\ref{fig:exp}b).
In the experiment, electro-optical devices and time synchronization techniques were used to quickly route photons through the quantum switch. The experiment observed an average reversal fidelity of over $95\%$.

\section{Loopholes and Criticisms of Current Experimental Demonstrations}
\label{sec:loopholes}
\begin{figure*}
\centering
\includegraphics[width=\linewidth]{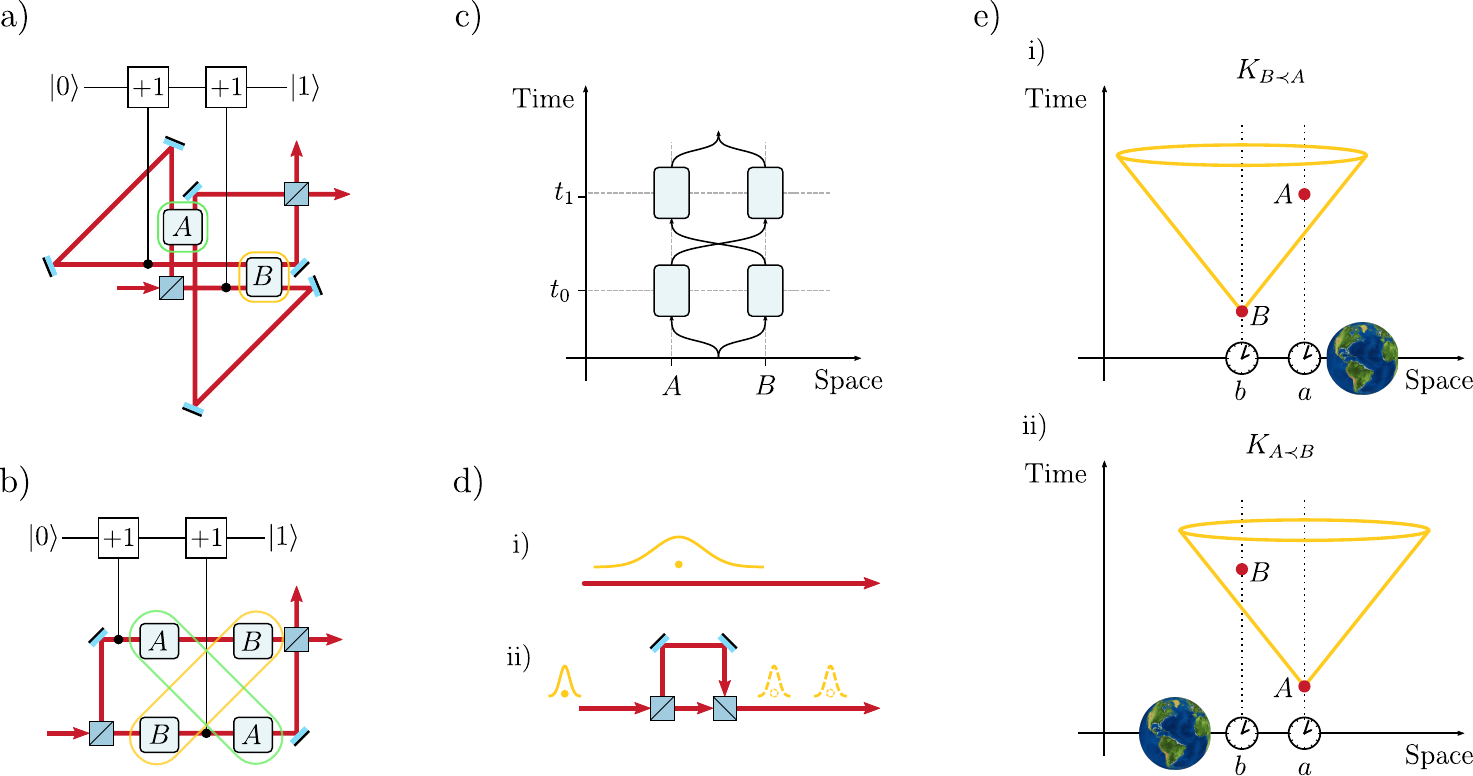}
\caption{Experimental Loopholes. (a) The quantum switch with a path control system, and a hypothetical counter system {(the upper system initialized in $\ket{0}$)} that is used to count the number of gate uses. After the experiment is run, the counter system is left in the state $\ket{1}$, indicating that the gate $A$ has only been used once.
(b) The so-called unfolded switch. Using two identical copies of the optical elements used to implement gates A and B, one can use a Mach-Zhender geometry to build a device that produces the same statistics as the quantum switch.  Even the {upper} counter system, will behave the same. 
However, in this device one cannot associate a single spatial region with gate A or B.
(c) Space-time diagram of the experimentally implemented quantum switch. At time $t_0$ the photon is in a superposition of being at both gates. At time $t_1$ it is still in a superposition of both locations, but swapped with respect to $t_0$. Even though the ``counter-arguement'' can be used to conclude that the photon only interacts with each gate once, the four distinct space-time events have led some to conclude that the quantum switch does not have a genuine indefinite causal order.
(d) Time-delocalized events. In practice, a quantum photon can never be perfectly localized in time. i) A photon must exist with some extended wavepacket, implying that it traverses an optical element in a superposition of different times. ii) Using an unbalanced interferometer a photon is placed in superposition of two time bins, and thus travels in a superposition of being centered at two distinct times.
(e) The gravitational quantum switch, recreated from Ref.~\cite{zych2019bell}. By placing a massive object in a quantum superposition one can implement the quantum switch.  In the branch of the superposition with the massive object on the left (right) the photon experiences event A (B) before event B (A).}
\label{fig:loopholes}
\end{figure*}

Although the field of experimental ICO has been extremely successful, there have been several noteworthy criticisms of the experimental implementations. Nevertheless, experimental work has continued to be defended by strong theoretical arguments from many in the ICO community \cite{oreshkov2019timeDelocalized,DeLaHemette2022quantumDiffeomorphisms,fellous2022comparing}.  Here we will briefly summarize this discussion from an experimental point of view.

The first criticism to appear in the literature can be considered as a `multiple use problem.' It states that since two paths traverse each optic, experimental quantum switches (Fig. \ref{fig:loopholes} a) are equivalent to an ``unfolded switch'' \cite{Maclean2017mixturesOfCausalRelations} (Fig. \ref{fig:loopholes} b).
Thus, quantum switch experiments use each gate twice.
One can attempt to simply count the number of times the photon traverses each optic \cite{Araujo2014ComputationalAdvantage,Procopio2015Experimental}.
However, as we discuss in Appendix B, counting gate uses in experimental settings is quite subtle, and most attempts to operationally count the gate uses support the point of view that the experiments only use each gate once.


The other major criticism can be termed the `multiple event problem' \cite{paunkovic2020distinguishing,ormrod2022sectorialConstraints,vilasini2022embedding}, illustrated in Fig \ref{fig:loopholes} c). It can be summarized as follows. In the experiment with two gates, each experiment consists of two time steps.  At time $t_1$, the photon is in a superposition of traversing gate 1 and gate 2, and at time $t_2$ it is in a superposition of traversing gate 2 and gate 1.
This corresponds to four different space-time events.
However, just as we saw that a photon can be in a superposition of spatial modes, a photon can be in a superposition of temporal modes.
In fact, photons are always in some form of a wave packet with a temporal extent \cite{SalehFundamentals} (Fig. \ref{fig:loopholes} d i).
When this delocalized particle passes some device, it seems more natural, to think of time-delocalized events.
In other words, the photon traverses the device in a superposition of times.
If this is the case for a photon in an extended wavefunction (Fig. \ref{fig:loopholes} d i), it also applies to a photon in a superposition of discrete times (as sketched in Fig \ref{fig:loopholes} ii). 
The full formal argument is presented in \cite{oreshkov2019timeDelocalized}.
From an experimental point of view, Goswami et al. used this `temporal uncertainty' to their advantage, using photons with a coherence time longer than the propagation time between the two gates \cite{goswami2018Indefinite}.
Thus, quantum mechanically, in their work it is impossible to define four unambiguous space-time events.
A different approach to overcome this critique has also been proposed by Felce et al. \cite{Felce2022superpositionsInTime}, where the spontaneous decay from two excited atoms could be used to implement the quantum switch in a table-top setting with only two space-time events. 

In spite of these arguments, some researchers maintain that photonic experiments performed to-date do not implement a true ICO \cite{paunkovic2020distinguishing}.
Vilasini and Renner argue that because the photon is in a superposition of different space-time locations, the gates must be applied not only on the photon itself but also on the vacuum system, this admits an alternative ``fine-grained'' causal order which does not have an ICO \cite{vilasini2022embedding}.
A very similar fine-graining argument was presented by Ormrod et al. 
\cite{ormrod2022sectorialConstraints}.
Another common theme among critiques of experimental implementations is that perhaps the gravitational quantum switch is required to implement a true ICO
\cite{paunkovic2020distinguishing,ormrod2022sectorialConstraints,vilasini2022embedding}.

The gravitational quantum switch was proposed for realising an ICO by creating a superposition of massive objects \cite{zych2019bell}.
Consider two events, A and B, placed between a massive object that is in a superposition being to the left of B (panel e ii) or to the right of A (panel e i).
Due to gravitational time dilation, time passes more slowly closer to the massive object, so in one case event B will occur first (panel e i), while in the other case event A will occur first (panel e ii).
This superposition of masses can thus be used to realize a quantum switch.
It was recently proven that the gravitational quantum switch and the photonic quantum switch are completely equivalent in terms of the causal order between the events \cite{DeLaHemette2022quantumDiffeomorphisms}. 

Given the discussion surrounding experiments, device- and theory-independent techniques to demonstrate ICO are clearly needed.  
As we discussed in Sec. \ref{Sec:Characterizing}, progress towards this goal is being made both on the experimental and theoretical fronts.
Indeed, a recent proposal for a device-independent verification,
\cite{van2023device} makes a significant step.
However, the identification of all experimental loopholes and finding a way to close them experimentally is still a work in progress.
One of the most significant loopholes is that both superposed parties act in a ``closed laboratory''  \cite{oreshkov2012quantum}.
This assumption is used to exclude trivial communication between the two parties; in other words, \hypersetup{linkcolor=.}\hyperlink{assumption3}{assumption (3)} essentially constrains the signaling direction. That is, one-way signaling is allowed, while bidirectional signaling is forbidden. 
For a fixed causal order, if Alice (Bob) acts first, then Bob's (Alice's) input can be perfectly correlated with Alice's (Bob's) output, but there can be no correlations between Alice's input and Bob's output.
In a standard Bell-like scenario, this could be enforced in a space-like way by separating the two parties.
However, in the quantum switch this is not possible since the two parties must exchange a quantum system.
Instead, we could attempt to enforce this by only allowing each party to act exactly once.
However, given the difficulty in even defining what it means for a party to act (or a gate to be applied) it is not clear how to verify this assumption experimentally, let alone in a device-independent way.

\section{Challenges and Future Perspectives}
\label{Sec:Challenges}
We have seen that the field of indefinite causal order is an exciting research topic that lies firmly at the intersection of quantum foundations and quantum information \cite{cavalcanti2023fresh}.
{This has resulted in a variety of innovative implementations of ICO processes, motivated by foundational questions and new quantum applications.}
In this Review, we have separated the experimental research into two parts: work attempting to verify ICO (Sec. 3) and work using ICO for applications (Sec. 4).
In both cases, the amount of theoretical research and number of proposals significantly outweigh the experimental tests and implementations, leaving an ample playground for experimentalists.

On one side, in order for the quantum switch to prove truly useful for applications, it should be scaled up.
Depending on the application, the dimensions of the target system, the control system, or both must be increased.
While the target scaling has been demonstrated with a $2^{12}$-dimensional system \cite{wei2019experimentalCommunication}, scaling up the control system remains a challenge.
In all current experimental implementations, each additional superposed causal order requires the dimension of the control system to increase by 1. Since the complete N-switch superposes $N!$ orders this means that $N!$ modes must traverse each optical element.
This difficulty has limited the number of experimentally superposed causal orders to four \cite{taddei2021computational}.

In order to overcome this challenge one could take two approaches. First, one could take the brute force approach and simply use a control DoF which may be easier to engineer. To this end, the approach of using a time-bin control may have some advantages.  While experimental demonstrations have thus far been limited to a two-gates  \cite{Antesberger2023tomography}, an N-gate switch proposal has been worked out \cite{rambo2016functional}.
This does not, however, deal with the scaling challenge: In fact, that proposal requires $N^N$ time bins to implement the full quantum N-switch.
Instead, a scalable (albeit more difficult) approach could require a multi-qubit register encoded in a different physical system to encode the control. In this case, a multi-qubit state would select the order to be implemented, and thus the physical resources could scale much more favourably in terms of the number of orders to be superposed.
However, if the target system is encoded in a photon, single-photon level nonlinearities are needed, which are still challenging for photonic experiments.
Nevertheless, this may be a promising new path for the next generation of ICO experiments.

In Sec. 5, we discussed various criticisms of experimental implementations of the quantum switch.
When it comes to applications, if a process that simulates an ICO or is inspired by an ICO can outperform standard techniques this could still lead to some practical advantages, as in recent ICO-inspired noise-cancellation proposals \cite{Lee23Error,RamiroSQEM2023,RamiroEnhancing2023}. However, this is of more concern when one claims that the advantage can \textit{only} be achieved with an ICO, as it may obfuscate the underlying physics.
These issues are, of course, of utmost importance for work verifying ICO.
In this respect, we believe that the push to verify ICO using device-independent techniques is necessary.
To this end, a recent theory has led to semi-device independent experimental tests of ICO \cite{Rubino2022experimentalEntanglement,cao2022Semideviceindependent}.
Moreover, recent proposals are opening the door for other types of experimental device-independent tests of ICO \cite{van2023device}.
Of course, any future experiments making such bold claims will need to be carefully examined for loopholes, with the closed laboratory assumption\cite{oreshkov2012quantum} likely being the most important.
Although a better theoretical understanding is still required, the current situation is not unlike the use of Bell's inequalities to verify entanglement before the seminal set of 2015 experiments \cite{Hensen2015Loophole,Giustina2015Significant, Shalm2015strong}.
Thus, we believe that there is much exciting and ground-breaking work that remains to be done in the burgeoning field of indefinite causal order in quantum mechanics.

\section*{Acknowledgements}
This research was funded in whole, or in part, by the European Union (ERC, GRAVITES, No 101071779) and its  Horizon 2020  and Horizon Europe research and innovation programme under grant agreement No 899368 (EPIQUS), and No 101135288 (EPIQUE), the Marie Skłodowska-Curie grant agreement No 956071 (AppQInfo). Further funding was received from the Austrian Science Fund (FWF) through 10.55776/COE1 (Quantum Science Austria), 10.55776/F71 (BeyondC) and  10.55776/FG5 (Research Group 5); and from AFOSR via FA9550-21- 1-0355 (QTRUST), and FA8655-23-1-7063 (TIQI); from the Austrian Federal Ministry for Digital and Economic Affairs, the National Foundation for Research, Technology and Development and the Christian Doppler Research Association. BHL and YG were supported by NSFC (No.~12374338 and No.~12204458) and China Postdoctoral Science Foundation (2021M700138 and BX2021289).

\bibliography{main}

\begin{thebibliography}{159}%
\makeatletter
\providecommand \@ifxundefined [1]{%
 \@ifx{#1\undefined}
}%
\providecommand \@ifnum [1]{%
 \ifnum #1\expandafter \@firstoftwo
 \else \expandafter \@secondoftwo
 \fi
}%
\providecommand \@ifx [1]{%
 \ifx #1\expandafter \@firstoftwo
 \else \expandafter \@secondoftwo
 \fi
}%
\providecommand \natexlab [1]{#1}%
\providecommand \enquote  [1]{``#1''}%
\providecommand \bibnamefont  [1]{#1}%
\providecommand \bibfnamefont [1]{#1}%
\providecommand \citenamefont [1]{#1}%
\providecommand \href@noop [0]{\@secondoftwo}%
\providecommand \href [0]{\begingroup \@sanitize@url \@href}%
\providecommand \@href[1]{\@@startlink{#1}\@@href}%
\providecommand \@@href[1]{\endgroup#1\@@endlink}%
\providecommand \@sanitize@url [0]{\catcode `\\12\catcode `\$12\catcode `\&12\catcode `\#12\catcode `\^12\catcode `\_12\catcode `\%12\relax}%
\providecommand \@@startlink[1]{}%
\providecommand \@@endlink[0]{}%
\providecommand \url  [0]{\begingroup\@sanitize@url \@url }%
\providecommand \@url [1]{\endgroup\@href {#1}{\urlprefix }}%
\providecommand \urlprefix  [0]{URL }%
\providecommand \Eprint [0]{\href }%
\providecommand \doibase [0]{https://doi.org/}%
\providecommand \selectlanguage [0]{\@gobble}%
\providecommand \bibinfo  [0]{\@secondoftwo}%
\providecommand \bibfield  [0]{\@secondoftwo}%
\providecommand \translation [1]{[#1]}%
\providecommand \BibitemOpen [0]{}%
\providecommand \bibitemStop [0]{}%
\providecommand \bibitemNoStop [0]{.\EOS\space}%
\providecommand \EOS [0]{\spacefactor3000\relax}%
\providecommand \BibitemShut  [1]{\csname bibitem#1\endcsname}%
\let\auto@bib@innerbib\@empty
\bibitem [{\citenamefont {{Ried}}\ \emph {et~al.}(2015)\citenamefont {{Ried}}, \citenamefont {{Agnew}}, \citenamefont {{Vermeyden}}, \citenamefont {{Janzing}}, \citenamefont {{Spekkens}},\ and\ \citenamefont {{Resch}}}]{Ried2015inferringCausal}%
  \BibitemOpen
  \bibfield  {author} {\bibinfo {author} {\bibfnamefont {K.}~\bibnamefont {{Ried}}}, \bibinfo {author} {\bibfnamefont {M.}~\bibnamefont {{Agnew}}}, \bibinfo {author} {\bibfnamefont {L.}~\bibnamefont {{Vermeyden}}}, \bibinfo {author} {\bibfnamefont {D.}~\bibnamefont {{Janzing}}}, \bibinfo {author} {\bibfnamefont {R.~W.}\ \bibnamefont {{Spekkens}}},\ and\ \bibinfo {author} {\bibfnamefont {K.~J.}\ \bibnamefont {{Resch}}},\ }\bibfield  {title} {\bibinfo {title} {{A quantum advantage for inferring causal structure}},\ }\href {https://doi.org/10.1038/nphys3266} {\bibfield  {journal} {\bibinfo  {journal} {Nature Physics}\ }\textbf {\bibinfo {volume} {11}},\ \bibinfo {pages} {414} (\bibinfo {year} {2015})},\ \Eprint {https://arxiv.org/abs/1406.5036} {arXiv:1406.5036 [quant-ph]} \BibitemShut {NoStop}%
\bibitem [{\citenamefont {{Maclean}}\ \emph {et~al.}(2017)\citenamefont {{Maclean}}, \citenamefont {{Ried}}, \citenamefont {{Spekkens}},\ and\ \citenamefont {{Resch}}}]{Maclean2017mixturesOfCausalRelations}%
  \BibitemOpen
  \bibfield  {author} {\bibinfo {author} {\bibfnamefont {J.-P.~W.}\ \bibnamefont {{Maclean}}}, \bibinfo {author} {\bibfnamefont {K.}~\bibnamefont {{Ried}}}, \bibinfo {author} {\bibfnamefont {R.~W.}\ \bibnamefont {{Spekkens}}},\ and\ \bibinfo {author} {\bibfnamefont {K.~J.}\ \bibnamefont {{Resch}}},\ }\bibfield  {title} {\bibinfo {title} {{Quantum-coherent mixtures of causal relations}},\ }\href {https://doi.org/10.1038/ncomms15149} {\bibfield  {journal} {\bibinfo  {journal} {Nature Communications}\ }\textbf {\bibinfo {volume} {8}},\ \bibinfo {eid} {15149} (\bibinfo {year} {2017})},\ \Eprint {https://arxiv.org/abs/1606.04523} {arXiv:1606.04523 [quant-ph]} \BibitemShut {NoStop}%
\bibitem [{\citenamefont {{Carvacho}}\ \emph {et~al.}(2017)\citenamefont {{Carvacho}}, \citenamefont {{Andreoli}}, \citenamefont {{Santodonato}}, \citenamefont {{Bentivegna}}, \citenamefont {{Chaves}},\ and\ \citenamefont {{Sciarrino}}}]{Carvacho2017Carvacho}%
  \BibitemOpen
  \bibfield  {author} {\bibinfo {author} {\bibfnamefont {G.}~\bibnamefont {{Carvacho}}}, \bibinfo {author} {\bibfnamefont {F.}~\bibnamefont {{Andreoli}}}, \bibinfo {author} {\bibfnamefont {L.}~\bibnamefont {{Santodonato}}}, \bibinfo {author} {\bibfnamefont {M.}~\bibnamefont {{Bentivegna}}}, \bibinfo {author} {\bibfnamefont {R.}~\bibnamefont {{Chaves}}},\ and\ \bibinfo {author} {\bibfnamefont {F.}~\bibnamefont {{Sciarrino}}},\ }\bibfield  {title} {\bibinfo {title} {{Experimental violation of local causality in a quantum network}},\ }\href {https://doi.org/10.1038/ncomms14775} {\bibfield  {journal} {\bibinfo  {journal} {Nature Communications}\ }\textbf {\bibinfo {volume} {8}},\ \bibinfo {eid} {14775} (\bibinfo {year} {2017})},\ \Eprint {https://arxiv.org/abs/1610.03327} {arXiv:1610.03327 [quant-ph]} \BibitemShut {NoStop}%
\bibitem [{\citenamefont {{Hardy}}(2007)}]{hardy2007quantumGravity}%
  \BibitemOpen
  \bibfield  {author} {\bibinfo {author} {\bibfnamefont {L.}~\bibnamefont {{Hardy}}},\ }\bibfield  {title} {\bibinfo {title} {{Towards quantum gravity: a framework for probabilistic theories with non-fixed causal structure}},\ }\href {https://doi.org/10.1088/1751-8113/40/12/S12} {\bibfield  {journal} {\bibinfo  {journal} {Journal of Physics A Mathematical General}\ }\textbf {\bibinfo {volume} {40}},\ \bibinfo {pages} {3081} (\bibinfo {year} {2007})},\ \Eprint {https://arxiv.org/abs/gr-qc/0608043} {arXiv:gr-qc/0608043 [gr-qc]} \BibitemShut {NoStop}%
\bibitem [{\citenamefont {Hardy}(2009)}]{hardy2009quantumGravComp}%
  \BibitemOpen
  \bibfield  {author} {\bibinfo {author} {\bibfnamefont {L.}~\bibnamefont {Hardy}},\ }\bibfield  {title} {\bibinfo {title} {Quantum gravity computers: On the theory of computation with indefinite causal structure},\ }in\ \href {https://doi.org/10.1007/978-1-4020-9107-0_21} {\emph {\bibinfo {booktitle} {Quantum reality, relativistic causality, and closing the epistemic circle}}}\ (\bibinfo  {publisher} {Springer},\ \bibinfo {year} {2009})\ pp.\ \bibinfo {pages} {379--401}\BibitemShut {NoStop}%
\bibitem [{\citenamefont {{Chiribella}}\ \emph {et~al.}(2013)\citenamefont {{Chiribella}}, \citenamefont {{D'Ariano}}, \citenamefont {{Perinotti}},\ and\ \citenamefont {{Valiron}}}]{Chiribella2013quantum}%
  \BibitemOpen
  \bibfield  {author} {\bibinfo {author} {\bibfnamefont {G.}~\bibnamefont {{Chiribella}}}, \bibinfo {author} {\bibfnamefont {G.~M.}\ \bibnamefont {{D'Ariano}}}, \bibinfo {author} {\bibfnamefont {P.}~\bibnamefont {{Perinotti}}},\ and\ \bibinfo {author} {\bibfnamefont {B.}~\bibnamefont {{Valiron}}},\ }\bibfield  {title} {\bibinfo {title} {{Quantum computations without definite causal structure}},\ }\href {https://doi.org/10.1103/PhysRevA.88.022318} {\bibfield  {journal} {\bibinfo  {journal} {Phys. Rev. A}\ }\textbf {\bibinfo {volume} {88}},\ \bibinfo {eid} {022318} (\bibinfo {year} {2013})},\ \Eprint {https://arxiv.org/abs/0912.0195} {arXiv:0912.0195 [quant-ph]} \BibitemShut {NoStop}%
\bibitem [{\citenamefont {{Procopio}}\ \emph {et~al.}(2015)\citenamefont {{Procopio}}, \citenamefont {{Moqanaki}}, \citenamefont {{Ara{\'u}jo}}, \citenamefont {{Costa}}, \citenamefont {{Alonso Calafell}}, \citenamefont {{Dowd}}, \citenamefont {{Hamel}}, \citenamefont {{Rozema}}, \citenamefont {{Brukner}},\ and\ \citenamefont {{Walther}}}]{Procopio2015Experimental}%
  \BibitemOpen
  \bibfield  {author} {\bibinfo {author} {\bibfnamefont {L.~M.}\ \bibnamefont {{Procopio}}}, \bibinfo {author} {\bibfnamefont {A.}~\bibnamefont {{Moqanaki}}}, \bibinfo {author} {\bibfnamefont {M.}~\bibnamefont {{Ara{\'u}jo}}}, \bibinfo {author} {\bibfnamefont {F.}~\bibnamefont {{Costa}}}, \bibinfo {author} {\bibfnamefont {I.}~\bibnamefont {{Alonso Calafell}}}, \bibinfo {author} {\bibfnamefont {E.~G.}\ \bibnamefont {{Dowd}}}, \bibinfo {author} {\bibfnamefont {D.~R.}\ \bibnamefont {{Hamel}}}, \bibinfo {author} {\bibfnamefont {L.~A.}\ \bibnamefont {{Rozema}}}, \bibinfo {author} {\bibfnamefont {{\v{C}}.}~\bibnamefont {{Brukner}}},\ and\ \bibinfo {author} {\bibfnamefont {P.}~\bibnamefont {{Walther}}},\ }\bibfield  {title} {\bibinfo {title} {{Experimental superposition of orders of quantum gates}},\ }\href {https://doi.org/10.1038/ncomms8913} {\bibfield  {journal} {\bibinfo  {journal} {Nature Communications}\ }\textbf {\bibinfo {volume} {6}},\ \bibinfo {eid} {7913} (\bibinfo {year} {2015})},\ \Eprint
  {https://arxiv.org/abs/1412.4006} {arXiv:1412.4006 [quant-ph]} \BibitemShut {NoStop}%
\bibitem [{\citenamefont {{Colnaghi}}\ \emph {et~al.}(2012)\citenamefont {{Colnaghi}}, \citenamefont {{D'Ariano}}, \citenamefont {{Facchini}},\ and\ \citenamefont {{Perinotti}}}]{Colnaghi2012quantumComputation}%
  \BibitemOpen
  \bibfield  {author} {\bibinfo {author} {\bibfnamefont {T.}~\bibnamefont {{Colnaghi}}}, \bibinfo {author} {\bibfnamefont {G.~M.}\ \bibnamefont {{D'Ariano}}}, \bibinfo {author} {\bibfnamefont {S.}~\bibnamefont {{Facchini}}},\ and\ \bibinfo {author} {\bibfnamefont {P.}~\bibnamefont {{Perinotti}}},\ }\bibfield  {title} {\bibinfo {title} {{Quantum computation with programmable connections between gates}},\ }\href {https://doi.org/10.1016/j.physleta.2012.08.028} {\bibfield  {journal} {\bibinfo  {journal} {Physics Letters A}\ }\textbf {\bibinfo {volume} {376}},\ \bibinfo {pages} {2940} (\bibinfo {year} {2012})},\ \Eprint {https://arxiv.org/abs/1109.5987} {arXiv:1109.5987 [quant-ph]} \BibitemShut {NoStop}%
\bibitem [{\citenamefont {{Oreshkov}}\ \emph {et~al.}(2012)\citenamefont {{Oreshkov}}, \citenamefont {{Costa}},\ and\ \citenamefont {{Brukner}}}]{oreshkov2012quantum}%
  \BibitemOpen
  \bibfield  {author} {\bibinfo {author} {\bibfnamefont {O.}~\bibnamefont {{Oreshkov}}}, \bibinfo {author} {\bibfnamefont {F.}~\bibnamefont {{Costa}}},\ and\ \bibinfo {author} {\bibfnamefont {{\v{C}}.}~\bibnamefont {{Brukner}}},\ }\bibfield  {title} {\bibinfo {title} {{Quantum correlations with no causal order}},\ }\href {https://doi.org/10.1038/ncomms2076} {\bibfield  {journal} {\bibinfo  {journal} {Nature Communications}\ }\textbf {\bibinfo {volume} {3}},\ \bibinfo {eid} {1092} (\bibinfo {year} {2012})},\ \Eprint {https://arxiv.org/abs/1105.4464} {arXiv:1105.4464 [quant-ph]} \BibitemShut {NoStop}%
\bibitem [{\citenamefont {{Ara{\'u}jo}}\ \emph {et~al.}(2015)\citenamefont {{Ara{\'u}jo}}, \citenamefont {{Branciard}}, \citenamefont {{Costa}}, \citenamefont {{Feix}}, \citenamefont {{Giarmatzi}},\ and\ \citenamefont {{Brukner}}}]{Araujo2015Witnessing}%
  \BibitemOpen
  \bibfield  {author} {\bibinfo {author} {\bibfnamefont {M.}~\bibnamefont {{Ara{\'u}jo}}}, \bibinfo {author} {\bibfnamefont {C.}~\bibnamefont {{Branciard}}}, \bibinfo {author} {\bibfnamefont {F.}~\bibnamefont {{Costa}}}, \bibinfo {author} {\bibfnamefont {A.}~\bibnamefont {{Feix}}}, \bibinfo {author} {\bibfnamefont {C.}~\bibnamefont {{Giarmatzi}}},\ and\ \bibinfo {author} {\bibfnamefont {{\v{C}}.}~\bibnamefont {{Brukner}}},\ }\bibfield  {title} {\bibinfo {title} {{Witnessing causal nonseparability}},\ }\href {https://doi.org/10.1088/1367-2630/17/10/102001} {\bibfield  {journal} {\bibinfo  {journal} {New Journal of Physics}\ }\textbf {\bibinfo {volume} {17}},\ \bibinfo {eid} {102001} (\bibinfo {year} {2015})},\ \Eprint {https://arxiv.org/abs/1506.03776} {arXiv:1506.03776 [quant-ph]} \BibitemShut {NoStop}%
\bibitem [{\citenamefont {{Rubino}}\ \emph {et~al.}(2017)\citenamefont {{Rubino}}, \citenamefont {{Rozema}}, \citenamefont {{Feix}}, \citenamefont {{Ara{\'u}jo}}, \citenamefont {{Zeuner}}, \citenamefont {{Procopio}}, \citenamefont {{Brukner}},\ and\ \citenamefont {{Walther}}}]{rubino2017ExperimentalVerification}%
  \BibitemOpen
  \bibfield  {author} {\bibinfo {author} {\bibfnamefont {G.}~\bibnamefont {{Rubino}}}, \bibinfo {author} {\bibfnamefont {L.~A.}\ \bibnamefont {{Rozema}}}, \bibinfo {author} {\bibfnamefont {A.}~\bibnamefont {{Feix}}}, \bibinfo {author} {\bibfnamefont {M.}~\bibnamefont {{Ara{\'u}jo}}}, \bibinfo {author} {\bibfnamefont {J.~M.}\ \bibnamefont {{Zeuner}}}, \bibinfo {author} {\bibfnamefont {L.~M.}\ \bibnamefont {{Procopio}}}, \bibinfo {author} {\bibfnamefont {{\v{C}}.}~\bibnamefont {{Brukner}}},\ and\ \bibinfo {author} {\bibfnamefont {P.}~\bibnamefont {{Walther}}},\ }\bibfield  {title} {\bibinfo {title} {{Experimental verification of an indefinite causal order}},\ }\href {https://doi.org/10.1126/sciadv.1602589} {\bibfield  {journal} {\bibinfo  {journal} {Science Advances}\ }\textbf {\bibinfo {volume} {3}},\ \bibinfo {pages} {e1602589} (\bibinfo {year} {2017})},\ \Eprint {https://arxiv.org/abs/1608.01683} {arXiv:1608.01683 [quant-ph]} \BibitemShut {NoStop}%
\bibitem [{\citenamefont {Goswami}\ \emph {et~al.}(2018)\citenamefont {Goswami}, \citenamefont {Giarmatzi}, \citenamefont {Kewming}, \citenamefont {Costa}, \citenamefont {Branciard}, \citenamefont {Romero},\ and\ \citenamefont {White}}]{goswami2018Indefinite}%
  \BibitemOpen
  \bibfield  {author} {\bibinfo {author} {\bibfnamefont {K.}~\bibnamefont {Goswami}}, \bibinfo {author} {\bibfnamefont {C.}~\bibnamefont {Giarmatzi}}, \bibinfo {author} {\bibfnamefont {M.}~\bibnamefont {Kewming}}, \bibinfo {author} {\bibfnamefont {F.}~\bibnamefont {Costa}}, \bibinfo {author} {\bibfnamefont {C.}~\bibnamefont {Branciard}}, \bibinfo {author} {\bibfnamefont {J.}~\bibnamefont {Romero}},\ and\ \bibinfo {author} {\bibfnamefont {A.~G.}\ \bibnamefont {White}},\ }\bibfield  {title} {\bibinfo {title} {Indefinite causal order in a quantum switch},\ }\href {https://doi.org/10.1103/PhysRevLett.121.090503} {\bibfield  {journal} {\bibinfo  {journal} {Phys. Rev. Lett.}\ }\textbf {\bibinfo {volume} {121}},\ \bibinfo {pages} {090503} (\bibinfo {year} {2018})}\BibitemShut {NoStop}%
\bibitem [{\citenamefont {Gu{\'e}rin}\ \emph {et~al.}(2016)\citenamefont {Gu{\'e}rin}, \citenamefont {Feix}, \citenamefont {Ara{\'u}jo},\ and\ \citenamefont {Brukner}}]{Guerin2016communicationComplexity}%
  \BibitemOpen
  \bibfield  {author} {\bibinfo {author} {\bibfnamefont {P.~A.}\ \bibnamefont {Gu{\'e}rin}}, \bibinfo {author} {\bibfnamefont {A.}~\bibnamefont {Feix}}, \bibinfo {author} {\bibfnamefont {M.}~\bibnamefont {Ara{\'u}jo}},\ and\ \bibinfo {author} {\bibfnamefont {{\v{C}}.}~\bibnamefont {Brukner}},\ }\bibfield  {title} {\bibinfo {title} {Exponential communication complexity advantage from quantum superposition of the direction of communication},\ }\href {https://doi.org/https://doi.org/10.1103/PhysRevLett.117.100502} {\bibfield  {journal} {\bibinfo  {journal} {Phys. Rev. Lett.}\ }\textbf {\bibinfo {volume} {117}},\ \bibinfo {pages} {100502} (\bibinfo {year} {2016})},\ \Eprint {https://arxiv.org/abs/1605.07372} {arXiv:1605.07372 [quant-ph]} \BibitemShut {NoStop}%
\bibitem [{\citenamefont {{Wei}}\ \emph {et~al.}(2019)\citenamefont {{Wei}}, \citenamefont {{Tischler}}, \citenamefont {{Zhao}}, \citenamefont {{Li}}, \citenamefont {{Arrazola}}, \citenamefont {{Liu}}, \citenamefont {{Zhang}}, \citenamefont {{Li}}, \citenamefont {{You}}, \citenamefont {{Wang}}, \citenamefont {{Chen}}, \citenamefont {{Sanders}}, \citenamefont {{Zhang}}, \citenamefont {{Pryde}}, \citenamefont {{Xu}},\ and\ \citenamefont {{Pan}}}]{wei2019experimentalCommunication}%
  \BibitemOpen
  \bibfield  {author} {\bibinfo {author} {\bibfnamefont {K.}~\bibnamefont {{Wei}}}, \bibinfo {author} {\bibfnamefont {N.}~\bibnamefont {{Tischler}}}, \bibinfo {author} {\bibfnamefont {S.-R.}\ \bibnamefont {{Zhao}}}, \bibinfo {author} {\bibfnamefont {Y.-H.}\ \bibnamefont {{Li}}}, \bibinfo {author} {\bibfnamefont {J.~M.}\ \bibnamefont {{Arrazola}}}, \bibinfo {author} {\bibfnamefont {Y.}~\bibnamefont {{Liu}}}, \bibinfo {author} {\bibfnamefont {W.}~\bibnamefont {{Zhang}}}, \bibinfo {author} {\bibfnamefont {H.}~\bibnamefont {{Li}}}, \bibinfo {author} {\bibfnamefont {L.}~\bibnamefont {{You}}}, \bibinfo {author} {\bibfnamefont {Z.}~\bibnamefont {{Wang}}}, \bibinfo {author} {\bibfnamefont {Y.-A.}\ \bibnamefont {{Chen}}}, \bibinfo {author} {\bibfnamefont {B.~C.}\ \bibnamefont {{Sanders}}}, \bibinfo {author} {\bibfnamefont {Q.}~\bibnamefont {{Zhang}}}, \bibinfo {author} {\bibfnamefont {G.~J.}\ \bibnamefont {{Pryde}}}, \bibinfo {author} {\bibfnamefont {F.}~\bibnamefont {{Xu}}},\ and\ \bibinfo {author} {\bibfnamefont
  {J.-W.}\ \bibnamefont {{Pan}}},\ }\bibfield  {title} {\bibinfo {title} {{Experimental Quantum Switching for Exponentially Superior Quantum Communication Complexity}},\ }\href {https://doi.org/10.1103/PhysRevLett.122.120504} {\bibfield  {journal} {\bibinfo  {journal} {Phys. Rev. Lett.}\ }\textbf {\bibinfo {volume} {122}},\ \bibinfo {eid} {120504} (\bibinfo {year} {2019})},\ \Eprint {https://arxiv.org/abs/1810.10238} {arXiv:1810.10238 [quant-ph]} \BibitemShut {NoStop}%
\bibitem [{\citenamefont {{Schiansky}}\ \emph {et~al.}(2023)\citenamefont {{Schiansky}}, \citenamefont {{Str{\"o}mberg}}, \citenamefont {{Trillo}}, \citenamefont {{Saggio}}, \citenamefont {{Dive}}, \citenamefont {{Navascu{\'e}s}},\ and\ \citenamefont {{Walther}}}]{Schiansky22TimeReversal}%
  \BibitemOpen
  \bibfield  {author} {\bibinfo {author} {\bibfnamefont {P.}~\bibnamefont {{Schiansky}}}, \bibinfo {author} {\bibfnamefont {T.}~\bibnamefont {{Str{\"o}mberg}}}, \bibinfo {author} {\bibfnamefont {D.}~\bibnamefont {{Trillo}}}, \bibinfo {author} {\bibfnamefont {V.}~\bibnamefont {{Saggio}}}, \bibinfo {author} {\bibfnamefont {B.}~\bibnamefont {{Dive}}}, \bibinfo {author} {\bibfnamefont {M.}~\bibnamefont {{Navascu{\'e}s}}},\ and\ \bibinfo {author} {\bibfnamefont {P.}~\bibnamefont {{Walther}}},\ }\bibfield  {title} {\bibinfo {title} {{Demonstration of universal time-reversal for qubit processes}},\ }\href {https://doi.org/10.1364/OPTICA.469109} {\bibfield  {journal} {\bibinfo  {journal} {Optica}\ }\textbf {\bibinfo {volume} {10}},\ \bibinfo {pages} {200} (\bibinfo {year} {2023})},\ \Eprint {https://arxiv.org/abs/2205.01122} {arXiv:2205.01122 [quant-ph]} \BibitemShut {NoStop}%
\bibitem [{\citenamefont {Yin}\ \emph {et~al.}(2023)\citenamefont {Yin}, \citenamefont {Zhao}, \citenamefont {Yang}, \citenamefont {Guo}, \citenamefont {Zhang}, \citenamefont {Li}, \citenamefont {Han}, \citenamefont {Liu}, \citenamefont {Xu}, \citenamefont {Chiribella} \emph {et~al.}}]{yin2023experimentalSuperHeisenberg}%
  \BibitemOpen
  \bibfield  {author} {\bibinfo {author} {\bibfnamefont {P.}~\bibnamefont {Yin}}, \bibinfo {author} {\bibfnamefont {X.}~\bibnamefont {Zhao}}, \bibinfo {author} {\bibfnamefont {Y.}~\bibnamefont {Yang}}, \bibinfo {author} {\bibfnamefont {Y.}~\bibnamefont {Guo}}, \bibinfo {author} {\bibfnamefont {W.-H.}\ \bibnamefont {Zhang}}, \bibinfo {author} {\bibfnamefont {G.-C.}\ \bibnamefont {Li}}, \bibinfo {author} {\bibfnamefont {Y.-J.}\ \bibnamefont {Han}}, \bibinfo {author} {\bibfnamefont {B.-H.}\ \bibnamefont {Liu}}, \bibinfo {author} {\bibfnamefont {J.-S.}\ \bibnamefont {Xu}}, \bibinfo {author} {\bibfnamefont {G.}~\bibnamefont {Chiribella}}, \emph {et~al.},\ }\bibfield  {title} {\bibinfo {title} {Experimental super-heisenberg quantum metrology with indefinite gate order},\ }\href@noop {} {\bibfield  {journal} {\bibinfo  {journal} {Nature Physics}\ ,\ \bibinfo {pages} {1}} (\bibinfo {year} {2023})}\BibitemShut {NoStop}%
\bibitem [{\citenamefont {{Cao}}\ \emph {et~al.}(2022)\citenamefont {{Cao}}, \citenamefont {{Wang}}, \citenamefont {{Jia}}, \citenamefont {{Zhang}}, \citenamefont {{Guo}}, \citenamefont {{Liu}}, \citenamefont {{Huang}}, \citenamefont {{Li}},\ and\ \citenamefont {{Guo}}}]{cao2022quantumSimulation}%
  \BibitemOpen
  \bibfield  {author} {\bibinfo {author} {\bibfnamefont {H.}~\bibnamefont {{Cao}}}, \bibinfo {author} {\bibfnamefont {N.-N.}\ \bibnamefont {{Wang}}}, \bibinfo {author} {\bibfnamefont {Z.}~\bibnamefont {{Jia}}}, \bibinfo {author} {\bibfnamefont {C.}~\bibnamefont {{Zhang}}}, \bibinfo {author} {\bibfnamefont {Y.}~\bibnamefont {{Guo}}}, \bibinfo {author} {\bibfnamefont {B.-H.}\ \bibnamefont {{Liu}}}, \bibinfo {author} {\bibfnamefont {Y.-F.}\ \bibnamefont {{Huang}}}, \bibinfo {author} {\bibfnamefont {C.-F.}\ \bibnamefont {{Li}}},\ and\ \bibinfo {author} {\bibfnamefont {G.-C.}\ \bibnamefont {{Guo}}},\ }\bibfield  {title} {\bibinfo {title} {{Quantum simulation of indefinite causal order induced quantum refrigeration}},\ }\href {https://doi.org/10.1103/PhysRevResearch.4.L032029} {\bibfield  {journal} {\bibinfo  {journal} {Physical Review Research}\ }\textbf {\bibinfo {volume} {4}},\ \bibinfo {eid} {L032029} (\bibinfo {year} {2022})},\ \Eprint {https://arxiv.org/abs/2101.07979} {arXiv:2101.07979 [quant-ph]}
  \BibitemShut {NoStop}%
\bibitem [{\citenamefont {{Goswami}}\ and\ \citenamefont {{Romero}}(2020)}]{goswami2020review}%
  \BibitemOpen
  \bibfield  {author} {\bibinfo {author} {\bibfnamefont {K.}~\bibnamefont {{Goswami}}}\ and\ \bibinfo {author} {\bibfnamefont {J.}~\bibnamefont {{Romero}}},\ }\bibfield  {title} {\bibinfo {title} {{Experiments on quantum causality}},\ }\href {https://doi.org/10.1116/5.0010747} {\bibfield  {journal} {\bibinfo  {journal} {AVS Quantum Science}\ }\textbf {\bibinfo {volume} {2}},\ \bibinfo {eid} {037101} (\bibinfo {year} {2020})},\ \Eprint {https://arxiv.org/abs/2009.00515} {arXiv:2009.00515 [quant-ph]} \BibitemShut {NoStop}%
\bibitem [{\citenamefont {Brukner}(2014)}]{Brukner2014quantumCausality}%
  \BibitemOpen
  \bibfield  {author} {\bibinfo {author} {\bibfnamefont {{\v{C}}.}~\bibnamefont {Brukner}},\ }\bibfield  {title} {\bibinfo {title} {Quantum causality},\ }\href {https://doi.org/10.1038/nphys2930} {\bibfield  {journal} {\bibinfo  {journal} {Nature Physics}\ }\textbf {\bibinfo {volume} {10}},\ \bibinfo {pages} {259} (\bibinfo {year} {2014})}\BibitemShut {NoStop}%
\bibitem [{\citenamefont {{Andersson}}\ \emph {et~al.}(2005)\citenamefont {{Andersson}}, \citenamefont {{Bergou}},\ and\ \citenamefont {{Jex}}}]{Andersson2005comparison}%
  \BibitemOpen
  \bibfield  {author} {\bibinfo {author} {\bibfnamefont {E.}~\bibnamefont {{Andersson}}}, \bibinfo {author} {\bibfnamefont {J.}~\bibnamefont {{Bergou}}},\ and\ \bibinfo {author} {\bibfnamefont {I.}~\bibnamefont {{Jex}}},\ }\bibfield  {title} {\bibinfo {title} {{Comparison of unitary transforms using Franson interferometry}},\ }\href {https://doi.org/10.1080/09500340500052911} {\bibfield  {journal} {\bibinfo  {journal} {Journal of Modern Optics}\ }\textbf {\bibinfo {volume} {52}},\ \bibinfo {pages} {1485} (\bibinfo {year} {2005})}\BibitemShut {NoStop}%
\bibitem [{\citenamefont {{Zhou}}\ \emph {et~al.}(2011)\citenamefont {{Zhou}}, \citenamefont {{Ralph}}, \citenamefont {{Kalasuwan}}, \citenamefont {{Zhang}}, \citenamefont {{Peruzzo}}, \citenamefont {{Lanyon}},\ and\ \citenamefont {{O'Brien}}}]{Zhou2011AddingControl}%
  \BibitemOpen
  \bibfield  {author} {\bibinfo {author} {\bibfnamefont {X.-Q.}\ \bibnamefont {{Zhou}}}, \bibinfo {author} {\bibfnamefont {T.~C.}\ \bibnamefont {{Ralph}}}, \bibinfo {author} {\bibfnamefont {P.}~\bibnamefont {{Kalasuwan}}}, \bibinfo {author} {\bibfnamefont {M.}~\bibnamefont {{Zhang}}}, \bibinfo {author} {\bibfnamefont {A.}~\bibnamefont {{Peruzzo}}}, \bibinfo {author} {\bibfnamefont {B.~P.}\ \bibnamefont {{Lanyon}}},\ and\ \bibinfo {author} {\bibfnamefont {J.~L.}\ \bibnamefont {{O'Brien}}},\ }\bibfield  {title} {\bibinfo {title} {{Adding control to arbitrary unknown quantum operations}},\ }\href {https://doi.org/10.1038/ncomms1392} {\bibfield  {journal} {\bibinfo  {journal} {Nature Communications}\ }\textbf {\bibinfo {volume} {2}},\ \bibinfo {eid} {413} (\bibinfo {year} {2011})},\ \Eprint {https://arxiv.org/abs/1006.2670} {arXiv:1006.2670 [quant-ph]} \BibitemShut {NoStop}%
\bibitem [{\citenamefont {{Ara{\'u}jo}}\ \emph {et~al.}(2014{\natexlab{a}})\citenamefont {{Ara{\'u}jo}}, \citenamefont {{Feix}}, \citenamefont {{Costa}},\ and\ \citenamefont {{Brukner}}}]{Araujo2014controlUnknown}%
  \BibitemOpen
  \bibfield  {author} {\bibinfo {author} {\bibfnamefont {M.}~\bibnamefont {{Ara{\'u}jo}}}, \bibinfo {author} {\bibfnamefont {A.}~\bibnamefont {{Feix}}}, \bibinfo {author} {\bibfnamefont {F.}~\bibnamefont {{Costa}}},\ and\ \bibinfo {author} {\bibfnamefont {{\v{C}}.}~\bibnamefont {{Brukner}}},\ }\bibfield  {title} {\bibinfo {title} {{Quantum circuits cannot control unknown operations}},\ }\href {https://doi.org/10.1088/1367-2630/16/9/093026} {\bibfield  {journal} {\bibinfo  {journal} {New Journal of Physics}\ }\textbf {\bibinfo {volume} {16}},\ \bibinfo {eid} {093026} (\bibinfo {year} {2014}{\natexlab{a}})},\ \Eprint {https://arxiv.org/abs/1309.7976} {arXiv:1309.7976 [quant-ph]} \BibitemShut {NoStop}%
\bibitem [{\citenamefont {{Ara{\'u}jo}}\ \emph {et~al.}(2014{\natexlab{b}})\citenamefont {{Ara{\'u}jo}}, \citenamefont {{Costa}},\ and\ \citenamefont {{Brukner}}}]{Araujo2014ComputationalAdvantage}%
  \BibitemOpen
  \bibfield  {author} {\bibinfo {author} {\bibfnamefont {M.}~\bibnamefont {{Ara{\'u}jo}}}, \bibinfo {author} {\bibfnamefont {F.}~\bibnamefont {{Costa}}},\ and\ \bibinfo {author} {\bibfnamefont {{\v{C}}.}~\bibnamefont {{Brukner}}},\ }\bibfield  {title} {\bibinfo {title} {{Computational Advantage from Quantum-Controlled Ordering of Gates}},\ }\href {https://doi.org/10.1103/PhysRevLett.113.250402} {\bibfield  {journal} {\bibinfo  {journal} {Phys. Rev. Lett.}\ }\textbf {\bibinfo {volume} {113}},\ \bibinfo {eid} {250402} (\bibinfo {year} {2014}{\natexlab{b}})},\ \Eprint {https://arxiv.org/abs/1401.8127} {arXiv:1401.8127 [quant-ph]} \BibitemShut {NoStop}%
\bibitem [{\citenamefont {{Friis}}\ \emph {et~al.}(2014)\citenamefont {{Friis}}, \citenamefont {{Dunjko}}, \citenamefont {{D{\"u}r}},\ and\ \citenamefont {{Briegel}}}]{Friis2014ImplementingQuantumControl}%
  \BibitemOpen
  \bibfield  {author} {\bibinfo {author} {\bibfnamefont {N.}~\bibnamefont {{Friis}}}, \bibinfo {author} {\bibfnamefont {V.}~\bibnamefont {{Dunjko}}}, \bibinfo {author} {\bibfnamefont {W.}~\bibnamefont {{D{\"u}r}}},\ and\ \bibinfo {author} {\bibfnamefont {H.~J.}\ \bibnamefont {{Briegel}}},\ }\bibfield  {title} {\bibinfo {title} {{Implementing quantum control for unknown subroutines}},\ }\href {https://doi.org/10.1103/PhysRevA.89.030303} {\bibfield  {journal} {\bibinfo  {journal} {Phys. Rev. A}\ }\textbf {\bibinfo {volume} {89}},\ \bibinfo {eid} {030303} (\bibinfo {year} {2014})},\ \Eprint {https://arxiv.org/abs/1401.8128} {arXiv:1401.8128 [quant-ph]} \BibitemShut {NoStop}%
\bibitem [{\citenamefont {Goswami}\ \emph {et~al.}(2020)\citenamefont {Goswami}, \citenamefont {Cao}, \citenamefont {Paz-Silva}, \citenamefont {Romero},\ and\ \citenamefont {White}}]{goswami2020IncreasingCommunication}%
  \BibitemOpen
  \bibfield  {author} {\bibinfo {author} {\bibfnamefont {K.}~\bibnamefont {Goswami}}, \bibinfo {author} {\bibfnamefont {Y.}~\bibnamefont {Cao}}, \bibinfo {author} {\bibfnamefont {G.~A.}\ \bibnamefont {Paz-Silva}}, \bibinfo {author} {\bibfnamefont {J.}~\bibnamefont {Romero}},\ and\ \bibinfo {author} {\bibfnamefont {A.~G.}\ \bibnamefont {White}},\ }\bibfield  {title} {\bibinfo {title} {Increasing communication capacity via superposition of order},\ }\href {https://doi.org/10.1103/PhysRevResearch.2.033292} {\bibfield  {journal} {\bibinfo  {journal} {Phys. Rev. Research}\ }\textbf {\bibinfo {volume} {2}},\ \bibinfo {pages} {033292} (\bibinfo {year} {2020})}\BibitemShut {NoStop}%
\bibitem [{\citenamefont {Str\"omberg}\ \emph {et~al.}(2023)\citenamefont {Str\"omberg}, \citenamefont {Schiansky}, \citenamefont {Peterson}, \citenamefont {Quintino},\ and\ \citenamefont {Walther}}]{stromberg23demonstration}%
  \BibitemOpen
  \bibfield  {author} {\bibinfo {author} {\bibfnamefont {T.}~\bibnamefont {Str\"omberg}}, \bibinfo {author} {\bibfnamefont {P.}~\bibnamefont {Schiansky}}, \bibinfo {author} {\bibfnamefont {R.~W.}\ \bibnamefont {Peterson}}, \bibinfo {author} {\bibfnamefont {M.~T.}\ \bibnamefont {Quintino}},\ and\ \bibinfo {author} {\bibfnamefont {P.}~\bibnamefont {Walther}},\ }\bibfield  {title} {\bibinfo {title} {Demonstration of a quantum switch in a sagnac configuration},\ }\href {https://doi.org/10.1103/PhysRevLett.131.060803} {\bibfield  {journal} {\bibinfo  {journal} {Phys. Rev. Lett.}\ }\textbf {\bibinfo {volume} {131}},\ \bibinfo {pages} {060803} (\bibinfo {year} {2023})}\BibitemShut {NoStop}%
\bibitem [{\citenamefont {{Liu}}\ \emph {et~al.}(2023{\natexlab{a}})\citenamefont {{Liu}}, \citenamefont {{Meng}}, \citenamefont {{Song}}, \citenamefont {{Li}}, \citenamefont {{Wu}}, \citenamefont {{Chen}}, \citenamefont {{Hong}}, \citenamefont {{Zhang}},\ and\ \citenamefont {{Yin}}}]{liu2023experimentally}%
  \BibitemOpen
  \bibfield  {author} {\bibinfo {author} {\bibfnamefont {W.-Q.}\ \bibnamefont {{Liu}}}, \bibinfo {author} {\bibfnamefont {Z.}~\bibnamefont {{Meng}}}, \bibinfo {author} {\bibfnamefont {B.-W.}\ \bibnamefont {{Song}}}, \bibinfo {author} {\bibfnamefont {J.}~\bibnamefont {{Li}}}, \bibinfo {author} {\bibfnamefont {Q.-Y.}\ \bibnamefont {{Wu}}}, \bibinfo {author} {\bibfnamefont {X.-X.}\ \bibnamefont {{Chen}}}, \bibinfo {author} {\bibfnamefont {J.-Y.}\ \bibnamefont {{Hong}}}, \bibinfo {author} {\bibfnamefont {A.-N.}\ \bibnamefont {{Zhang}}},\ and\ \bibinfo {author} {\bibfnamefont {Z.-q.}\ \bibnamefont {{Yin}}},\ }\bibfield  {title} {\bibinfo {title} {{Experimentally demonstrating indefinite causal order algorithms to solve the generalized Deutsch's problem}},\ }\href@noop {} {\bibfield  {journal} {\bibinfo  {journal} {arXiv e-prints}\ } (\bibinfo {year} {2023}{\natexlab{a}})},\ \Eprint {https://arxiv.org/abs/2305.05416} {arXiv:2305.05416 [quant-ph]} \BibitemShut {NoStop}%
\bibitem [{\citenamefont {Antesberger}\ \emph {et~al.}(2024)\citenamefont {Antesberger}, \citenamefont {Quintino}, \citenamefont {Walther},\ and\ \citenamefont {Rozema}}]{Antesberger2023tomography}%
  \BibitemOpen
  \bibfield  {author} {\bibinfo {author} {\bibfnamefont {M.}~\bibnamefont {Antesberger}}, \bibinfo {author} {\bibfnamefont {M.~T.}\ \bibnamefont {Quintino}}, \bibinfo {author} {\bibfnamefont {P.}~\bibnamefont {Walther}},\ and\ \bibinfo {author} {\bibfnamefont {L.~A.}\ \bibnamefont {Rozema}},\ }\bibfield  {title} {\bibinfo {title} {Higher-order process matrix tomography of a passively-stable quantum switch},\ }\href {https://doi.org/10.1103/PRXQuantum.5.010325} {\bibfield  {journal} {\bibinfo  {journal} {PRX Quantum}\ }\textbf {\bibinfo {volume} {5}},\ \bibinfo {pages} {010325} (\bibinfo {year} {2024})}\BibitemShut {NoStop}%
\bibitem [{\citenamefont {{Rubino}}\ \emph {et~al.}(2021{\natexlab{a}})\citenamefont {{Rubino}}, \citenamefont {{Rozema}}, \citenamefont {{Ebler}}, \citenamefont {{Kristj{\'a}nsson}}, \citenamefont {{Salek}}, \citenamefont {{Allard Gu{\'e}rin}}, \citenamefont {{Abbott}}, \citenamefont {{Branciard}}, \citenamefont {{Brukner}}, \citenamefont {{Chiribella}},\ and\ \citenamefont {{Walther}}}]{Rubino2021Communication}%
  \BibitemOpen
  \bibfield  {author} {\bibinfo {author} {\bibfnamefont {G.}~\bibnamefont {{Rubino}}}, \bibinfo {author} {\bibfnamefont {L.~A.}\ \bibnamefont {{Rozema}}}, \bibinfo {author} {\bibfnamefont {D.}~\bibnamefont {{Ebler}}}, \bibinfo {author} {\bibfnamefont {H.}~\bibnamefont {{Kristj{\'a}nsson}}}, \bibinfo {author} {\bibfnamefont {S.}~\bibnamefont {{Salek}}}, \bibinfo {author} {\bibfnamefont {P.}~\bibnamefont {{Allard Gu{\'e}rin}}}, \bibinfo {author} {\bibfnamefont {A.~A.}\ \bibnamefont {{Abbott}}}, \bibinfo {author} {\bibfnamefont {C.}~\bibnamefont {{Branciard}}}, \bibinfo {author} {\bibfnamefont {{\v{C}}.}~\bibnamefont {{Brukner}}}, \bibinfo {author} {\bibfnamefont {G.}~\bibnamefont {{Chiribella}}},\ and\ \bibinfo {author} {\bibfnamefont {P.}~\bibnamefont {{Walther}}},\ }\bibfield  {title} {\bibinfo {title} {{Experimental quantum communication enhancement by superposing trajectories}},\ }\href {https://doi.org/10.1103/PhysRevResearch.3.013093} {\bibfield  {journal} {\bibinfo  {journal} {Physical Review Research}\
  }\textbf {\bibinfo {volume} {3}},\ \bibinfo {eid} {013093} (\bibinfo {year} {2021}{\natexlab{a}})},\ \Eprint {https://arxiv.org/abs/2007.05005} {arXiv:2007.05005 [quant-ph]} \BibitemShut {NoStop}%
\bibitem [{\citenamefont {Rubino}\ \emph {et~al.}(2022)\citenamefont {Rubino}, \citenamefont {Rozema}, \citenamefont {Massa}, \citenamefont {Ara{\'{u}}jo}, \citenamefont {Zych}, \citenamefont {Brukner},\ and\ \citenamefont {Walther}}]{Rubino2022experimentalEntanglement}%
  \BibitemOpen
  \bibfield  {author} {\bibinfo {author} {\bibfnamefont {G.}~\bibnamefont {Rubino}}, \bibinfo {author} {\bibfnamefont {L.~A.}\ \bibnamefont {Rozema}}, \bibinfo {author} {\bibfnamefont {F.}~\bibnamefont {Massa}}, \bibinfo {author} {\bibfnamefont {M.}~\bibnamefont {Ara{\'{u}}jo}}, \bibinfo {author} {\bibfnamefont {M.}~\bibnamefont {Zych}}, \bibinfo {author} {\bibfnamefont {{\v{C}}.}~\bibnamefont {Brukner}},\ and\ \bibinfo {author} {\bibfnamefont {P.}~\bibnamefont {Walther}},\ }\bibfield  {title} {\bibinfo {title} {Experimental entanglement of temporal order},\ }\href {https://doi.org/10.22331/q-2022-01-11-621} {\bibfield  {journal} {\bibinfo  {journal} {{Quantum}}\ }\textbf {\bibinfo {volume} {6}},\ \bibinfo {pages} {621} (\bibinfo {year} {2022})},\ \Eprint {https://arxiv.org/abs/1712.06884} {arXiv:1712.06884 [quant-ph]} \BibitemShut {NoStop}%
\bibitem [{\citenamefont {Guo}\ \emph {et~al.}(2020)\citenamefont {Guo}, \citenamefont {Hu}, \citenamefont {Hou}, \citenamefont {Cao}, \citenamefont {Cui}, \citenamefont {Liu}, \citenamefont {Huang}, \citenamefont {Li}, \citenamefont {Guo},\ and\ \citenamefont {Chiribella}}]{guo2020experimental}%
  \BibitemOpen
  \bibfield  {author} {\bibinfo {author} {\bibfnamefont {Y.}~\bibnamefont {Guo}}, \bibinfo {author} {\bibfnamefont {X.-M.}\ \bibnamefont {Hu}}, \bibinfo {author} {\bibfnamefont {Z.-B.}\ \bibnamefont {Hou}}, \bibinfo {author} {\bibfnamefont {H.}~\bibnamefont {Cao}}, \bibinfo {author} {\bibfnamefont {J.-M.}\ \bibnamefont {Cui}}, \bibinfo {author} {\bibfnamefont {B.-H.}\ \bibnamefont {Liu}}, \bibinfo {author} {\bibfnamefont {Y.-F.}\ \bibnamefont {Huang}}, \bibinfo {author} {\bibfnamefont {C.-F.}\ \bibnamefont {Li}}, \bibinfo {author} {\bibfnamefont {G.-C.}\ \bibnamefont {Guo}},\ and\ \bibinfo {author} {\bibfnamefont {G.}~\bibnamefont {Chiribella}},\ }\bibfield  {title} {\bibinfo {title} {Experimental transmission of quantum information using a superposition of causal orders},\ }\href {https://doi.org/10.1103/PhysRevLett.124.030502} {\bibfield  {journal} {\bibinfo  {journal} {Phys. Rev. Lett.}\ }\textbf {\bibinfo {volume} {124}},\ \bibinfo {pages} {030502} (\bibinfo {year} {2020})},\ \Eprint
  {https://arxiv.org/abs/1811.07526} {arXiv:1811.07526 [quant-ph]} \BibitemShut {NoStop}%
\bibitem [{\citenamefont {{Cao}}\ \emph {et~al.}(2023)\citenamefont {{Cao}}, \citenamefont {{Bavaresco}}, \citenamefont {{Wang}}, \citenamefont {{Rozema}}, \citenamefont {{Zhang}}, \citenamefont {{Huang}}, \citenamefont {{Liu}}, \citenamefont {{Li}}, \citenamefont {{Guo}},\ and\ \citenamefont {{Walther}}}]{cao2022Semideviceindependent}%
  \BibitemOpen
  \bibfield  {author} {\bibinfo {author} {\bibfnamefont {H.}~\bibnamefont {{Cao}}}, \bibinfo {author} {\bibfnamefont {J.}~\bibnamefont {{Bavaresco}}}, \bibinfo {author} {\bibfnamefont {N.-N.}\ \bibnamefont {{Wang}}}, \bibinfo {author} {\bibfnamefont {L.~A.}\ \bibnamefont {{Rozema}}}, \bibinfo {author} {\bibfnamefont {C.}~\bibnamefont {{Zhang}}}, \bibinfo {author} {\bibfnamefont {Y.-F.}\ \bibnamefont {{Huang}}}, \bibinfo {author} {\bibfnamefont {B.-H.}\ \bibnamefont {{Liu}}}, \bibinfo {author} {\bibfnamefont {C.-F.}\ \bibnamefont {{Li}}}, \bibinfo {author} {\bibfnamefont {G.-C.}\ \bibnamefont {{Guo}}},\ and\ \bibinfo {author} {\bibfnamefont {P.}~\bibnamefont {{Walther}}},\ }\bibfield  {title} {\bibinfo {title} {{Semi-device-independent certification of indefinite causal order in a photonic quantum switch}},\ }\href {https://doi.org/10.1364/OPTICA.483876} {\bibfield  {journal} {\bibinfo  {journal} {Optica}\ }\textbf {\bibinfo {volume} {10}},\ \bibinfo {pages} {561} (\bibinfo {year} {2023})},\ \Eprint
  {https://arxiv.org/abs/2202.05346} {arXiv:2202.05346 [quant-ph]} \BibitemShut {NoStop}%
\bibitem [{\citenamefont {Zhu}\ \emph {et~al.}(2023)\citenamefont {Zhu}, \citenamefont {Chen}, \citenamefont {Hasegawa},\ and\ \citenamefont {Xue}}]{zhu2023prl}%
  \BibitemOpen
  \bibfield  {author} {\bibinfo {author} {\bibfnamefont {G.}~\bibnamefont {Zhu}}, \bibinfo {author} {\bibfnamefont {Y.}~\bibnamefont {Chen}}, \bibinfo {author} {\bibfnamefont {Y.}~\bibnamefont {Hasegawa}},\ and\ \bibinfo {author} {\bibfnamefont {P.}~\bibnamefont {Xue}},\ }\bibfield  {title} {\bibinfo {title} {Charging quantum batteries via indefinite causal order: Theory and experiment},\ }\href {https://doi.org/10.1103/PhysRevLett.131.240401} {\bibfield  {journal} {\bibinfo  {journal} {Phys. Rev. Lett.}\ }\textbf {\bibinfo {volume} {131}},\ \bibinfo {pages} {240401} (\bibinfo {year} {2023})}\BibitemShut {NoStop}%
\bibitem [{\citenamefont {An}\ \emph {et~al.}(2024)\citenamefont {An}, \citenamefont {Ru}, \citenamefont {Wang}, \citenamefont {Yang}, \citenamefont {Wang}, \citenamefont {Zhang},\ and\ \citenamefont {Li}}]{Min23}%
  \BibitemOpen
  \bibfield  {author} {\bibinfo {author} {\bibfnamefont {M.}~\bibnamefont {An}}, \bibinfo {author} {\bibfnamefont {S.}~\bibnamefont {Ru}}, \bibinfo {author} {\bibfnamefont {Y.}~\bibnamefont {Wang}}, \bibinfo {author} {\bibfnamefont {Y.}~\bibnamefont {Yang}}, \bibinfo {author} {\bibfnamefont {F.}~\bibnamefont {Wang}}, \bibinfo {author} {\bibfnamefont {P.}~\bibnamefont {Zhang}},\ and\ \bibinfo {author} {\bibfnamefont {F.}~\bibnamefont {Li}},\ }\bibfield  {title} {\bibinfo {title} {Noisy quantum parameter estimation with indefinite causal order},\ }\href {https://doi.org/10.1103/PhysRevA.109.012603} {\bibfield  {journal} {\bibinfo  {journal} {Phys. Rev. A}\ }\textbf {\bibinfo {volume} {109}},\ \bibinfo {pages} {012603} (\bibinfo {year} {2024})}\BibitemShut {NoStop}%
\bibitem [{\citenamefont {Reed}\ and\ \citenamefont {Simon}(1980)}]{reed1980methods}%
  \BibitemOpen
  \bibfield  {author} {\bibinfo {author} {\bibfnamefont {M.}~\bibnamefont {Reed}}\ and\ \bibinfo {author} {\bibfnamefont {B.}~\bibnamefont {Simon}},\ }\href@noop {} {\emph {\bibinfo {title} {Methods of modern mathematical physics: Functional analysis}}},\ Vol.~\bibinfo {volume} {1}\ (\bibinfo  {publisher} {Gulf Professional Publishing},\ \bibinfo {year} {1980})\BibitemShut {NoStop}%
\bibitem [{\citenamefont {Rambo}\ \emph {et~al.}(2016)\citenamefont {Rambo}, \citenamefont {Altepeter}, \citenamefont {Kumar},\ and\ \citenamefont {D'Ariano}}]{rambo2016functional}%
  \BibitemOpen
  \bibfield  {author} {\bibinfo {author} {\bibfnamefont {T.~M.}\ \bibnamefont {Rambo}}, \bibinfo {author} {\bibfnamefont {J.~B.}\ \bibnamefont {Altepeter}}, \bibinfo {author} {\bibfnamefont {P.}~\bibnamefont {Kumar}},\ and\ \bibinfo {author} {\bibfnamefont {G.~M.}\ \bibnamefont {D'Ariano}},\ }\bibfield  {title} {\bibinfo {title} {Functional quantum computing: An optical approach},\ }\href {https://doi.org/10.1103/PhysRevA.93.052321} {\bibfield  {journal} {\bibinfo  {journal} {Phys. Rev. A}\ }\textbf {\bibinfo {volume} {93}},\ \bibinfo {pages} {052321} (\bibinfo {year} {2016})},\ \Eprint {https://arxiv.org/abs/1211.1257} {arXiv:1211.1257 [quant-ph]} \BibitemShut {NoStop}%
\bibitem [{\citenamefont {{Dong}}\ \emph {et~al.}(2023)\citenamefont {{Dong}}, \citenamefont {{Quintino}}, \citenamefont {{Soeda}},\ and\ \citenamefont {{Murao}}}]{dong2023quantum}%
  \BibitemOpen
  \bibfield  {author} {\bibinfo {author} {\bibfnamefont {Q.}~\bibnamefont {{Dong}}}, \bibinfo {author} {\bibfnamefont {M.~T.}\ \bibnamefont {{Quintino}}}, \bibinfo {author} {\bibfnamefont {A.}~\bibnamefont {{Soeda}}},\ and\ \bibinfo {author} {\bibfnamefont {M.}~\bibnamefont {{Murao}}},\ }\bibfield  {title} {\bibinfo {title} {{The quantum switch is uniquely defined by its action on unitary operations}},\ }\href {https://doi.org/10.22331/q-2023-11-07-1169} {\bibfield  {journal} {\bibinfo  {journal} {Quantum}\ }\textbf {\bibinfo {volume} {7}},\ \bibinfo {pages} {1169} (\bibinfo {year} {2023})},\ \Eprint {https://arxiv.org/abs/2106.00034} {arXiv:2106.00034 [quant-ph]} \BibitemShut {NoStop}%
\bibitem [{\citenamefont {{Abbott}}\ \emph {et~al.}(2016)\citenamefont {{Abbott}}, \citenamefont {{Giarmatzi}}, \citenamefont {{Costa}},\ and\ \citenamefont {{Branciard}}}]{Abbott2016MultipartiteCausalCorrelations}%
  \BibitemOpen
  \bibfield  {author} {\bibinfo {author} {\bibfnamefont {A.~A.}\ \bibnamefont {{Abbott}}}, \bibinfo {author} {\bibfnamefont {C.}~\bibnamefont {{Giarmatzi}}}, \bibinfo {author} {\bibfnamefont {F.}~\bibnamefont {{Costa}}},\ and\ \bibinfo {author} {\bibfnamefont {C.}~\bibnamefont {{Branciard}}},\ }\bibfield  {title} {\bibinfo {title} {{Multipartite causal correlations: Polytopes and inequalities}},\ }\href {https://doi.org/10.1103/PhysRevA.94.032131} {\bibfield  {journal} {\bibinfo  {journal} {Phys. Rev. A}\ }\textbf {\bibinfo {volume} {94}},\ \bibinfo {eid} {032131} (\bibinfo {year} {2016})},\ \Eprint {https://arxiv.org/abs/1608.01528} {arXiv:1608.01528 [quant-ph]} \BibitemShut {NoStop}%
\bibitem [{\citenamefont {{Abbott}}\ \emph {et~al.}(2017)\citenamefont {{Abbott}}, \citenamefont {{Wechs}}, \citenamefont {{Costa}},\ and\ \citenamefont {{Branciard}}}]{Abbott2017GenuinelyMultipartite}%
  \BibitemOpen
  \bibfield  {author} {\bibinfo {author} {\bibfnamefont {A.~A.}\ \bibnamefont {{Abbott}}}, \bibinfo {author} {\bibfnamefont {J.}~\bibnamefont {{Wechs}}}, \bibinfo {author} {\bibfnamefont {F.}~\bibnamefont {{Costa}}},\ and\ \bibinfo {author} {\bibfnamefont {C.}~\bibnamefont {{Branciard}}},\ }\bibfield  {title} {\bibinfo {title} {{Genuinely multipartite noncausality}},\ }\href {https://doi.org/10.22331/q-2017-12-14-39} {\bibfield  {journal} {\bibinfo  {journal} {Quantum}\ }\textbf {\bibinfo {volume} {1}},\ \bibinfo {pages} {39} (\bibinfo {year} {2017})},\ \Eprint {https://arxiv.org/abs/1708.07663} {arXiv:1708.07663 [quant-ph]} \BibitemShut {NoStop}%
\bibitem [{\citenamefont {{Wechs}}\ \emph {et~al.}(2019)\citenamefont {{Wechs}}, \citenamefont {{Abbott}},\ and\ \citenamefont {{Branciard}}}]{Wechs2019definition}%
  \BibitemOpen
  \bibfield  {author} {\bibinfo {author} {\bibfnamefont {J.}~\bibnamefont {{Wechs}}}, \bibinfo {author} {\bibfnamefont {A.~A.}\ \bibnamefont {{Abbott}}},\ and\ \bibinfo {author} {\bibfnamefont {C.}~\bibnamefont {{Branciard}}},\ }\bibfield  {title} {\bibinfo {title} {{On the definition and characterisation of multipartite causal (non)separability}},\ }\href {https://doi.org/10.1088/1367-2630/aaf352} {\bibfield  {journal} {\bibinfo  {journal} {New Journal of Physics}\ }\textbf {\bibinfo {volume} {21}},\ \bibinfo {eid} {013027} (\bibinfo {year} {2019})},\ \Eprint {https://arxiv.org/abs/1807.10557} {arXiv:1807.10557 [quant-ph]} \BibitemShut {NoStop}%
\bibitem [{\citenamefont {Taddei}\ \emph {et~al.}(2021)\citenamefont {Taddei}, \citenamefont {Cari\~ne}, \citenamefont {Mart\'{\i}nez}, \citenamefont {Garc\'{\i}a}, \citenamefont {Guerrero}, \citenamefont {Abbott}, \citenamefont {Ara\'ujo}, \citenamefont {Branciard}, \citenamefont {G\'omez}, \citenamefont {Walborn}, \citenamefont {Aolita},\ and\ \citenamefont {Lima}}]{taddei2021computational}%
  \BibitemOpen
  \bibfield  {author} {\bibinfo {author} {\bibfnamefont {M.~M.}\ \bibnamefont {Taddei}}, \bibinfo {author} {\bibfnamefont {J.}~\bibnamefont {Cari\~ne}}, \bibinfo {author} {\bibfnamefont {D.}~\bibnamefont {Mart\'{\i}nez}}, \bibinfo {author} {\bibfnamefont {T.}~\bibnamefont {Garc\'{\i}a}}, \bibinfo {author} {\bibfnamefont {N.}~\bibnamefont {Guerrero}}, \bibinfo {author} {\bibfnamefont {A.~A.}\ \bibnamefont {Abbott}}, \bibinfo {author} {\bibfnamefont {M.}~\bibnamefont {Ara\'ujo}}, \bibinfo {author} {\bibfnamefont {C.}~\bibnamefont {Branciard}}, \bibinfo {author} {\bibfnamefont {E.~S.}\ \bibnamefont {G\'omez}}, \bibinfo {author} {\bibfnamefont {S.~P.}\ \bibnamefont {Walborn}}, \bibinfo {author} {\bibfnamefont {L.}~\bibnamefont {Aolita}},\ and\ \bibinfo {author} {\bibfnamefont {G.}~\bibnamefont {Lima}},\ }\bibfield  {title} {\bibinfo {title} {Computational advantage from the quantum superposition of multiple temporal orders of photonic gates},\ }\href {https://doi.org/10.1103/PRXQuantum.2.010320} {\bibfield
  {journal} {\bibinfo  {journal} {PRX Quantum}\ }\textbf {\bibinfo {volume} {2}},\ \bibinfo {pages} {010320} (\bibinfo {year} {2021})},\ \Eprint {https://arxiv.org/abs/2002.07817} {arXiv:2002.07817 [quant-ph]} \BibitemShut {NoStop}%
\bibitem [{\citenamefont {{Procopio}}\ \emph {et~al.}(2020)\citenamefont {{Procopio}}, \citenamefont {{Delgado}}, \citenamefont {{Enr{\'\i}quez}}, \citenamefont {{Belabas}},\ and\ \citenamefont {{Levenson}}}]{procopio2020sending}%
  \BibitemOpen
  \bibfield  {author} {\bibinfo {author} {\bibfnamefont {L.~M.}\ \bibnamefont {{Procopio}}}, \bibinfo {author} {\bibfnamefont {F.}~\bibnamefont {{Delgado}}}, \bibinfo {author} {\bibfnamefont {M.}~\bibnamefont {{Enr{\'\i}quez}}}, \bibinfo {author} {\bibfnamefont {N.}~\bibnamefont {{Belabas}}},\ and\ \bibinfo {author} {\bibfnamefont {J.~A.}\ \bibnamefont {{Levenson}}},\ }\bibfield  {title} {\bibinfo {title} {{Sending classical information via three noisy channels in superposition of causal orders}},\ }\href {https://doi.org/10.1103/PhysRevA.101.012346} {\bibfield  {journal} {\bibinfo  {journal} {Phys. Rev. A}\ }\textbf {\bibinfo {volume} {101}},\ \bibinfo {eid} {012346} (\bibinfo {year} {2020})},\ \Eprint {https://arxiv.org/abs/1910.11137} {arXiv:1910.11137 [quant-ph]} \BibitemShut {NoStop}%
\bibitem [{\citenamefont {{Cari{\~n}e}}\ \emph {et~al.}(2020)\citenamefont {{Cari{\~n}e}}, \citenamefont {{Ca{\~n}as}}, \citenamefont {{Skrzypczyk}}, \citenamefont {{{\v{S}}upi{\'c}}}, \citenamefont {{Guerrero}}, \citenamefont {{Garcia}}, \citenamefont {{Pereira}}, \citenamefont {{Prosser}}, \citenamefont {{Xavier}}, \citenamefont {{Delgado}}, \citenamefont {{Walborn}}, \citenamefont {{Cavalcanti}},\ and\ \citenamefont {{Lima}}}]{carine2020multi}%
  \BibitemOpen
  \bibfield  {author} {\bibinfo {author} {\bibfnamefont {J.}~\bibnamefont {{Cari{\~n}e}}}, \bibinfo {author} {\bibfnamefont {G.}~\bibnamefont {{Ca{\~n}as}}}, \bibinfo {author} {\bibfnamefont {P.}~\bibnamefont {{Skrzypczyk}}}, \bibinfo {author} {\bibfnamefont {I.}~\bibnamefont {{{\v{S}}upi{\'c}}}}, \bibinfo {author} {\bibfnamefont {N.}~\bibnamefont {{Guerrero}}}, \bibinfo {author} {\bibfnamefont {T.}~\bibnamefont {{Garcia}}}, \bibinfo {author} {\bibfnamefont {L.}~\bibnamefont {{Pereira}}}, \bibinfo {author} {\bibfnamefont {M.~A.~S.}\ \bibnamefont {{Prosser}}}, \bibinfo {author} {\bibfnamefont {G.~B.}\ \bibnamefont {{Xavier}}}, \bibinfo {author} {\bibfnamefont {A.}~\bibnamefont {{Delgado}}}, \bibinfo {author} {\bibfnamefont {S.~P.}\ \bibnamefont {{Walborn}}}, \bibinfo {author} {\bibfnamefont {D.}~\bibnamefont {{Cavalcanti}}},\ and\ \bibinfo {author} {\bibfnamefont {G.}~\bibnamefont {{Lima}}},\ }\bibfield  {title} {\bibinfo {title} {{Multi-core fiber integrated multi-port beam splitters for quantum information
  processing}},\ }\href {https://doi.org/10.1364/OPTICA.388912} {\bibfield  {journal} {\bibinfo  {journal} {Optica}\ }\textbf {\bibinfo {volume} {7}},\ \bibinfo {pages} {542} (\bibinfo {year} {2020})},\ \Eprint {https://arxiv.org/abs/2001.11056} {arXiv:2001.11056 [quant-ph]} \BibitemShut {NoStop}%
\bibitem [{\citenamefont {{Felce}}\ and\ \citenamefont {{Vedral}}(2020)}]{felce2020QuantumRefrigeration}%
  \BibitemOpen
  \bibfield  {author} {\bibinfo {author} {\bibfnamefont {D.}~\bibnamefont {{Felce}}}\ and\ \bibinfo {author} {\bibfnamefont {V.}~\bibnamefont {{Vedral}}},\ }\bibfield  {title} {\bibinfo {title} {{Quantum Refrigeration with Indefinite Causal Order}},\ }\href {https://doi.org/10.1103/PhysRevLett.125.070603} {\bibfield  {journal} {\bibinfo  {journal} {Phys. Rev. Lett.}\ }\textbf {\bibinfo {volume} {125}},\ \bibinfo {eid} {070603} (\bibinfo {year} {2020})},\ \Eprint {https://arxiv.org/abs/2003.00794} {arXiv:2003.00794 [quant-ph]} \BibitemShut {NoStop}%
\bibitem [{\citenamefont {{Felce}}\ \emph {et~al.}(2021)\citenamefont {{Felce}}, \citenamefont {{Vedral}},\ and\ \citenamefont {{Tennie}}}]{Felce2021IBMswitch}%
  \BibitemOpen
  \bibfield  {author} {\bibinfo {author} {\bibfnamefont {D.}~\bibnamefont {{Felce}}}, \bibinfo {author} {\bibfnamefont {V.}~\bibnamefont {{Vedral}}},\ and\ \bibinfo {author} {\bibfnamefont {F.}~\bibnamefont {{Tennie}}},\ }\bibfield  {title} {\bibinfo {title} {{Refrigeration with Indefinite Causal Orders on a Cloud Quantum Computer}},\ }\href {https://doi.org/10.48550/arXiv.2107.12413} {\bibfield  {journal} {\bibinfo  {journal} {arXiv e-prints}\ ,\ \bibinfo {eid} {arXiv:2107.12413}} (\bibinfo {year} {2021})},\ \Eprint {https://arxiv.org/abs/2107.12413} {arXiv:2107.12413 [quant-ph]} \BibitemShut {NoStop}%
\bibitem [{\citenamefont {{Nie}}\ \emph {et~al.}(2022)\citenamefont {{Nie}}, \citenamefont {{Zhu}}, \citenamefont {{Huang}}, \citenamefont {{Tang}}, \citenamefont {{Long}}, \citenamefont {{Lin}}, \citenamefont {{Tian}}, \citenamefont {{Qiu}}, \citenamefont {{Xi}}, \citenamefont {{Yang}}, \citenamefont {{Li}}, \citenamefont {{Dong}}, \citenamefont {{Xin}},\ and\ \citenamefont {{Lu}}}]{nei2022NMRswitch}%
  \BibitemOpen
  \bibfield  {author} {\bibinfo {author} {\bibfnamefont {X.}~\bibnamefont {{Nie}}}, \bibinfo {author} {\bibfnamefont {X.}~\bibnamefont {{Zhu}}}, \bibinfo {author} {\bibfnamefont {K.}~\bibnamefont {{Huang}}}, \bibinfo {author} {\bibfnamefont {K.}~\bibnamefont {{Tang}}}, \bibinfo {author} {\bibfnamefont {X.}~\bibnamefont {{Long}}}, \bibinfo {author} {\bibfnamefont {Z.}~\bibnamefont {{Lin}}}, \bibinfo {author} {\bibfnamefont {Y.}~\bibnamefont {{Tian}}}, \bibinfo {author} {\bibfnamefont {C.}~\bibnamefont {{Qiu}}}, \bibinfo {author} {\bibfnamefont {C.}~\bibnamefont {{Xi}}}, \bibinfo {author} {\bibfnamefont {X.}~\bibnamefont {{Yang}}}, \bibinfo {author} {\bibfnamefont {J.}~\bibnamefont {{Li}}}, \bibinfo {author} {\bibfnamefont {Y.}~\bibnamefont {{Dong}}}, \bibinfo {author} {\bibfnamefont {T.}~\bibnamefont {{Xin}}},\ and\ \bibinfo {author} {\bibfnamefont {D.}~\bibnamefont {{Lu}}},\ }\bibfield  {title} {\bibinfo {title} {{Experimental Realization of a Quantum Refrigerator Driven by Indefinite Causal Orders}},\ }\href
  {https://doi.org/10.1103/PhysRevLett.129.100603} {\bibfield  {journal} {\bibinfo  {journal} {Phys. Rev. Lett.}\ }\textbf {\bibinfo {volume} {129}},\ \bibinfo {eid} {100603} (\bibinfo {year} {2022})}\BibitemShut {NoStop}%
\bibitem [{\citenamefont {{Chiribella}}\ and\ \citenamefont {{Liu}}(2022)}]{chiribella2022indefiniteTimeDirection}%
  \BibitemOpen
  \bibfield  {author} {\bibinfo {author} {\bibfnamefont {G.}~\bibnamefont {{Chiribella}}}\ and\ \bibinfo {author} {\bibfnamefont {Z.}~\bibnamefont {{Liu}}},\ }\bibfield  {title} {\bibinfo {title} {{Quantum operations with indefinite time direction}},\ }\href {https://doi.org/10.1038/s42005-022-00967-3} {\bibfield  {journal} {\bibinfo  {journal} {Communications Physics}\ }\textbf {\bibinfo {volume} {5}},\ \bibinfo {eid} {190} (\bibinfo {year} {2022})},\ \Eprint {https://arxiv.org/abs/2012.03859} {arXiv:2012.03859 [quant-ph]} \BibitemShut {NoStop}%
\bibitem [{\citenamefont {{Rubino}}\ \emph {et~al.}(2021{\natexlab{b}})\citenamefont {{Rubino}}, \citenamefont {{Manzano}},\ and\ \citenamefont {{Brukner}}}]{Rubino2021superpositionThermodynamicEvolutions}%
  \BibitemOpen
  \bibfield  {author} {\bibinfo {author} {\bibfnamefont {G.}~\bibnamefont {{Rubino}}}, \bibinfo {author} {\bibfnamefont {G.}~\bibnamefont {{Manzano}}},\ and\ \bibinfo {author} {\bibfnamefont {{\v{C}}.}~\bibnamefont {{Brukner}}},\ }\bibfield  {title} {\bibinfo {title} {{Quantum superposition of thermodynamic evolutions with opposing time's arrows}},\ }\href {https://doi.org/10.1038/s42005-021-00759-1} {\bibfield  {journal} {\bibinfo  {journal} {Communications Physics}\ }\textbf {\bibinfo {volume} {4}},\ \bibinfo {eid} {251} (\bibinfo {year} {2021}{\natexlab{b}})},\ \Eprint {https://arxiv.org/abs/2008.02818} {arXiv:2008.02818 [quant-ph]} \BibitemShut {NoStop}%
\bibitem [{\citenamefont {{Str{\"o}mberg}}\ \emph {et~al.}(2022)\citenamefont {{Str{\"o}mberg}}, \citenamefont {{Schiansky}}, \citenamefont {{T{\'u}lio Quintino}}, \citenamefont {{Antesberger}}, \citenamefont {{Rozema}}, \citenamefont {{Agresti}}, \citenamefont {{Brukner}},\ and\ \citenamefont {{Walther}}}]{stromberg2022timeDirections}%
  \BibitemOpen
  \bibfield  {author} {\bibinfo {author} {\bibfnamefont {T.}~\bibnamefont {{Str{\"o}mberg}}}, \bibinfo {author} {\bibfnamefont {P.}~\bibnamefont {{Schiansky}}}, \bibinfo {author} {\bibfnamefont {M.}~\bibnamefont {{T{\'u}lio Quintino}}}, \bibinfo {author} {\bibfnamefont {M.}~\bibnamefont {{Antesberger}}}, \bibinfo {author} {\bibfnamefont {L.}~\bibnamefont {{Rozema}}}, \bibinfo {author} {\bibfnamefont {I.}~\bibnamefont {{Agresti}}}, \bibinfo {author} {\bibfnamefont {{\v{C}}.}~\bibnamefont {{Brukner}}},\ and\ \bibinfo {author} {\bibfnamefont {P.}~\bibnamefont {{Walther}}},\ }\bibfield  {title} {\bibinfo {title} {{Experimental superposition of time directions}},\ }\href@noop {} {\bibfield  {journal} {\bibinfo  {journal} {arXiv e-prints}\ } (\bibinfo {year} {2022})},\ \Eprint {https://arxiv.org/abs/2211.01283} {arXiv:2211.01283 [quant-ph]} \BibitemShut {NoStop}%
\bibitem [{\citenamefont {{Guo}}\ \emph {et~al.}(2022)\citenamefont {{Guo}}, \citenamefont {{Liu}}, \citenamefont {{Tang}}, \citenamefont {{Hu}}, \citenamefont {{Liu}}, \citenamefont {{Huang}}, \citenamefont {{Li}}, \citenamefont {{Guo}},\ and\ \citenamefont {{Chiribella}}}]{guo2022inputOutput}%
  \BibitemOpen
  \bibfield  {author} {\bibinfo {author} {\bibfnamefont {Y.}~\bibnamefont {{Guo}}}, \bibinfo {author} {\bibfnamefont {Z.}~\bibnamefont {{Liu}}}, \bibinfo {author} {\bibfnamefont {H.}~\bibnamefont {{Tang}}}, \bibinfo {author} {\bibfnamefont {X.-M.}\ \bibnamefont {{Hu}}}, \bibinfo {author} {\bibfnamefont {B.-H.}\ \bibnamefont {{Liu}}}, \bibinfo {author} {\bibfnamefont {Y.-F.}\ \bibnamefont {{Huang}}}, \bibinfo {author} {\bibfnamefont {C.-F.}\ \bibnamefont {{Li}}}, \bibinfo {author} {\bibfnamefont {G.-C.}\ \bibnamefont {{Guo}}},\ and\ \bibinfo {author} {\bibfnamefont {G.}~\bibnamefont {{Chiribella}}},\ }\bibfield  {title} {\bibinfo {title} {{Experimental demonstration of input-output indefiniteness in a single quantum device}},\ }\href@noop {} {\bibfield  {journal} {\bibinfo  {journal} {arXiv e-prints}\ } (\bibinfo {year} {2022})},\ \Eprint {https://arxiv.org/abs/2210.17046} {arXiv:2210.17046 [quant-ph]} \BibitemShut {NoStop}%
\bibitem [{\citenamefont {{Liu}}\ \emph {et~al.}(2023{\natexlab{b}})\citenamefont {{Liu}}, \citenamefont {{Yang}},\ and\ \citenamefont {{Chiribella}}}]{Zixuan2023indefiniteIO}%
  \BibitemOpen
  \bibfield  {author} {\bibinfo {author} {\bibfnamefont {Z.}~\bibnamefont {{Liu}}}, \bibinfo {author} {\bibfnamefont {M.}~\bibnamefont {{Yang}}},\ and\ \bibinfo {author} {\bibfnamefont {G.}~\bibnamefont {{Chiribella}}},\ }\bibfield  {title} {\bibinfo {title} {{Quantum communication through devices with indefinite input-output direction}},\ }\href {https://doi.org/10.1088/1367-2630/acc8f2} {\bibfield  {journal} {\bibinfo  {journal} {New Journal of Physics}\ }\textbf {\bibinfo {volume} {25}},\ \bibinfo {eid} {043017} (\bibinfo {year} {2023}{\natexlab{b}})},\ \Eprint {https://arxiv.org/abs/2212.08265} {arXiv:2212.08265 [quant-ph]} \BibitemShut {NoStop}%
\bibitem [{\citenamefont {{Rubino}}\ \emph {et~al.}(2022)\citenamefont {{Rubino}}, \citenamefont {{Manzano}}, \citenamefont {{Rozema}}, \citenamefont {{Walther}}, \citenamefont {{Parrondo}},\ and\ \citenamefont {{Brukner}}}]{Rubino2022InferringWork}%
  \BibitemOpen
  \bibfield  {author} {\bibinfo {author} {\bibfnamefont {G.}~\bibnamefont {{Rubino}}}, \bibinfo {author} {\bibfnamefont {G.}~\bibnamefont {{Manzano}}}, \bibinfo {author} {\bibfnamefont {L.~A.}\ \bibnamefont {{Rozema}}}, \bibinfo {author} {\bibfnamefont {P.}~\bibnamefont {{Walther}}}, \bibinfo {author} {\bibfnamefont {J.~M.~R.}\ \bibnamefont {{Parrondo}}},\ and\ \bibinfo {author} {\bibfnamefont {{\v{C}}.}~\bibnamefont {{Brukner}}},\ }\bibfield  {title} {\bibinfo {title} {{Inferring work by quantum superposing forward and time-reversal evolutions}},\ }\href {https://doi.org/10.1103/PhysRevResearch.4.013208} {\bibfield  {journal} {\bibinfo  {journal} {Physical Review Research}\ }\textbf {\bibinfo {volume} {4}},\ \bibinfo {eid} {013208} (\bibinfo {year} {2022})},\ \Eprint {https://arxiv.org/abs/2107.02201} {arXiv:2107.02201 [quant-ph]} \BibitemShut {NoStop}%
\bibitem [{\citenamefont {{Branciard}}(2016)}]{Branciard2016witnesses}%
  \BibitemOpen
  \bibfield  {author} {\bibinfo {author} {\bibfnamefont {C.}~\bibnamefont {{Branciard}}},\ }\bibfield  {title} {\bibinfo {title} {{Witnesses of causal nonseparability: an introduction and a few case studies}},\ }\href {https://doi.org/10.1038/srep26018} {\bibfield  {journal} {\bibinfo  {journal} {Scientific Reports}\ }\textbf {\bibinfo {volume} {6}},\ \bibinfo {eid} {26018} (\bibinfo {year} {2016})},\ \Eprint {https://arxiv.org/abs/1603.00043} {arXiv:1603.00043 [quant-ph]} \BibitemShut {NoStop}%
\bibitem [{\citenamefont {{Bavaresco}}\ \emph {et~al.}(2021)\citenamefont {{Bavaresco}}, \citenamefont {{Murao}},\ and\ \citenamefont {{Quintino}}}]{Bavaresco2021StrictHierarchy}%
  \BibitemOpen
  \bibfield  {author} {\bibinfo {author} {\bibfnamefont {J.}~\bibnamefont {{Bavaresco}}}, \bibinfo {author} {\bibfnamefont {M.}~\bibnamefont {{Murao}}},\ and\ \bibinfo {author} {\bibfnamefont {M.~T.}\ \bibnamefont {{Quintino}}},\ }\bibfield  {title} {\bibinfo {title} {{Strict Hierarchy between Parallel, Sequential, and Indefinite-Causal-Order Strategies for Channel Discrimination}},\ }\href {https://doi.org/10.1103/PhysRevLett.127.200504} {\bibfield  {journal} {\bibinfo  {journal} {Phys. Rev. Lett.}\ }\textbf {\bibinfo {volume} {127}},\ \bibinfo {eid} {200504} (\bibinfo {year} {2021})},\ \Eprint {https://arxiv.org/abs/2011.08300} {arXiv:2011.08300 [quant-ph]} \BibitemShut {NoStop}%
\bibitem [{\citenamefont {Svetlichny}(1987)}]{svetlichny1987distinguishing}%
  \BibitemOpen
  \bibfield  {author} {\bibinfo {author} {\bibfnamefont {G.}~\bibnamefont {Svetlichny}},\ }\bibfield  {title} {\bibinfo {title} {Distinguishing three-body from two-body nonseparability by a bell-type inequality},\ }\href {https://doi.org/10.1103/PhysRevD.35.3066} {\bibfield  {journal} {\bibinfo  {journal} {Phys. Rev. D}\ }\textbf {\bibinfo {volume} {35}},\ \bibinfo {pages} {3066} (\bibinfo {year} {1987})}\BibitemShut {NoStop}%
\bibitem [{\citenamefont {Seevinck}\ and\ \citenamefont {Svetlichny}(2002)}]{seevinck2002bell}%
  \BibitemOpen
  \bibfield  {author} {\bibinfo {author} {\bibfnamefont {M.}~\bibnamefont {Seevinck}}\ and\ \bibinfo {author} {\bibfnamefont {G.}~\bibnamefont {Svetlichny}},\ }\bibfield  {title} {\bibinfo {title} {Bell-type inequalities for partial separability in $n$-particle systems and quantum mechanical violations},\ }\href {https://doi.org/10.1103/PhysRevLett.89.060401} {\bibfield  {journal} {\bibinfo  {journal} {Phys. Rev. Lett.}\ }\textbf {\bibinfo {volume} {89}},\ \bibinfo {pages} {060401} (\bibinfo {year} {2002})}\BibitemShut {NoStop}%
\bibitem [{\citenamefont {{Giarmatzi}}\ \emph {et~al.}(2023)\citenamefont {{Giarmatzi}}, \citenamefont {{Jones}}, \citenamefont {{Gilchrist}}, \citenamefont {{Pakkiam}}, \citenamefont {{Fedorov}},\ and\ \citenamefont {{Costa}}}]{Giarmatzi2023MultiTimeTomo}%
  \BibitemOpen
  \bibfield  {author} {\bibinfo {author} {\bibfnamefont {C.}~\bibnamefont {{Giarmatzi}}}, \bibinfo {author} {\bibfnamefont {T.}~\bibnamefont {{Jones}}}, \bibinfo {author} {\bibfnamefont {A.}~\bibnamefont {{Gilchrist}}}, \bibinfo {author} {\bibfnamefont {P.}~\bibnamefont {{Pakkiam}}}, \bibinfo {author} {\bibfnamefont {A.}~\bibnamefont {{Fedorov}}},\ and\ \bibinfo {author} {\bibfnamefont {F.}~\bibnamefont {{Costa}}},\ }\bibfield  {title} {\bibinfo {title} {{Multi-time quantum process tomography of a superconducting qubit}},\ }\href {https://doi.org/10.48550/arXiv.2308.00750} {\bibfield  {journal} {\bibinfo  {journal} {arXiv e-prints}\ ,\ \bibinfo {eid} {arXiv:2308.00750}} (\bibinfo {year} {2023})},\ \Eprint {https://arxiv.org/abs/2308.00750} {arXiv:2308.00750 [quant-ph]} \BibitemShut {NoStop}%
\bibitem [{\citenamefont {{White}}\ \emph {et~al.}(2020)\citenamefont {{White}}, \citenamefont {{Hill}}, \citenamefont {{Pollock}}, \citenamefont {{Hollenberg}},\ and\ \citenamefont {{Modi}}}]{Kavan_experimental}%
  \BibitemOpen
  \bibfield  {author} {\bibinfo {author} {\bibfnamefont {G.~A.~L.}\ \bibnamefont {{White}}}, \bibinfo {author} {\bibfnamefont {C.~D.}\ \bibnamefont {{Hill}}}, \bibinfo {author} {\bibfnamefont {F.~A.}\ \bibnamefont {{Pollock}}}, \bibinfo {author} {\bibfnamefont {L.~C.~L.}\ \bibnamefont {{Hollenberg}}},\ and\ \bibinfo {author} {\bibfnamefont {K.}~\bibnamefont {{Modi}}},\ }\bibfield  {title} {\bibinfo {title} {{Demonstration of non-Markovian process characterisation and control on a quantum processor}},\ }\href {https://doi.org/10.1038/s41467-020-20113-3} {\bibfield  {journal} {\bibinfo  {journal} {Nature Communications}\ }\textbf {\bibinfo {volume} {11}},\ \bibinfo {eid} {6301} (\bibinfo {year} {2020})},\ \Eprint {https://arxiv.org/abs/2004.14018} {arXiv:2004.14018 [quant-ph]} \BibitemShut {NoStop}%
\bibitem [{\citenamefont {{White}}\ \emph {et~al.}(2021)\citenamefont {{White}}, \citenamefont {{Pollock}}, \citenamefont {{Hollenberg}}, \citenamefont {{Hill}},\ and\ \citenamefont {{Modi}}}]{Kavan_experimental2}%
  \BibitemOpen
  \bibfield  {author} {\bibinfo {author} {\bibfnamefont {G.~A.~L.}\ \bibnamefont {{White}}}, \bibinfo {author} {\bibfnamefont {F.~A.}\ \bibnamefont {{Pollock}}}, \bibinfo {author} {\bibfnamefont {L.~C.~L.}\ \bibnamefont {{Hollenberg}}}, \bibinfo {author} {\bibfnamefont {C.~D.}\ \bibnamefont {{Hill}}},\ and\ \bibinfo {author} {\bibfnamefont {K.}~\bibnamefont {{Modi}}},\ }\bibfield  {title} {\bibinfo {title} {{From many-body to many-time physics}},\ }\href@noop {} {\bibfield  {journal} {\bibinfo  {journal} {arXiv e-prints}\ } (\bibinfo {year} {2021})},\ \Eprint {https://arxiv.org/abs/2107.13934} {arXiv:2107.13934 [quant-ph]} \BibitemShut {NoStop}%
\bibitem [{\citenamefont {Guo}\ \emph {et~al.}(2021)\citenamefont {Guo}, \citenamefont {Taranto}, \citenamefont {Liu}, \citenamefont {Hu}, \citenamefont {Huang}, \citenamefont {Li},\ and\ \citenamefont {Guo}}]{guo21Markovorder}%
  \BibitemOpen
  \bibfield  {author} {\bibinfo {author} {\bibfnamefont {Y.}~\bibnamefont {Guo}}, \bibinfo {author} {\bibfnamefont {P.}~\bibnamefont {Taranto}}, \bibinfo {author} {\bibfnamefont {B.-H.}\ \bibnamefont {Liu}}, \bibinfo {author} {\bibfnamefont {X.-M.}\ \bibnamefont {Hu}}, \bibinfo {author} {\bibfnamefont {Y.-F.}\ \bibnamefont {Huang}}, \bibinfo {author} {\bibfnamefont {C.-F.}\ \bibnamefont {Li}},\ and\ \bibinfo {author} {\bibfnamefont {G.-C.}\ \bibnamefont {Guo}},\ }\bibfield  {title} {\bibinfo {title} {Experimental demonstration of instrument-specific quantum memory effects and non-markovian process recovery for common-cause processes},\ }\href {https://doi.org/10.1103/PhysRevLett.126.230401} {\bibfield  {journal} {\bibinfo  {journal} {Phys. Rev. Lett.}\ }\textbf {\bibinfo {volume} {126}},\ \bibinfo {pages} {230401} (\bibinfo {year} {2021})}\BibitemShut {NoStop}%
\bibitem [{\citenamefont {Zych}\ \emph {et~al.}(2019)\citenamefont {Zych}, \citenamefont {Costa}, \citenamefont {Pikovski},\ and\ \citenamefont {Brukner}}]{zych2019bell}%
  \BibitemOpen
  \bibfield  {author} {\bibinfo {author} {\bibfnamefont {M.}~\bibnamefont {Zych}}, \bibinfo {author} {\bibfnamefont {F.}~\bibnamefont {Costa}}, \bibinfo {author} {\bibfnamefont {I.}~\bibnamefont {Pikovski}},\ and\ \bibinfo {author} {\bibfnamefont {{\v{C}}.}~\bibnamefont {Brukner}},\ }\bibfield  {title} {\bibinfo {title} {Bell’s theorem for temporal order},\ }\href {https://doi.org/10.1038/s41467-019-11579-x} {\bibfield  {journal} {\bibinfo  {journal} {Nature Communications}\ }\textbf {\bibinfo {volume} {10}},\ \bibinfo {pages} {1} (\bibinfo {year} {2019})}\BibitemShut {NoStop}%
\bibitem [{\citenamefont {Wiseman}\ \emph {et~al.}(2007)\citenamefont {Wiseman}, \citenamefont {Jones},\ and\ \citenamefont {Doherty}}]{wiseman2007steering}%
  \BibitemOpen
  \bibfield  {author} {\bibinfo {author} {\bibfnamefont {H.~M.}\ \bibnamefont {Wiseman}}, \bibinfo {author} {\bibfnamefont {S.~J.}\ \bibnamefont {Jones}},\ and\ \bibinfo {author} {\bibfnamefont {A.~C.}\ \bibnamefont {Doherty}},\ }\bibfield  {title} {\bibinfo {title} {Steering, entanglement, nonlocality, and the einstein-podolsky-rosen paradox},\ }\href {https://doi.org/10.1103/PhysRevLett.98.140402} {\bibfield  {journal} {\bibinfo  {journal} {Phys. Rev. Lett.}\ }\textbf {\bibinfo {volume} {98}},\ \bibinfo {pages} {140402} (\bibinfo {year} {2007})}\BibitemShut {NoStop}%
\bibitem [{\citenamefont {Uola}\ \emph {et~al.}(2014)\citenamefont {Uola}, \citenamefont {Moroder},\ and\ \citenamefont {G\"uhne}}]{uola2014joint}%
  \BibitemOpen
  \bibfield  {author} {\bibinfo {author} {\bibfnamefont {R.}~\bibnamefont {Uola}}, \bibinfo {author} {\bibfnamefont {T.}~\bibnamefont {Moroder}},\ and\ \bibinfo {author} {\bibfnamefont {O.}~\bibnamefont {G\"uhne}},\ }\bibfield  {title} {\bibinfo {title} {Joint measurability of generalized measurements implies classicality},\ }\href {https://doi.org/10.1103/PhysRevLett.113.160403} {\bibfield  {journal} {\bibinfo  {journal} {Phys. Rev. Lett.}\ }\textbf {\bibinfo {volume} {113}},\ \bibinfo {pages} {160403} (\bibinfo {year} {2014})}\BibitemShut {NoStop}%
\bibitem [{\citenamefont {Quintino}\ \emph {et~al.}(2014)\citenamefont {Quintino}, \citenamefont {V\'ertesi},\ and\ \citenamefont {Brunner}}]{quintino2014joint}%
  \BibitemOpen
  \bibfield  {author} {\bibinfo {author} {\bibfnamefont {M.~T.}\ \bibnamefont {Quintino}}, \bibinfo {author} {\bibfnamefont {T.}~\bibnamefont {V\'ertesi}},\ and\ \bibinfo {author} {\bibfnamefont {N.}~\bibnamefont {Brunner}},\ }\bibfield  {title} {\bibinfo {title} {Joint measurability, einstein-podolsky-rosen steering, and bell nonlocality},\ }\href {https://doi.org/10.1103/PhysRevLett.113.160402} {\bibfield  {journal} {\bibinfo  {journal} {Phys. Rev. Lett.}\ }\textbf {\bibinfo {volume} {113}},\ \bibinfo {pages} {160402} (\bibinfo {year} {2014})}\BibitemShut {NoStop}%
\bibitem [{\citenamefont {Branciard}\ \emph {et~al.}(2012)\citenamefont {Branciard}, \citenamefont {Cavalcanti}, \citenamefont {Walborn}, \citenamefont {Scarani},\ and\ \citenamefont {Wiseman}}]{branciard2012one}%
  \BibitemOpen
  \bibfield  {author} {\bibinfo {author} {\bibfnamefont {C.}~\bibnamefont {Branciard}}, \bibinfo {author} {\bibfnamefont {E.~G.}\ \bibnamefont {Cavalcanti}}, \bibinfo {author} {\bibfnamefont {S.~P.}\ \bibnamefont {Walborn}}, \bibinfo {author} {\bibfnamefont {V.}~\bibnamefont {Scarani}},\ and\ \bibinfo {author} {\bibfnamefont {H.~M.}\ \bibnamefont {Wiseman}},\ }\bibfield  {title} {\bibinfo {title} {One-sided device-independent quantum key distribution: Security, feasibility, and the connection with steering},\ }\href {https://doi.org/10.1103/PhysRevA.85.010301} {\bibfield  {journal} {\bibinfo  {journal} {Phys. Rev. A}\ }\textbf {\bibinfo {volume} {85}},\ \bibinfo {pages} {010301} (\bibinfo {year} {2012})}\BibitemShut {NoStop}%
\bibitem [{\citenamefont {Bavaresco}\ \emph {et~al.}(2019)\citenamefont {Bavaresco}, \citenamefont {Ara{\'{u}}jo}, \citenamefont {Brukner},\ and\ \citenamefont {Quintino}}]{bavaresco2019Semideviceindependent}%
  \BibitemOpen
  \bibfield  {author} {\bibinfo {author} {\bibfnamefont {J.}~\bibnamefont {Bavaresco}}, \bibinfo {author} {\bibfnamefont {M.}~\bibnamefont {Ara{\'{u}}jo}}, \bibinfo {author} {\bibfnamefont {{\v{C}}.}~\bibnamefont {Brukner}},\ and\ \bibinfo {author} {\bibfnamefont {M.~T.}\ \bibnamefont {Quintino}},\ }\bibfield  {title} {\bibinfo {title} {Semi-device-independent certification of indefinite causal order},\ }\href {https://doi.org/10.22331/q-2019-08-19-176} {\bibfield  {journal} {\bibinfo  {journal} {{Quantum}}\ }\textbf {\bibinfo {volume} {3}},\ \bibinfo {pages} {176} (\bibinfo {year} {2019})},\ \Eprint {https://arxiv.org/abs/1903.10526} {arXiv:1903.10526 [quant-ph]} \BibitemShut {NoStop}%
\bibitem [{\citenamefont {{Dourdent}}\ \emph {et~al.}(2022)\citenamefont {{Dourdent}}, \citenamefont {{Abbott}}, \citenamefont {{Brunner}}, \citenamefont {{{\v{S}}upi{\'c}}},\ and\ \citenamefont {{Branciard}}}]{Dourdent2022SemiDeviceIndependent}%
  \BibitemOpen
  \bibfield  {author} {\bibinfo {author} {\bibfnamefont {H.}~\bibnamefont {{Dourdent}}}, \bibinfo {author} {\bibfnamefont {A.~A.}\ \bibnamefont {{Abbott}}}, \bibinfo {author} {\bibfnamefont {N.}~\bibnamefont {{Brunner}}}, \bibinfo {author} {\bibfnamefont {I.}~\bibnamefont {{{\v{S}}upi{\'c}}}},\ and\ \bibinfo {author} {\bibfnamefont {C.}~\bibnamefont {{Branciard}}},\ }\bibfield  {title} {\bibinfo {title} {{Semi-Device-Independent Certification of Causal Nonseparability with Trusted Quantum Inputs}},\ }\href {https://doi.org/10.1103/PhysRevLett.129.090402} {\bibfield  {journal} {\bibinfo  {journal} {Phys. Rev. Lett.}\ }\textbf {\bibinfo {volume} {129}},\ \bibinfo {eid} {090402} (\bibinfo {year} {2022})},\ \Eprint {https://arxiv.org/abs/2107.10877} {arXiv:2107.10877 [quant-ph]} \BibitemShut {NoStop}%
\bibitem [{\citenamefont {Aspect}(1999)}]{aspect1999bell}%
  \BibitemOpen
  \bibfield  {author} {\bibinfo {author} {\bibfnamefont {A.}~\bibnamefont {Aspect}},\ }\bibfield  {title} {\bibinfo {title} {Bell's inequality test: more ideal than ever},\ }\href {https://doi.org/https://doi.org/10.1038/18296} {\bibfield  {journal} {\bibinfo  {journal} {Nature}\ }\textbf {\bibinfo {volume} {398}},\ \bibinfo {pages} {189} (\bibinfo {year} {1999})}\BibitemShut {NoStop}%
\bibitem [{\citenamefont {Clauser}\ \emph {et~al.}(1969)\citenamefont {Clauser}, \citenamefont {Horne}, \citenamefont {Shimony},\ and\ \citenamefont {Holt}}]{clauser1969proposed}%
  \BibitemOpen
  \bibfield  {author} {\bibinfo {author} {\bibfnamefont {J.~F.}\ \bibnamefont {Clauser}}, \bibinfo {author} {\bibfnamefont {M.~A.}\ \bibnamefont {Horne}}, \bibinfo {author} {\bibfnamefont {A.}~\bibnamefont {Shimony}},\ and\ \bibinfo {author} {\bibfnamefont {R.~A.}\ \bibnamefont {Holt}},\ }\bibfield  {title} {\bibinfo {title} {Proposed experiment to test local hidden-variable theories},\ }\href {https://doi.org/10.1103/PhysRevLett.23.880} {\bibfield  {journal} {\bibinfo  {journal} {Physical Review Letters}\ }\textbf {\bibinfo {volume} {23}},\ \bibinfo {pages} {880} (\bibinfo {year} {1969})}\BibitemShut {NoStop}%
\bibitem [{\citenamefont {Bell}(1964)}]{bell1964einstein}%
  \BibitemOpen
  \bibfield  {author} {\bibinfo {author} {\bibfnamefont {J.~S.}\ \bibnamefont {Bell}},\ }\bibfield  {title} {\bibinfo {title} {On the einstein podolsky rosen paradox},\ }\href@noop {} {\bibfield  {journal} {\bibinfo  {journal} {Physics Physique Fizika}\ }\textbf {\bibinfo {volume} {1}},\ \bibinfo {pages} {195} (\bibinfo {year} {1964})}\BibitemShut {NoStop}%
\bibitem [{\citenamefont {Bell}(1966)}]{bell1966problem}%
  \BibitemOpen
  \bibfield  {author} {\bibinfo {author} {\bibfnamefont {J.~S.}\ \bibnamefont {Bell}},\ }\bibfield  {title} {\bibinfo {title} {On the problem of hidden variables in quantum mechanics},\ }\href {https://doi.org/10.1103/RevModPhys.38.447} {\bibfield  {journal} {\bibinfo  {journal} {Reviews of Modern Physics}\ }\textbf {\bibinfo {volume} {38}},\ \bibinfo {pages} {447} (\bibinfo {year} {1966})}\BibitemShut {NoStop}%
\bibitem [{\citenamefont {Weihs}\ \emph {et~al.}(1998)\citenamefont {Weihs}, \citenamefont {Jennewein}, \citenamefont {Simon}, \citenamefont {Weinfurter},\ and\ \citenamefont {Zeilinger}}]{weihs1998violation}%
  \BibitemOpen
  \bibfield  {author} {\bibinfo {author} {\bibfnamefont {G.}~\bibnamefont {Weihs}}, \bibinfo {author} {\bibfnamefont {T.}~\bibnamefont {Jennewein}}, \bibinfo {author} {\bibfnamefont {C.}~\bibnamefont {Simon}}, \bibinfo {author} {\bibfnamefont {H.}~\bibnamefont {Weinfurter}},\ and\ \bibinfo {author} {\bibfnamefont {A.}~\bibnamefont {Zeilinger}},\ }\bibfield  {title} {\bibinfo {title} {Violation of bell's inequality under strict einstein locality conditions},\ }\href {https://doi.org/10.1103/PhysRevLett.81.5039} {\bibfield  {journal} {\bibinfo  {journal} {Phys. Rev. Lett.}\ }\textbf {\bibinfo {volume} {81}},\ \bibinfo {pages} {5039} (\bibinfo {year} {1998})}\BibitemShut {NoStop}%
\bibitem [{\citenamefont {Aspect}\ \emph {et~al.}(1981)\citenamefont {Aspect}, \citenamefont {Grangier},\ and\ \citenamefont {Roger}}]{aspect1981experimental}%
  \BibitemOpen
  \bibfield  {author} {\bibinfo {author} {\bibfnamefont {A.}~\bibnamefont {Aspect}}, \bibinfo {author} {\bibfnamefont {P.}~\bibnamefont {Grangier}},\ and\ \bibinfo {author} {\bibfnamefont {G.}~\bibnamefont {Roger}},\ }\bibfield  {title} {\bibinfo {title} {Experimental tests of realistic local theories via bell's theorem},\ }\href {https://doi.org/10.1103/PhysRevLett.81.5039} {\bibfield  {journal} {\bibinfo  {journal} {Phys. Rev. Lett.}\ }\textbf {\bibinfo {volume} {47}},\ \bibinfo {pages} {460} (\bibinfo {year} {1981})}\BibitemShut {NoStop}%
\bibitem [{\citenamefont {Freedman}\ and\ \citenamefont {Clauser}(1972)}]{freedman1972experimental}%
  \BibitemOpen
  \bibfield  {author} {\bibinfo {author} {\bibfnamefont {S.~J.}\ \bibnamefont {Freedman}}\ and\ \bibinfo {author} {\bibfnamefont {J.~F.}\ \bibnamefont {Clauser}},\ }\bibfield  {title} {\bibinfo {title} {Experimental test of local hidden-variable theories},\ }\href {https://doi.org/10.1103/PhysRevLett.28.938} {\bibfield  {journal} {\bibinfo  {journal} {Phys. Rev. Lett.}\ }\textbf {\bibinfo {volume} {28}},\ \bibinfo {pages} {938} (\bibinfo {year} {1972})}\BibitemShut {NoStop}%
\bibitem [{\citenamefont {T\'oth}\ and\ \citenamefont {G\"uhne}(2005)}]{toth2005detecting}%
  \BibitemOpen
  \bibfield  {author} {\bibinfo {author} {\bibfnamefont {G.}~\bibnamefont {T\'oth}}\ and\ \bibinfo {author} {\bibfnamefont {O.}~\bibnamefont {G\"uhne}},\ }\bibfield  {title} {\bibinfo {title} {Detecting genuine multipartite entanglement with two local measurements},\ }\href {https://doi.org/10.1103/PhysRevLett.94.060501} {\bibfield  {journal} {\bibinfo  {journal} {Phys. Rev. Lett.}\ }\textbf {\bibinfo {volume} {94}},\ \bibinfo {pages} {060501} (\bibinfo {year} {2005})}\BibitemShut {NoStop}%
\bibitem [{\citenamefont {G\"uhne}\ \emph {et~al.}(2007)\citenamefont {G\"uhne}, \citenamefont {Lu}, \citenamefont {Gao},\ and\ \citenamefont {Pan}}]{guhne2007toolbox}%
  \BibitemOpen
  \bibfield  {author} {\bibinfo {author} {\bibfnamefont {O.}~\bibnamefont {G\"uhne}}, \bibinfo {author} {\bibfnamefont {C.-Y.}\ \bibnamefont {Lu}}, \bibinfo {author} {\bibfnamefont {W.-B.}\ \bibnamefont {Gao}},\ and\ \bibinfo {author} {\bibfnamefont {J.-W.}\ \bibnamefont {Pan}},\ }\bibfield  {title} {\bibinfo {title} {Toolbox for entanglement detection and fidelity estimation},\ }\href {https://doi.org/10.1103/PhysRevA.76.030305} {\bibfield  {journal} {\bibinfo  {journal} {Phys. Rev. A}\ }\textbf {\bibinfo {volume} {76}},\ \bibinfo {pages} {030305} (\bibinfo {year} {2007})}\BibitemShut {NoStop}%
\bibitem [{\citenamefont {{Brukner}}(2015)}]{Brukner2015Bounding}%
  \BibitemOpen
  \bibfield  {author} {\bibinfo {author} {\bibfnamefont {{\v{C}}.}~\bibnamefont {{Brukner}}},\ }\bibfield  {title} {\bibinfo {title} {{Bounding quantum correlations with indefinite causal order}},\ }\href {https://doi.org/10.1088/1367-2630/17/8/083034} {\bibfield  {journal} {\bibinfo  {journal} {New Journal of Physics}\ }\textbf {\bibinfo {volume} {17}},\ \bibinfo {eid} {083034} (\bibinfo {year} {2015})},\ \Eprint {https://arxiv.org/abs/1404.0721} {arXiv:1404.0721 [quant-ph]} \BibitemShut {NoStop}%
\bibitem [{\citenamefont {{Branciard}}\ \emph {et~al.}(2016)\citenamefont {{Branciard}}, \citenamefont {{Ara{\'u}jo}}, \citenamefont {{Feix}}, \citenamefont {{Costa}},\ and\ \citenamefont {{Brukner}}}]{Branciard2016simplestCausalInequalities}%
  \BibitemOpen
  \bibfield  {author} {\bibinfo {author} {\bibfnamefont {C.}~\bibnamefont {{Branciard}}}, \bibinfo {author} {\bibfnamefont {M.}~\bibnamefont {{Ara{\'u}jo}}}, \bibinfo {author} {\bibfnamefont {A.}~\bibnamefont {{Feix}}}, \bibinfo {author} {\bibfnamefont {F.}~\bibnamefont {{Costa}}},\ and\ \bibinfo {author} {\bibfnamefont {{\v{C}}.}~\bibnamefont {{Brukner}}},\ }\bibfield  {title} {\bibinfo {title} {{The simplest causal inequalities and their violation}},\ }\href {https://doi.org/10.1088/1367-2630/18/1/013008} {\bibfield  {journal} {\bibinfo  {journal} {New Journal of Physics}\ }\textbf {\bibinfo {volume} {18}},\ \bibinfo {eid} {013008} (\bibinfo {year} {2016})},\ \Eprint {https://arxiv.org/abs/1508.01704} {arXiv:1508.01704 [quant-ph]} \BibitemShut {NoStop}%
\bibitem [{\citenamefont {{Oreshkov}}\ and\ \citenamefont {{Giarmatzi}}(2016)}]{Oreshkov2016causal}%
  \BibitemOpen
  \bibfield  {author} {\bibinfo {author} {\bibfnamefont {O.}~\bibnamefont {{Oreshkov}}}\ and\ \bibinfo {author} {\bibfnamefont {C.}~\bibnamefont {{Giarmatzi}}},\ }\bibfield  {title} {\bibinfo {title} {{Causal and causally separable processes}},\ }\href {https://doi.org/10.1088/1367-2630/18/9/093020} {\bibfield  {journal} {\bibinfo  {journal} {New Journal of Physics}\ }\textbf {\bibinfo {volume} {18}},\ \bibinfo {eid} {093020} (\bibinfo {year} {2016})},\ \Eprint {https://arxiv.org/abs/1506.05449} {arXiv:1506.05449 [quant-ph]} \BibitemShut {NoStop}%
\bibitem [{\citenamefont {{Miklin}}\ \emph {et~al.}(2017)\citenamefont {{Miklin}}, \citenamefont {{Abbott}}, \citenamefont {{Branciard}}, \citenamefont {{Chaves}},\ and\ \citenamefont {{Budroni}}}]{Miklin2017entropicApproach}%
  \BibitemOpen
  \bibfield  {author} {\bibinfo {author} {\bibfnamefont {N.}~\bibnamefont {{Miklin}}}, \bibinfo {author} {\bibfnamefont {A.~A.}\ \bibnamefont {{Abbott}}}, \bibinfo {author} {\bibfnamefont {C.}~\bibnamefont {{Branciard}}}, \bibinfo {author} {\bibfnamefont {R.}~\bibnamefont {{Chaves}}},\ and\ \bibinfo {author} {\bibfnamefont {C.}~\bibnamefont {{Budroni}}},\ }\bibfield  {title} {\bibinfo {title} {{The entropic approach to causal correlations}},\ }\href {https://doi.org/10.1088/1367-2630/aa8f9f} {\bibfield  {journal} {\bibinfo  {journal} {New Journal of Physics}\ }\textbf {\bibinfo {volume} {19}},\ \bibinfo {eid} {113041} (\bibinfo {year} {2017})},\ \Eprint {https://arxiv.org/abs/1706.10270} {arXiv:1706.10270 [quant-ph]} \BibitemShut {NoStop}%
\bibitem [{\citenamefont {{Wechs}}\ \emph {et~al.}(2023)\citenamefont {{Wechs}}, \citenamefont {{Branciard}},\ and\ \citenamefont {{Oreshkov}}}]{Wechs2023Existence}%
  \BibitemOpen
  \bibfield  {author} {\bibinfo {author} {\bibfnamefont {J.}~\bibnamefont {{Wechs}}}, \bibinfo {author} {\bibfnamefont {C.}~\bibnamefont {{Branciard}}},\ and\ \bibinfo {author} {\bibfnamefont {O.}~\bibnamefont {{Oreshkov}}},\ }\bibfield  {title} {\bibinfo {title} {{Existence of processes violating causal inequalities on time-delocalised subsystems}},\ }\href {https://doi.org/10.1038/s41467-023-36893-3} {\bibfield  {journal} {\bibinfo  {journal} {Nature Communications}\ }\textbf {\bibinfo {volume} {14}},\ \bibinfo {eid} {1471} (\bibinfo {year} {2023})},\ \Eprint {https://arxiv.org/abs/2201.11832} {arXiv:2201.11832 [quant-ph]} \BibitemShut {NoStop}%
\bibitem [{\citenamefont {{Purves}}\ and\ \citenamefont {{Short}}(2021)}]{Purves2021CannotViolate}%
  \BibitemOpen
  \bibfield  {author} {\bibinfo {author} {\bibfnamefont {T.}~\bibnamefont {{Purves}}}\ and\ \bibinfo {author} {\bibfnamefont {A.~J.}\ \bibnamefont {{Short}}},\ }\bibfield  {title} {\bibinfo {title} {{Quantum Theory Cannot Violate a Causal Inequality}},\ }\href {https://doi.org/10.1103/PhysRevLett.127.110402} {\bibfield  {journal} {\bibinfo  {journal} {Phys. Rev. Lett.}\ }\textbf {\bibinfo {volume} {127}},\ \bibinfo {eid} {110402} (\bibinfo {year} {2021})},\ \Eprint {https://arxiv.org/abs/2101.09107} {arXiv:2101.09107 [quant-ph]} \BibitemShut {NoStop}%
\bibitem [{\citenamefont {{Ara{\'u}jo}}\ \emph {et~al.}(2017{\natexlab{a}})\citenamefont {{Ara{\'u}jo}}, \citenamefont {{Feix}}, \citenamefont {{Navascu{\'e}s}},\ and\ \citenamefont {{Brukner}}}]{Araujo2017purificationPostulate}%
  \BibitemOpen
  \bibfield  {author} {\bibinfo {author} {\bibfnamefont {M.}~\bibnamefont {{Ara{\'u}jo}}}, \bibinfo {author} {\bibfnamefont {A.}~\bibnamefont {{Feix}}}, \bibinfo {author} {\bibfnamefont {M.}~\bibnamefont {{Navascu{\'e}s}}},\ and\ \bibinfo {author} {\bibfnamefont {{\v{C}}.}~\bibnamefont {{Brukner}}},\ }\bibfield  {title} {\bibinfo {title} {{A purification postulate for quantum mechanics with indefinite causal order}},\ }\href {https://doi.org/10.22331/q-2017-04-26-10} {\bibfield  {journal} {\bibinfo  {journal} {Quantum}\ }\textbf {\bibinfo {volume} {1}},\ \bibinfo {pages} {10} (\bibinfo {year} {2017}{\natexlab{a}})},\ \Eprint {https://arxiv.org/abs/1611.08535} {arXiv:1611.08535 [quant-ph]} \BibitemShut {NoStop}%
\bibitem [{\citenamefont {{Silva}}\ \emph {et~al.}(2017)\citenamefont {{Silva}}, \citenamefont {{Guryanova}}, \citenamefont {{Short}}, \citenamefont {{Skrzypczyk}}, \citenamefont {{Brunner}},\ and\ \citenamefont {{Popescu}}}]{Silva2017Connecting}%
  \BibitemOpen
  \bibfield  {author} {\bibinfo {author} {\bibfnamefont {R.}~\bibnamefont {{Silva}}}, \bibinfo {author} {\bibfnamefont {Y.}~\bibnamefont {{Guryanova}}}, \bibinfo {author} {\bibfnamefont {A.~J.}\ \bibnamefont {{Short}}}, \bibinfo {author} {\bibfnamefont {P.}~\bibnamefont {{Skrzypczyk}}}, \bibinfo {author} {\bibfnamefont {N.}~\bibnamefont {{Brunner}}},\ and\ \bibinfo {author} {\bibfnamefont {S.}~\bibnamefont {{Popescu}}},\ }\bibfield  {title} {\bibinfo {title} {{Connecting processes with indefinite causal order and multi-time quantum states}},\ }\href {https://doi.org/10.1088/1367-2630/aa84fe} {\bibfield  {journal} {\bibinfo  {journal} {New Journal of Physics}\ }\textbf {\bibinfo {volume} {19}},\ \bibinfo {eid} {103022} (\bibinfo {year} {2017})},\ \Eprint {https://arxiv.org/abs/1701.08638} {arXiv:1701.08638 [quant-ph]} \BibitemShut {NoStop}%
\bibitem [{\citenamefont {{Dimi{\'c}}}\ \emph {et~al.}(2020)\citenamefont {{Dimi{\'c}}}, \citenamefont {{Milivojevi{\'c}}}, \citenamefont {{Go{\v{c}}anin}}, \citenamefont {{M{\'o}ller}},\ and\ \citenamefont {{Brukner}}}]{Dimic2020SimulatingICO}%
  \BibitemOpen
  \bibfield  {author} {\bibinfo {author} {\bibfnamefont {A.}~\bibnamefont {{Dimi{\'c}}}}, \bibinfo {author} {\bibfnamefont {M.}~\bibnamefont {{Milivojevi{\'c}}}}, \bibinfo {author} {\bibfnamefont {D.}~\bibnamefont {{Go{\v{c}}anin}}}, \bibinfo {author} {\bibfnamefont {N.~S.}\ \bibnamefont {{M{\'o}ller}}},\ and\ \bibinfo {author} {\bibfnamefont {{\v{C}}.}~\bibnamefont {{Brukner}}},\ }\bibfield  {title} {\bibinfo {title} {{Simulating indefinite causal order with Rindler observers}},\ }\href {https://doi.org/10.3389/fphy.2020.525333} {\bibfield  {journal} {\bibinfo  {journal} {Frontiers in Physics}\ }\textbf {\bibinfo {volume} {8}},\ \bibinfo {eid} {470} (\bibinfo {year} {2020})},\ \Eprint {https://arxiv.org/abs/1712.02689} {arXiv:1712.02689 [quant-ph]} \BibitemShut {NoStop}%
\bibitem [{\citenamefont {van~der Lugt}\ \emph {et~al.}(2023)\citenamefont {van~der Lugt}, \citenamefont {Barrett},\ and\ \citenamefont {Chiribella}}]{van2023device}%
  \BibitemOpen
  \bibfield  {author} {\bibinfo {author} {\bibfnamefont {T.}~\bibnamefont {van~der Lugt}}, \bibinfo {author} {\bibfnamefont {J.}~\bibnamefont {Barrett}},\ and\ \bibinfo {author} {\bibfnamefont {G.}~\bibnamefont {Chiribella}},\ }\bibfield  {title} {\bibinfo {title} {Device-independent certification of indefinite causal order in the quantum switch},\ }\href {https://doi.org/10.1038/s41467-023-40162-8} {\bibfield  {journal} {\bibinfo  {journal} {Nature Communications}\ }\textbf {\bibinfo {volume} {14}},\ \bibinfo {pages} {5811} (\bibinfo {year} {2023})}\BibitemShut {NoStop}%
\bibitem [{\citenamefont {Gogioso}\ and\ \citenamefont {Pinzani}(2023)}]{gogioso2023geometry}%
  \BibitemOpen
  \bibfield  {author} {\bibinfo {author} {\bibfnamefont {S.}~\bibnamefont {Gogioso}}\ and\ \bibinfo {author} {\bibfnamefont {N.}~\bibnamefont {Pinzani}},\ }\bibfield  {title} {\bibinfo {title} {The geometry of causality},\ }\bibfield  {journal} {\bibinfo  {journal} {arXiv preprint arXiv:2303.09017}\ }\href {https://doi.org/arXiv.2303.09017} {arXiv.2303.09017} (\bibinfo {year} {2023})\BibitemShut {NoStop}%
\bibitem [{\citenamefont {Bong}\ \emph {et~al.}(2020)\citenamefont {Bong}, \citenamefont {Utreras-Alarc{\'o}n}, \citenamefont {Ghafari}, \citenamefont {Liang}, \citenamefont {Tischler}, \citenamefont {Cavalcanti}, \citenamefont {Pryde},\ and\ \citenamefont {Wiseman}}]{bong2020strong}%
  \BibitemOpen
  \bibfield  {author} {\bibinfo {author} {\bibfnamefont {K.-W.}\ \bibnamefont {Bong}}, \bibinfo {author} {\bibfnamefont {A.}~\bibnamefont {Utreras-Alarc{\'o}n}}, \bibinfo {author} {\bibfnamefont {F.}~\bibnamefont {Ghafari}}, \bibinfo {author} {\bibfnamefont {Y.-C.}\ \bibnamefont {Liang}}, \bibinfo {author} {\bibfnamefont {N.}~\bibnamefont {Tischler}}, \bibinfo {author} {\bibfnamefont {E.~G.}\ \bibnamefont {Cavalcanti}}, \bibinfo {author} {\bibfnamefont {G.~J.}\ \bibnamefont {Pryde}},\ and\ \bibinfo {author} {\bibfnamefont {H.~M.}\ \bibnamefont {Wiseman}},\ }\bibfield  {title} {\bibinfo {title} {A strong no-go theorem on the wigner’s friend paradox},\ }\href {https://doi.org/https://doi.org/10.1038/s41567-020-0990-x} {\bibfield  {journal} {\bibinfo  {journal} {Nature Physics}\ }\textbf {\bibinfo {volume} {16}},\ \bibinfo {pages} {1199} (\bibinfo {year} {2020})}\BibitemShut {NoStop}%
\bibitem [{\citenamefont {van~der Lugt}\ and\ \citenamefont {Ormrod}(2023)}]{van2023possibilistic}%
  \BibitemOpen
  \bibfield  {author} {\bibinfo {author} {\bibfnamefont {T.}~\bibnamefont {van~der Lugt}}\ and\ \bibinfo {author} {\bibfnamefont {N.}~\bibnamefont {Ormrod}},\ }\bibfield  {title} {\bibinfo {title} {Possibilistic and maximal indefinite causal order in the quantum switch},\ }\bibfield  {journal} {\bibinfo  {journal} {arXiv preprint arXiv:2311.00557}\ }\href {https://doi.org/arXiv:2311.00557} {arXiv:2311.00557} (\bibinfo {year} {2023})\BibitemShut {NoStop}%
\bibitem [{\citenamefont {{Taddei}}\ \emph {et~al.}(2019)\citenamefont {{Taddei}}, \citenamefont {{Nery}},\ and\ \citenamefont {{Aolita}}}]{Taddei2019operationalResource}%
  \BibitemOpen
  \bibfield  {author} {\bibinfo {author} {\bibfnamefont {M.~M.}\ \bibnamefont {{Taddei}}}, \bibinfo {author} {\bibfnamefont {R.~V.}\ \bibnamefont {{Nery}}},\ and\ \bibinfo {author} {\bibfnamefont {L.}~\bibnamefont {{Aolita}}},\ }\bibfield  {title} {\bibinfo {title} {{Quantum superpositions of causal orders as an operational resource}},\ }\href {https://doi.org/10.1103/PhysRevResearch.1.033174} {\bibfield  {journal} {\bibinfo  {journal} {Physical Review Research}\ }\textbf {\bibinfo {volume} {1}},\ \bibinfo {eid} {033174} (\bibinfo {year} {2019})},\ \Eprint {https://arxiv.org/abs/1903.06180} {arXiv:1903.06180 [quant-ph]} \BibitemShut {NoStop}%
\bibitem [{\citenamefont {Chitambar}\ and\ \citenamefont {Gour}(2019)}]{RevModPhys.91.025001}%
  \BibitemOpen
  \bibfield  {author} {\bibinfo {author} {\bibfnamefont {E.}~\bibnamefont {Chitambar}}\ and\ \bibinfo {author} {\bibfnamefont {G.}~\bibnamefont {Gour}},\ }\bibfield  {title} {\bibinfo {title} {Quantum resource theories},\ }\href {https://doi.org/10.1103/RevModPhys.91.025001} {\bibfield  {journal} {\bibinfo  {journal} {Rev. Mod. Phys.}\ }\textbf {\bibinfo {volume} {91}},\ \bibinfo {pages} {025001} (\bibinfo {year} {2019})}\BibitemShut {NoStop}%
\bibitem [{\citenamefont {{Ara{\'u}jo}}\ \emph {et~al.}(2017{\natexlab{b}})\citenamefont {{Ara{\'u}jo}}, \citenamefont {{Gu{\'e}rin}},\ and\ \citenamefont {{Baumeler}}}]{Araujo2017QuantumComputation}%
  \BibitemOpen
  \bibfield  {author} {\bibinfo {author} {\bibfnamefont {M.}~\bibnamefont {{Ara{\'u}jo}}}, \bibinfo {author} {\bibfnamefont {P.~A.}\ \bibnamefont {{Gu{\'e}rin}}},\ and\ \bibinfo {author} {\bibfnamefont {A.}~\bibnamefont {{Baumeler}}},\ }\bibfield  {title} {\bibinfo {title} {{Quantum computation with indefinite causal structures}},\ }\href {https://doi.org/10.1103/PhysRevA.96.052315} {\bibfield  {journal} {\bibinfo  {journal} {Phys. Rev. A}\ }\textbf {\bibinfo {volume} {96}},\ \bibinfo {eid} {052315} (\bibinfo {year} {2017}{\natexlab{b}})},\ \Eprint {https://arxiv.org/abs/1706.09854} {arXiv:1706.09854 [quant-ph]} \BibitemShut {NoStop}%
\bibitem [{\citenamefont {{Chiribella}}(2012)}]{Chiribella2012PerfectDiscrimination}%
  \BibitemOpen
  \bibfield  {author} {\bibinfo {author} {\bibfnamefont {G.}~\bibnamefont {{Chiribella}}},\ }\bibfield  {title} {\bibinfo {title} {{Perfect discrimination of no-signalling channels via quantum superposition of causal structures}},\ }\href {https://doi.org/10.1103/PhysRevA.86.040301} {\bibfield  {journal} {\bibinfo  {journal} {Phys. Rev. A}\ }\textbf {\bibinfo {volume} {86}},\ \bibinfo {eid} {040301} (\bibinfo {year} {2012})},\ \Eprint {https://arxiv.org/abs/1109.5154} {arXiv:1109.5154 [quant-ph]} \BibitemShut {NoStop}%
\bibitem [{\citenamefont {{Renner}}\ and\ \citenamefont {{Brukner}}(2021)}]{Renner2021Reassessing}%
  \BibitemOpen
  \bibfield  {author} {\bibinfo {author} {\bibfnamefont {M.~J.}\ \bibnamefont {{Renner}}}\ and\ \bibinfo {author} {\bibfnamefont {{\v{C}}.}~\bibnamefont {{Brukner}}},\ }\bibfield  {title} {\bibinfo {title} {{Reassessing the computational advantage of quantum-controlled ordering of gates}},\ }\href {https://doi.org/10.1103/PhysRevResearch.3.043012} {\bibfield  {journal} {\bibinfo  {journal} {Physical Review Research}\ }\textbf {\bibinfo {volume} {3}},\ \bibinfo {eid} {043012} (\bibinfo {year} {2021})},\ \Eprint {https://arxiv.org/abs/2102.11293} {arXiv:2102.11293 [quant-ph]} \BibitemShut {NoStop}%
\bibitem [{\citenamefont {{Renner}}\ and\ \citenamefont {{Brukner}}(2022)}]{renner2022advantage}%
  \BibitemOpen
  \bibfield  {author} {\bibinfo {author} {\bibfnamefont {M.~J.}\ \bibnamefont {{Renner}}}\ and\ \bibinfo {author} {\bibfnamefont {{\v{C}}.}~\bibnamefont {{Brukner}}},\ }\bibfield  {title} {\bibinfo {title} {{Computational Advantage from a Quantum Superposition of Qubit Gate Orders}},\ }\href {https://doi.org/10.1103/PhysRevLett.128.230503} {\bibfield  {journal} {\bibinfo  {journal} {Phys. Rev. Lett.}\ }\textbf {\bibinfo {volume} {128}},\ \bibinfo {eid} {230503} (\bibinfo {year} {2022})},\ \Eprint {https://arxiv.org/abs/2112.14541} {arXiv:2112.14541 [quant-ph]} \BibitemShut {NoStop}%
\bibitem [{\citenamefont {Buhrman}\ \emph {et~al.}(2010)\citenamefont {Buhrman}, \citenamefont {Cleve}, \citenamefont {Massar},\ and\ \citenamefont {de~Wolf}}]{RevModPhys.82.665}%
  \BibitemOpen
  \bibfield  {author} {\bibinfo {author} {\bibfnamefont {H.}~\bibnamefont {Buhrman}}, \bibinfo {author} {\bibfnamefont {R.}~\bibnamefont {Cleve}}, \bibinfo {author} {\bibfnamefont {S.}~\bibnamefont {Massar}},\ and\ \bibinfo {author} {\bibfnamefont {R.}~\bibnamefont {de~Wolf}},\ }\bibfield  {title} {\bibinfo {title} {Nonlocality and communication complexity},\ }\href {https://doi.org/10.1103/RevModPhys.82.665} {\bibfield  {journal} {\bibinfo  {journal} {Rev. Mod. Phys.}\ }\textbf {\bibinfo {volume} {82}},\ \bibinfo {pages} {665} (\bibinfo {year} {2010})}\BibitemShut {NoStop}%
\bibitem [{\citenamefont {Buhrman}\ \emph {et~al.}(1999)\citenamefont {Buhrman}, \citenamefont {Cleve},\ and\ \citenamefont {Wigderson}}]{DJproblem}%
  \BibitemOpen
  \bibfield  {author} {\bibinfo {author} {\bibfnamefont {H.~R.}\ \bibnamefont {Buhrman}}, \bibinfo {author} {\bibfnamefont {R.}~\bibnamefont {Cleve}},\ and\ \bibinfo {author} {\bibfnamefont {A.}~\bibnamefont {Wigderson}},\ }in\ \href@noop {} {\emph {\bibinfo {booktitle} {Proceedings of the 30-th annual ACM symposium on theory of computing}}}\ (\bibinfo {year} {1999})\ pp.\ \bibinfo {pages} {63--68}\BibitemShut {NoStop}%
\bibitem [{\citenamefont {Raz}(1999)}]{raz1999exponential}%
  \BibitemOpen
  \bibfield  {author} {\bibinfo {author} {\bibfnamefont {R.}~\bibnamefont {Raz}},\ }in\ \href@noop {} {\emph {\bibinfo {booktitle} {Proceedings of the thirty-first annual ACM symposium on theory of computing}}}\ (\bibinfo {year} {1999})\ pp.\ \bibinfo {pages} {358--367}\BibitemShut {NoStop}%
\bibitem [{\citenamefont {Baumeler}\ and\ \citenamefont {Wolf}(2014)}]{baumeler2014perfectSignaling}%
  \BibitemOpen
  \bibfield  {author} {\bibinfo {author} {\bibfnamefont {{\"A}.}~\bibnamefont {Baumeler}}\ and\ \bibinfo {author} {\bibfnamefont {S.}~\bibnamefont {Wolf}},\ }\bibfield  {title} {\bibinfo {title} {Perfect signaling among three parties violating predefined causal order},\ }in\ \href {https://doi.org/10.1109/ISIT.2014.6874888} {\emph {\bibinfo {booktitle} {2014 IEEE International Symposium on Information Theory}}}\ (\bibinfo {organization} {IEEE},\ \bibinfo {year} {2014})\ pp.\ \bibinfo {pages} {526--530},\ \Eprint {https://arxiv.org/abs/1312.5916} {arXiv:1312.5916 [quant-ph]} \BibitemShut {NoStop}%
\bibitem [{\citenamefont {{Feix}}\ \emph {et~al.}(2015)\citenamefont {{Feix}}, \citenamefont {{Ara{\'u}jo}},\ and\ \citenamefont {{Brukner}}}]{feix2015communicationResource}%
  \BibitemOpen
  \bibfield  {author} {\bibinfo {author} {\bibfnamefont {A.}~\bibnamefont {{Feix}}}, \bibinfo {author} {\bibfnamefont {M.}~\bibnamefont {{Ara{\'u}jo}}},\ and\ \bibinfo {author} {\bibfnamefont {{\v{C}}.}~\bibnamefont {{Brukner}}},\ }\bibfield  {title} {\bibinfo {title} {{Quantum superposition of the order of parties as a communication resource}},\ }\href {https://doi.org/10.1103/PhysRevA.92.052326} {\bibfield  {journal} {\bibinfo  {journal} {Phys. Rev. A}\ }\textbf {\bibinfo {volume} {92}},\ \bibinfo {eid} {052326} (\bibinfo {year} {2015})},\ \Eprint {https://arxiv.org/abs/1508.07840} {arXiv:1508.07840 [quant-ph]} \BibitemShut {NoStop}%
\bibitem [{\citenamefont {{Ebler}}\ \emph {et~al.}(2018)\citenamefont {{Ebler}}, \citenamefont {{Salek}},\ and\ \citenamefont {{Chiribella}}}]{Ebler2018EnhancedCommunication}%
  \BibitemOpen
  \bibfield  {author} {\bibinfo {author} {\bibfnamefont {D.}~\bibnamefont {{Ebler}}}, \bibinfo {author} {\bibfnamefont {S.}~\bibnamefont {{Salek}}},\ and\ \bibinfo {author} {\bibfnamefont {G.}~\bibnamefont {{Chiribella}}},\ }\bibfield  {title} {\bibinfo {title} {{Enhanced Communication with the Assistance of Indefinite Causal Order}},\ }\href {https://doi.org/10.1103/PhysRevLett.120.120502} {\bibfield  {journal} {\bibinfo  {journal} {Phys. Rev. Lett.}\ }\textbf {\bibinfo {volume} {120}},\ \bibinfo {eid} {120502} (\bibinfo {year} {2018})},\ \Eprint {https://arxiv.org/abs/1711.10165} {arXiv:1711.10165 [quant-ph]} \BibitemShut {NoStop}%
\bibitem [{\citenamefont {{Salek}}\ \emph {et~al.}(2018)\citenamefont {{Salek}}, \citenamefont {{Ebler}},\ and\ \citenamefont {{Chiribella}}}]{Salek2018QuantumCommunication}%
  \BibitemOpen
  \bibfield  {author} {\bibinfo {author} {\bibfnamefont {S.}~\bibnamefont {{Salek}}}, \bibinfo {author} {\bibfnamefont {D.}~\bibnamefont {{Ebler}}},\ and\ \bibinfo {author} {\bibfnamefont {G.}~\bibnamefont {{Chiribella}}},\ }\bibfield  {title} {\bibinfo {title} {{Quantum communication in a superposition of causal orders}},\ }\href {https://doi.org/10.48550/arXiv.1809.06655} {\bibfield  {journal} {\bibinfo  {journal} {arXiv e-prints}\ ,\ \bibinfo {eid} {arXiv:1809.06655}} (\bibinfo {year} {2018})},\ \Eprint {https://arxiv.org/abs/1809.06655} {arXiv:1809.06655 [quant-ph]} \BibitemShut {NoStop}%
\bibitem [{\citenamefont {Wilde}(2013)}]{wilde2013quantum}%
  \BibitemOpen
  \bibfield  {author} {\bibinfo {author} {\bibfnamefont {M.~M.}\ \bibnamefont {Wilde}},\ }\href@noop {} {\emph {\bibinfo {title} {Quantum information theory}}}\ (\bibinfo  {publisher} {Cambridge university press},\ \bibinfo {year} {2013})\BibitemShut {NoStop}%
\bibitem [{\citenamefont {{Chiribella}}\ \emph {et~al.}(2021)\citenamefont {{Chiribella}}, \citenamefont {{Banik}}, \citenamefont {{Bhattacharya}}, \citenamefont {{Guha}}, \citenamefont {{Alimuddin}}, \citenamefont {{Roy}}, \citenamefont {{Saha}}, \citenamefont {{Agrawal}},\ and\ \citenamefont {{Kar}}}]{Chiribella2021perfectQuantumCommunication}%
  \BibitemOpen
  \bibfield  {author} {\bibinfo {author} {\bibfnamefont {G.}~\bibnamefont {{Chiribella}}}, \bibinfo {author} {\bibfnamefont {M.}~\bibnamefont {{Banik}}}, \bibinfo {author} {\bibfnamefont {S.~S.}\ \bibnamefont {{Bhattacharya}}}, \bibinfo {author} {\bibfnamefont {T.}~\bibnamefont {{Guha}}}, \bibinfo {author} {\bibfnamefont {M.}~\bibnamefont {{Alimuddin}}}, \bibinfo {author} {\bibfnamefont {A.}~\bibnamefont {{Roy}}}, \bibinfo {author} {\bibfnamefont {S.}~\bibnamefont {{Saha}}}, \bibinfo {author} {\bibfnamefont {S.}~\bibnamefont {{Agrawal}}},\ and\ \bibinfo {author} {\bibfnamefont {G.}~\bibnamefont {{Kar}}},\ }\bibfield  {title} {\bibinfo {title} {{Indefinite causal order enables perfect quantum communication with zero capacity channels}},\ }\href {https://doi.org/10.1088/1367-2630/abe7a0} {\bibfield  {journal} {\bibinfo  {journal} {New Journal of Physics}\ }\textbf {\bibinfo {volume} {23}},\ \bibinfo {eid} {033039} (\bibinfo {year} {2021})},\ \Eprint {https://arxiv.org/abs/1810.10457} {arXiv:1810.10457
  [quant-ph]} \BibitemShut {NoStop}%
\bibitem [{\citenamefont {{Kristj{\'a}nsson}}\ \emph {et~al.}(2020)\citenamefont {{Kristj{\'a}nsson}}, \citenamefont {{Chiribella}}, \citenamefont {{Salek}}, \citenamefont {{Ebler}},\ and\ \citenamefont {{Wilson}}}]{kristjansson2019ResourceTheories}%
  \BibitemOpen
  \bibfield  {author} {\bibinfo {author} {\bibfnamefont {H.}~\bibnamefont {{Kristj{\'a}nsson}}}, \bibinfo {author} {\bibfnamefont {G.}~\bibnamefont {{Chiribella}}}, \bibinfo {author} {\bibfnamefont {S.}~\bibnamefont {{Salek}}}, \bibinfo {author} {\bibfnamefont {D.}~\bibnamefont {{Ebler}}},\ and\ \bibinfo {author} {\bibfnamefont {M.}~\bibnamefont {{Wilson}}},\ }\bibfield  {title} {\bibinfo {title} {{Resource theories of communication}},\ }\href {https://doi.org/10.1088/1367-2630/ab8ef7} {\bibfield  {journal} {\bibinfo  {journal} {New Journal of Physics}\ }\textbf {\bibinfo {volume} {22}},\ \bibinfo {eid} {073014} (\bibinfo {year} {2020})},\ \Eprint {https://arxiv.org/abs/1910.08197} {arXiv:1910.08197 [quant-ph]} \BibitemShut {NoStop}%
\bibitem [{\citenamefont {{Chiribella}}\ and\ \citenamefont {{Kristj{\'a}nsson}}(2019)}]{chiribella2019quantumShannon}%
  \BibitemOpen
  \bibfield  {author} {\bibinfo {author} {\bibfnamefont {G.}~\bibnamefont {{Chiribella}}}\ and\ \bibinfo {author} {\bibfnamefont {H.}~\bibnamefont {{Kristj{\'a}nsson}}},\ }\bibfield  {title} {\bibinfo {title} {{Quantum Shannon theory with superpositions of trajectories}},\ }\href {https://doi.org/10.1098/rspa.2018.0903} {\bibfield  {journal} {\bibinfo  {journal} {Proceedings of the Royal Society of London Series A}\ }\textbf {\bibinfo {volume} {475}},\ \bibinfo {pages} {20180903} (\bibinfo {year} {2019})},\ \Eprint {https://arxiv.org/abs/1812.05292} {arXiv:1812.05292 [quant-ph]} \BibitemShut {NoStop}%
\bibitem [{\citenamefont {{Procopio}}\ \emph {et~al.}(2019)\citenamefont {{Procopio}}, \citenamefont {{Delgado}}, \citenamefont {{Enr{\'\i}quez}}, \citenamefont {{Belabas}},\ and\ \citenamefont {{Levenson}}}]{Procopio2019CommunicationEnhancement}%
  \BibitemOpen
  \bibfield  {author} {\bibinfo {author} {\bibfnamefont {L.~M.}\ \bibnamefont {{Procopio}}}, \bibinfo {author} {\bibfnamefont {F.}~\bibnamefont {{Delgado}}}, \bibinfo {author} {\bibfnamefont {M.}~\bibnamefont {{Enr{\'\i}quez}}}, \bibinfo {author} {\bibfnamefont {N.}~\bibnamefont {{Belabas}}},\ and\ \bibinfo {author} {\bibfnamefont {J.~A.}\ \bibnamefont {{Levenson}}},\ }\bibfield  {title} {\bibinfo {title} {{Communication Enhancement through Quantum Coherent Control of N Channels in an Indefinite Causal-Order Scenario}},\ }\href {https://doi.org/10.3390/e21101012} {\bibfield  {journal} {\bibinfo  {journal} {Entropy}\ }\textbf {\bibinfo {volume} {21}},\ \bibinfo {pages} {1012} (\bibinfo {year} {2019})},\ \Eprint {https://arxiv.org/abs/1902.01807} {arXiv:1902.01807 [quant-ph]} \BibitemShut {NoStop}%
\bibitem [{\citenamefont {{Sazim}}\ \emph {et~al.}(2021)\citenamefont {{Sazim}}, \citenamefont {{Sedlak}}, \citenamefont {{Singh}},\ and\ \citenamefont {{Pati}}}]{Sazim2021ClassicalCommunication}%
  \BibitemOpen
  \bibfield  {author} {\bibinfo {author} {\bibfnamefont {S.}~\bibnamefont {{Sazim}}}, \bibinfo {author} {\bibfnamefont {M.}~\bibnamefont {{Sedlak}}}, \bibinfo {author} {\bibfnamefont {K.}~\bibnamefont {{Singh}}},\ and\ \bibinfo {author} {\bibfnamefont {A.~K.}\ \bibnamefont {{Pati}}},\ }\bibfield  {title} {\bibinfo {title} {{Classical communication with indefinite causal order for N completely depolarizing channels}},\ }\href {https://doi.org/10.1103/PhysRevA.103.062610} {\bibfield  {journal} {\bibinfo  {journal} {Phys. Rev. A}\ }\textbf {\bibinfo {volume} {103}},\ \bibinfo {eid} {062610} (\bibinfo {year} {2021})},\ \Eprint {https://arxiv.org/abs/2004.14339} {arXiv:2004.14339 [quant-ph]} \BibitemShut {NoStop}%
\bibitem [{\citenamefont {Caleffi}\ and\ \citenamefont {Cacciapuoti}(2020)}]{caleffi2020NoiselessCommunications}%
  \BibitemOpen
  \bibfield  {author} {\bibinfo {author} {\bibfnamefont {M.}~\bibnamefont {Caleffi}}\ and\ \bibinfo {author} {\bibfnamefont {A.~S.}\ \bibnamefont {Cacciapuoti}},\ }\bibfield  {title} {\bibinfo {title} {Quantum switch for the quantum internet: Noiseless communications through noisy channels},\ }\href {https://doi.org/10.1109/JSAC.2020.2969035} {\bibfield  {journal} {\bibinfo  {journal} {IEEE Journal on Selected Areas in Communications}\ }\textbf {\bibinfo {volume} {38}},\ \bibinfo {pages} {575} (\bibinfo {year} {2020})}\BibitemShut {NoStop}%
\bibitem [{\citenamefont {{Abbott}}\ \emph {et~al.}(2020)\citenamefont {{Abbott}}, \citenamefont {{Wechs}}, \citenamefont {{Horsman}}, \citenamefont {{Mhalla}},\ and\ \citenamefont {{Branciard}}}]{Abbott2020coherentControl}%
  \BibitemOpen
  \bibfield  {author} {\bibinfo {author} {\bibfnamefont {A.~A.}\ \bibnamefont {{Abbott}}}, \bibinfo {author} {\bibfnamefont {J.}~\bibnamefont {{Wechs}}}, \bibinfo {author} {\bibfnamefont {D.}~\bibnamefont {{Horsman}}}, \bibinfo {author} {\bibfnamefont {M.}~\bibnamefont {{Mhalla}}},\ and\ \bibinfo {author} {\bibfnamefont {C.}~\bibnamefont {{Branciard}}},\ }\bibfield  {title} {\bibinfo {title} {{Communication through coherent control of quantum channels}},\ }\href {https://doi.org/10.22331/q-2020-09-24-333} {\bibfield  {journal} {\bibinfo  {journal} {Quantum}\ }\textbf {\bibinfo {volume} {4}},\ \bibinfo {pages} {333} (\bibinfo {year} {2020})}\BibitemShut {NoStop}%
\bibitem [{\citenamefont {{Gu{\'e}rin}}\ \emph {et~al.}(2019)\citenamefont {{Gu{\'e}rin}}, \citenamefont {{Rubino}},\ and\ \citenamefont {{Brukner}}}]{Guerin2019quantumControlledNoise}%
  \BibitemOpen
  \bibfield  {author} {\bibinfo {author} {\bibfnamefont {P.~A.}\ \bibnamefont {{Gu{\'e}rin}}}, \bibinfo {author} {\bibfnamefont {G.}~\bibnamefont {{Rubino}}},\ and\ \bibinfo {author} {\bibfnamefont {{\v{C}}.}~\bibnamefont {{Brukner}}},\ }\bibfield  {title} {\bibinfo {title} {{Communication through quantum-controlled noise}},\ }\href {https://doi.org/10.1103/PhysRevA.99.062317} {\bibfield  {journal} {\bibinfo  {journal} {Phys. Rev. A}\ }\textbf {\bibinfo {volume} {99}},\ \bibinfo {eid} {062317} (\bibinfo {year} {2019})}\BibitemShut {NoStop}%
\bibitem [{\citenamefont {{Gisin}}\ \emph {et~al.}(2005)\citenamefont {{Gisin}}, \citenamefont {{Linden}}, \citenamefont {{Massar}},\ and\ \citenamefont {{Popescu}}}]{Gisin2005ErrorFiltration}%
  \BibitemOpen
  \bibfield  {author} {\bibinfo {author} {\bibfnamefont {N.}~\bibnamefont {{Gisin}}}, \bibinfo {author} {\bibfnamefont {N.}~\bibnamefont {{Linden}}}, \bibinfo {author} {\bibfnamefont {S.}~\bibnamefont {{Massar}}},\ and\ \bibinfo {author} {\bibfnamefont {S.}~\bibnamefont {{Popescu}}},\ }\bibfield  {title} {\bibinfo {title} {{Error filtration and entanglement purification for quantum communication}},\ }\href {https://doi.org/10.1103/PhysRevA.72.012338} {\bibfield  {journal} {\bibinfo  {journal} {Phys. Rev. A}\ }\textbf {\bibinfo {volume} {72}},\ \bibinfo {eid} {012338} (\bibinfo {year} {2005})},\ \Eprint {https://arxiv.org/abs/quant-ph/0407021} {arXiv:quant-ph/0407021 [quant-ph]} \BibitemShut {NoStop}%
\bibitem [{\citenamefont {Pang}\ \emph {et~al.}(2023)\citenamefont {Pang}, \citenamefont {Lupu-Gladstein}, \citenamefont {Ferretti}, \citenamefont {Yilmaz}, \citenamefont {Brodutch},\ and\ \citenamefont {Steinberg}}]{pang2023experimental}%
  \BibitemOpen
  \bibfield  {author} {\bibinfo {author} {\bibfnamefont {A.~O.}\ \bibnamefont {Pang}}, \bibinfo {author} {\bibfnamefont {N.}~\bibnamefont {Lupu-Gladstein}}, \bibinfo {author} {\bibfnamefont {H.}~\bibnamefont {Ferretti}}, \bibinfo {author} {\bibfnamefont {Y.~B.}\ \bibnamefont {Yilmaz}}, \bibinfo {author} {\bibfnamefont {A.}~\bibnamefont {Brodutch}},\ and\ \bibinfo {author} {\bibfnamefont {A.~M.}\ \bibnamefont {Steinberg}},\ }\bibfield  {title} {\bibinfo {title} {Experimental communication through superposition of quantum channels},\ }\bibfield  {journal} {\bibinfo  {journal} {arXiv preprint arXiv:2302.14820}\ }\href {https://doi.org/10.48550/arXiv.2302.14820} {10.48550/arXiv.2302.14820} (\bibinfo {year} {2023})\BibitemShut {NoStop}%
\bibitem [{\citenamefont {Lee}\ \emph {et~al.}(2023)\citenamefont {Lee}, \citenamefont {Hann}, \citenamefont {Puri}, \citenamefont {Girvin},\ and\ \citenamefont {Jiang}}]{Lee23Error}%
  \BibitemOpen
  \bibfield  {author} {\bibinfo {author} {\bibfnamefont {G.}~\bibnamefont {Lee}}, \bibinfo {author} {\bibfnamefont {C.~T.}\ \bibnamefont {Hann}}, \bibinfo {author} {\bibfnamefont {S.}~\bibnamefont {Puri}}, \bibinfo {author} {\bibfnamefont {S.~M.}\ \bibnamefont {Girvin}},\ and\ \bibinfo {author} {\bibfnamefont {L.}~\bibnamefont {Jiang}},\ }\bibfield  {title} {\bibinfo {title} {Error suppression for arbitrary-size black box quantum operations},\ }\href {https://doi.org/10.1103/PhysRevLett.131.190601} {\bibfield  {journal} {\bibinfo  {journal} {Phys. Rev. Lett.}\ }\textbf {\bibinfo {volume} {131}},\ \bibinfo {pages} {190601} (\bibinfo {year} {2023})}\BibitemShut {NoStop}%
\bibitem [{\citenamefont {Miguel-Ramiro}\ \emph {et~al.}(2023)\citenamefont {Miguel-Ramiro}, \citenamefont {Shi}, \citenamefont {Dellantonio}, \citenamefont {Chan}, \citenamefont {Muschik},\ and\ \citenamefont {D\"ur}}]{RamiroSQEM2023}%
  \BibitemOpen
  \bibfield  {author} {\bibinfo {author} {\bibfnamefont {J.}~\bibnamefont {Miguel-Ramiro}}, \bibinfo {author} {\bibfnamefont {Z.}~\bibnamefont {Shi}}, \bibinfo {author} {\bibfnamefont {L.}~\bibnamefont {Dellantonio}}, \bibinfo {author} {\bibfnamefont {A.}~\bibnamefont {Chan}}, \bibinfo {author} {\bibfnamefont {C.~A.}\ \bibnamefont {Muschik}},\ and\ \bibinfo {author} {\bibfnamefont {W.}~\bibnamefont {D\"ur}},\ }\bibfield  {title} {\bibinfo {title} {Superposed quantum error mitigation},\ }\href {https://doi.org/10.1103/PhysRevLett.131.230601} {\bibfield  {journal} {\bibinfo  {journal} {Phys. Rev. Lett.}\ }\textbf {\bibinfo {volume} {131}},\ \bibinfo {pages} {230601} (\bibinfo {year} {2023})}\BibitemShut {NoStop}%
\bibitem [{\citenamefont {{Miguel-Ramiro}}\ \emph {et~al.}(2023)\citenamefont {{Miguel-Ramiro}}, \citenamefont {{Shi}}, \citenamefont {{Dellantonio}}, \citenamefont {{Chan}}, \citenamefont {{Muschik}},\ and\ \citenamefont {{D{\"u}r}}}]{RamiroEnhancing2023}%
  \BibitemOpen
  \bibfield  {author} {\bibinfo {author} {\bibfnamefont {J.}~\bibnamefont {{Miguel-Ramiro}}}, \bibinfo {author} {\bibfnamefont {Z.}~\bibnamefont {{Shi}}}, \bibinfo {author} {\bibfnamefont {L.}~\bibnamefont {{Dellantonio}}}, \bibinfo {author} {\bibfnamefont {A.}~\bibnamefont {{Chan}}}, \bibinfo {author} {\bibfnamefont {C.~A.}\ \bibnamefont {{Muschik}}},\ and\ \bibinfo {author} {\bibfnamefont {W.}~\bibnamefont {{D{\"u}r}}},\ }\bibfield  {title} {\bibinfo {title} {{Enhancing Quantum Computation via Superposition of Quantum Gates}},\ }\href {https://doi.org/10.48550/arXiv.2304.08529} {\bibfield  {journal} {\bibinfo  {journal} {arXiv e-prints}\ ,\ \bibinfo {eid} {arXiv:2304.08529}} (\bibinfo {year} {2023})},\ \Eprint {https://arxiv.org/abs/2304.08529} {arXiv:2304.08529 [quant-ph]} \BibitemShut {NoStop}%
\bibitem [{\citenamefont {{Spencer-Wood}}(2023)}]{SpencerWood2023qkd}%
  \BibitemOpen
  \bibfield  {author} {\bibinfo {author} {\bibfnamefont {H.}~\bibnamefont {{Spencer-Wood}}},\ }\bibfield  {title} {\bibinfo {title} {{Indefinite causal key distribution}},\ }\href {https://doi.org/10.48550/arXiv.2303.03893} {\bibfield  {journal} {\bibinfo  {journal} {arXiv e-prints}\ ,\ \bibinfo {eid} {arXiv:2303.03893}} (\bibinfo {year} {2023})},\ \Eprint {https://arxiv.org/abs/2303.03893} {arXiv:2303.03893 [quant-ph]} \BibitemShut {NoStop}%
\bibitem [{\citenamefont {Koudia}\ \emph {et~al.}(2023)\citenamefont {Koudia}, \citenamefont {Cacciapuoti},\ and\ \citenamefont {Caleffi}}]{Koudia2023EntanglementGeneration}%
  \BibitemOpen
  \bibfield  {author} {\bibinfo {author} {\bibfnamefont {S.}~\bibnamefont {Koudia}}, \bibinfo {author} {\bibfnamefont {A.~S.}\ \bibnamefont {Cacciapuoti}},\ and\ \bibinfo {author} {\bibfnamefont {M.}~\bibnamefont {Caleffi}},\ }\bibfield  {title} {\bibinfo {title} {Deterministic generation of multipartite entanglement via causal activation in the quantum internet},\ }\href {https://doi.org/10.1109/ACCESS.2023.3296587} {\bibfield  {journal} {\bibinfo  {journal} {IEEE Access}\ }\textbf {\bibinfo {volume} {11}},\ \bibinfo {pages} {73863} (\bibinfo {year} {2023})},\ \Eprint {https://arxiv.org/abs/2112.00543} {arXiv:2112.00543 [quant-ph]} \BibitemShut {NoStop}%
\bibitem [{\citenamefont {{Dey}}\ and\ \citenamefont {{Marchetti}}(2023)}]{Dey2023EntanglementDistribution}%
  \BibitemOpen
  \bibfield  {author} {\bibinfo {author} {\bibfnamefont {I.}~\bibnamefont {{Dey}}}\ and\ \bibinfo {author} {\bibfnamefont {N.}~\bibnamefont {{Marchetti}}},\ }\bibfield  {title} {\bibinfo {title} {{Entanglement Distribution and Quantum Teleportation in Higher Dimension over the Superposition of Causal Orders of Quantum Channels}},\ }\href {https://doi.org/10.48550/arXiv.2303.10683} {\bibfield  {journal} {\bibinfo  {journal} {arXiv e-prints}\ ,\ \bibinfo {eid} {arXiv:2303.10683}} (\bibinfo {year} {2023})},\ \Eprint {https://arxiv.org/abs/2303.10683} {arXiv:2303.10683 [quant-ph]} \BibitemShut {NoStop}%
\bibitem [{\citenamefont {Simonov}\ \emph {et~al.}(2023)\citenamefont {Simonov}, \citenamefont {Caleffi}, \citenamefont {Illiano},\ and\ \citenamefont {Cacciapuoti}}]{simonov2023universal}%
  \BibitemOpen
  \bibfield  {author} {\bibinfo {author} {\bibfnamefont {K.}~\bibnamefont {Simonov}}, \bibinfo {author} {\bibfnamefont {M.}~\bibnamefont {Caleffi}}, \bibinfo {author} {\bibfnamefont {J.}~\bibnamefont {Illiano}},\ and\ \bibinfo {author} {\bibfnamefont {A.~S.}\ \bibnamefont {Cacciapuoti}},\ }\href@noop {} {\bibinfo {title} {Universal quantum computation via superposed orders of single-qubit gates}} (\bibinfo {year} {2023}),\ \Eprint {https://arxiv.org/abs/2311.13654} {arXiv:2311.13654 [quant-ph]} \BibitemShut {NoStop}%
\bibitem [{\citenamefont {{Zuo}}\ \emph {et~al.}(2023)\citenamefont {{Zuo}}, \citenamefont {{Hanks}},\ and\ \citenamefont {{Kim}}}]{Zuo2023EntanglementDistillation}%
  \BibitemOpen
  \bibfield  {author} {\bibinfo {author} {\bibfnamefont {Z.}~\bibnamefont {{Zuo}}}, \bibinfo {author} {\bibfnamefont {M.}~\bibnamefont {{Hanks}}},\ and\ \bibinfo {author} {\bibfnamefont {M.~S.}\ \bibnamefont {{Kim}}},\ }\bibfield  {title} {\bibinfo {title} {{Coherent Control of Causal Order of Entanglement Distillation}},\ }\href {https://doi.org/10.48550/arXiv.2302.13990} {\bibfield  {journal} {\bibinfo  {journal} {arXiv e-prints}\ ,\ \bibinfo {eid} {arXiv:2302.13990}} (\bibinfo {year} {2023})},\ \Eprint {https://arxiv.org/abs/2302.13990} {arXiv:2302.13990 [quant-ph]} \BibitemShut {NoStop}%
\bibitem [{\citenamefont {Chen}\ and\ \citenamefont {Hasegawa}(2021)}]{chen2021indefinite}%
  \BibitemOpen
  \bibfield  {author} {\bibinfo {author} {\bibfnamefont {Y.}~\bibnamefont {Chen}}\ and\ \bibinfo {author} {\bibfnamefont {Y.}~\bibnamefont {Hasegawa}},\ }\href@noop {} {\bibinfo {title} {Indefinite causal order in quantum batteries}} (\bibinfo {year} {2021}),\ \Eprint {https://arxiv.org/abs/2105.12466} {arXiv:2105.12466 [quant-ph]} \BibitemShut {NoStop}%
\bibitem [{\citenamefont {{Guha}}\ \emph {et~al.}(2020)\citenamefont {{Guha}}, \citenamefont {{Alimuddin}},\ and\ \citenamefont {{Parashar}}}]{Guha2020Thermodynamic}%
  \BibitemOpen
  \bibfield  {author} {\bibinfo {author} {\bibfnamefont {T.}~\bibnamefont {{Guha}}}, \bibinfo {author} {\bibfnamefont {M.}~\bibnamefont {{Alimuddin}}},\ and\ \bibinfo {author} {\bibfnamefont {P.}~\bibnamefont {{Parashar}}},\ }\bibfield  {title} {\bibinfo {title} {{Thermodynamic advancement in the causally inseparable occurrence of thermal maps}},\ }\href {https://doi.org/10.1103/PhysRevA.102.032215} {\bibfield  {journal} {\bibinfo  {journal} {Phys. Rev. A}\ }\textbf {\bibinfo {volume} {102}},\ \bibinfo {eid} {032215} (\bibinfo {year} {2020})},\ \Eprint {https://arxiv.org/abs/2003.01464} {arXiv:2003.01464 [quant-ph]} \BibitemShut {NoStop}%
\bibitem [{\citenamefont {{Guha}}\ \emph {et~al.}(2022)\citenamefont {{Guha}}, \citenamefont {{Roy}}, \citenamefont {{Simonov}},\ and\ \citenamefont {{Zimbor{\'a}s}}}]{Guha2022ActivationThermalStates}%
  \BibitemOpen
  \bibfield  {author} {\bibinfo {author} {\bibfnamefont {T.}~\bibnamefont {{Guha}}}, \bibinfo {author} {\bibfnamefont {S.}~\bibnamefont {{Roy}}}, \bibinfo {author} {\bibfnamefont {K.}~\bibnamefont {{Simonov}}},\ and\ \bibinfo {author} {\bibfnamefont {Z.}~\bibnamefont {{Zimbor{\'a}s}}},\ }\bibfield  {title} {\bibinfo {title} {{Activation of thermal states by quantum SWITCH-driven thermalization and its limits}},\ }\href {https://doi.org/10.48550/arXiv.2208.04034} {\bibfield  {journal} {\bibinfo  {journal} {arXiv e-prints}\ ,\ \bibinfo {eid} {arXiv:2208.04034}} (\bibinfo {year} {2022})},\ \Eprint {https://arxiv.org/abs/2208.04034} {arXiv:2208.04034 [quant-ph]} \BibitemShut {NoStop}%
\bibitem [{\citenamefont {{Simonov}}\ \emph {et~al.}(2022)\citenamefont {{Simonov}}, \citenamefont {{Francica}}, \citenamefont {{Guarnieri}},\ and\ \citenamefont {{Paternostro}}}]{Simonov2022WorkExtraction}%
  \BibitemOpen
  \bibfield  {author} {\bibinfo {author} {\bibfnamefont {K.}~\bibnamefont {{Simonov}}}, \bibinfo {author} {\bibfnamefont {G.}~\bibnamefont {{Francica}}}, \bibinfo {author} {\bibfnamefont {G.}~\bibnamefont {{Guarnieri}}},\ and\ \bibinfo {author} {\bibfnamefont {M.}~\bibnamefont {{Paternostro}}},\ }\bibfield  {title} {\bibinfo {title} {{Work extraction from coherently activated maps via quantum switch}},\ }\href {https://doi.org/10.1103/PhysRevA.105.032217} {\bibfield  {journal} {\bibinfo  {journal} {Phys. Rev. A}\ }\textbf {\bibinfo {volume} {105}},\ \bibinfo {eid} {032217} (\bibinfo {year} {2022})}\BibitemShut {NoStop}%
\bibitem [{\citenamefont {Mancino}\ \emph {et~al.}(2017)\citenamefont {Mancino}, \citenamefont {Sbroscia}, \citenamefont {Gianani}, \citenamefont {Roccia},\ and\ \citenamefont {Barbieri}}]{Mancino17quantum}%
  \BibitemOpen
  \bibfield  {author} {\bibinfo {author} {\bibfnamefont {L.}~\bibnamefont {Mancino}}, \bibinfo {author} {\bibfnamefont {M.}~\bibnamefont {Sbroscia}}, \bibinfo {author} {\bibfnamefont {I.}~\bibnamefont {Gianani}}, \bibinfo {author} {\bibfnamefont {E.}~\bibnamefont {Roccia}},\ and\ \bibinfo {author} {\bibfnamefont {M.}~\bibnamefont {Barbieri}},\ }\bibfield  {title} {\bibinfo {title} {Quantum simulation of single-qubit thermometry using linear optics},\ }\href {https://doi.org/10.1103/PhysRevLett.118.130502} {\bibfield  {journal} {\bibinfo  {journal} {Phys. Rev. Lett.}\ }\textbf {\bibinfo {volume} {118}},\ \bibinfo {pages} {130502} (\bibinfo {year} {2017})}\BibitemShut {NoStop}%
\bibitem [{\citenamefont {Mancino}\ \emph {et~al.}(2018)\citenamefont {Mancino}, \citenamefont {Cavina}, \citenamefont {De~Pasquale}, \citenamefont {Sbroscia}, \citenamefont {Booth}, \citenamefont {Roccia}, \citenamefont {Gianani}, \citenamefont {Giovannetti},\ and\ \citenamefont {Barbieri}}]{Mancino18geometrical}%
  \BibitemOpen
  \bibfield  {author} {\bibinfo {author} {\bibfnamefont {L.}~\bibnamefont {Mancino}}, \bibinfo {author} {\bibfnamefont {V.}~\bibnamefont {Cavina}}, \bibinfo {author} {\bibfnamefont {A.}~\bibnamefont {De~Pasquale}}, \bibinfo {author} {\bibfnamefont {M.}~\bibnamefont {Sbroscia}}, \bibinfo {author} {\bibfnamefont {R.~I.}\ \bibnamefont {Booth}}, \bibinfo {author} {\bibfnamefont {E.}~\bibnamefont {Roccia}}, \bibinfo {author} {\bibfnamefont {I.}~\bibnamefont {Gianani}}, \bibinfo {author} {\bibfnamefont {V.}~\bibnamefont {Giovannetti}},\ and\ \bibinfo {author} {\bibfnamefont {M.}~\bibnamefont {Barbieri}},\ }\bibfield  {title} {\bibinfo {title} {Geometrical bounds on irreversibility in open quantum systems},\ }\href {https://doi.org/10.1103/PhysRevLett.121.160602} {\bibfield  {journal} {\bibinfo  {journal} {Phys. Rev. Lett.}\ }\textbf {\bibinfo {volume} {121}},\ \bibinfo {pages} {160602} (\bibinfo {year} {2018})}\BibitemShut {NoStop}%
\bibitem [{\citenamefont {{Ball}}(2022)}]{Ball2022fridge}%
  \BibitemOpen
  \bibfield  {author} {\bibinfo {author} {\bibfnamefont {P.}~\bibnamefont {{Ball}}},\ }\bibfield  {title} {\bibinfo {title} {{A fridge without a cause}},\ }\href {https://doi.org/10.1038/s41563-022-01377-0} {\bibfield  {journal} {\bibinfo  {journal} {Nature Materials}\ }\textbf {\bibinfo {volume} {21}},\ \bibinfo {pages} {1099} (\bibinfo {year} {2022})}\BibitemShut {NoStop}%
\bibitem [{\citenamefont {{Capela}}\ \emph {et~al.}(2023)\citenamefont {{Capela}}, \citenamefont {{Verma}}, \citenamefont {{Costa}},\ and\ \citenamefont {{C{\'e}leri}}}]{Capela2023ReassessingThermodynamic}%
  \BibitemOpen
  \bibfield  {author} {\bibinfo {author} {\bibfnamefont {M.}~\bibnamefont {{Capela}}}, \bibinfo {author} {\bibfnamefont {H.}~\bibnamefont {{Verma}}}, \bibinfo {author} {\bibfnamefont {F.}~\bibnamefont {{Costa}}},\ and\ \bibinfo {author} {\bibfnamefont {L.~C.}\ \bibnamefont {{C{\'e}leri}}},\ }\bibfield  {title} {\bibinfo {title} {{Reassessing thermodynamic advantage from indefinite causal order}},\ }\href {https://doi.org/10.1103/PhysRevA.107.062208} {\bibfield  {journal} {\bibinfo  {journal} {Phys. Rev. A}\ }\textbf {\bibinfo {volume} {107}},\ \bibinfo {eid} {062208} (\bibinfo {year} {2023})},\ \Eprint {https://arxiv.org/abs/2208.03205} {arXiv:2208.03205 [quant-ph]} \BibitemShut {NoStop}%
\bibitem [{\citenamefont {Liu}\ \emph {et~al.}(2022)\citenamefont {Liu}, \citenamefont {Ebler},\ and\ \citenamefont {Dahlsten}}]{Liu23thermodynamics}%
  \BibitemOpen
  \bibfield  {author} {\bibinfo {author} {\bibfnamefont {X.}~\bibnamefont {Liu}}, \bibinfo {author} {\bibfnamefont {D.}~\bibnamefont {Ebler}},\ and\ \bibinfo {author} {\bibfnamefont {O.}~\bibnamefont {Dahlsten}},\ }\bibfield  {title} {\bibinfo {title} {Thermodynamics of quantum switch information capacity activation},\ }\href {https://doi.org/10.1103/PhysRevLett.129.230604} {\bibfield  {journal} {\bibinfo  {journal} {Phys. Rev. Lett.}\ }\textbf {\bibinfo {volume} {129}},\ \bibinfo {pages} {230604} (\bibinfo {year} {2022})}\BibitemShut {NoStop}%
\bibitem [{\citenamefont {{Frey}}(2019)}]{Frey2019depolarizingChannelIdentification}%
  \BibitemOpen
  \bibfield  {author} {\bibinfo {author} {\bibfnamefont {M.}~\bibnamefont {{Frey}}},\ }\bibfield  {title} {\bibinfo {title} {{Indefinite causal order aids quantum depolarizing channel identification}},\ }\href {https://doi.org/10.1007/s11128-019-2186-9} {\bibfield  {journal} {\bibinfo  {journal} {Quantum Information Processing}\ }\textbf {\bibinfo {volume} {18}},\ \bibinfo {eid} {96} (\bibinfo {year} {2019})}\BibitemShut {NoStop}%
\bibitem [{\citenamefont {{Ban}}(2023{\natexlab{a}})}]{Ban2023FisherInformation}%
  \BibitemOpen
  \bibfield  {author} {\bibinfo {author} {\bibfnamefont {M.}~\bibnamefont {{Ban}}},\ }\bibfield  {title} {\bibinfo {title} {{Quantum Fisher information of phase estimation in the presence of indefinite causal order}},\ }\href {https://doi.org/10.1016/j.physleta.2023.128749} {\bibfield  {journal} {\bibinfo  {journal} {Physics Letters A}\ }\textbf {\bibinfo {volume} {468}},\ \bibinfo {eid} {128749} (\bibinfo {year} {2023}{\natexlab{a}})}\BibitemShut {NoStop}%
\bibitem [{\citenamefont {{Chapeau-Blondeau}}(2021)}]{Chapeau2021NoisyQuantumMetrology}%
  \BibitemOpen
  \bibfield  {author} {\bibinfo {author} {\bibfnamefont {F.}~\bibnamefont {{Chapeau-Blondeau}}},\ }\bibfield  {title} {\bibinfo {title} {{Noisy quantum metrology with the assistance of indefinite causal order}},\ }\href {https://doi.org/10.1103/PhysRevA.103.032615} {\bibfield  {journal} {\bibinfo  {journal} {Phys. Rev. A}\ }\textbf {\bibinfo {volume} {103}},\ \bibinfo {eid} {032615} (\bibinfo {year} {2021})},\ \Eprint {https://arxiv.org/abs/2104.06284} {arXiv:2104.06284 [quant-ph]} \BibitemShut {NoStop}%
\bibitem [{\citenamefont {{Delgado}}(2023)}]{Delgado2023EstimationPauliChannels}%
  \BibitemOpen
  \bibfield  {author} {\bibinfo {author} {\bibfnamefont {F.}~\bibnamefont {{Delgado}}},\ }\bibfield  {title} {\bibinfo {title} {{Parametric Symmetries in Architectures Involving Indefinite Causal Order and Path Superposition for Quantum Parameter Estimation of Pauli Channels}},\ }\href {https://doi.org/10.3390/sym15051097} {\bibfield  {journal} {\bibinfo  {journal} {Symmetry}\ }\textbf {\bibinfo {volume} {15}},\ \bibinfo {pages} {1097} (\bibinfo {year} {2023})}\BibitemShut {NoStop}%
\bibitem [{\citenamefont {Kurdzia\l{}ek}\ \emph {et~al.}(2023)\citenamefont {Kurdzia\l{}ek}, \citenamefont {G\'orecki}, \citenamefont {Albarelli},\ and\ \citenamefont {Demkowicz-Dobrza\ifmmode~\acute{n}\else \'{n}\fi{}ski}}]{Kurdzialek23}%
  \BibitemOpen
  \bibfield  {author} {\bibinfo {author} {\bibfnamefont {S.}~\bibnamefont {Kurdzia\l{}ek}}, \bibinfo {author} {\bibfnamefont {W.}~\bibnamefont {G\'orecki}}, \bibinfo {author} {\bibfnamefont {F.}~\bibnamefont {Albarelli}},\ and\ \bibinfo {author} {\bibfnamefont {R.}~\bibnamefont {Demkowicz-Dobrza\ifmmode~\acute{n}\else \'{n}\fi{}ski}},\ }\bibfield  {title} {\bibinfo {title} {Using adaptiveness and causal superpositions against noise in quantum metrology},\ }\href {https://doi.org/10.1103/PhysRevLett.131.090801} {\bibfield  {journal} {\bibinfo  {journal} {Phys. Rev. Lett.}\ }\textbf {\bibinfo {volume} {131}},\ \bibinfo {pages} {090801} (\bibinfo {year} {2023})}\BibitemShut {NoStop}%
\bibitem [{\citenamefont {{Zhao}}\ \emph {et~al.}(2020)\citenamefont {{Zhao}}, \citenamefont {{Yang}},\ and\ \citenamefont {{Chiribella}}}]{zhao2019QuantumMetrology}%
  \BibitemOpen
  \bibfield  {author} {\bibinfo {author} {\bibfnamefont {X.}~\bibnamefont {{Zhao}}}, \bibinfo {author} {\bibfnamefont {Y.}~\bibnamefont {{Yang}}},\ and\ \bibinfo {author} {\bibfnamefont {G.}~\bibnamefont {{Chiribella}}},\ }\bibfield  {title} {\bibinfo {title} {{Quantum Metrology with Indefinite Causal Order}},\ }\href {https://doi.org/10.1103/PhysRevLett.124.190503} {\bibfield  {journal} {\bibinfo  {journal} {Phys. Rev. Lett.}\ }\textbf {\bibinfo {volume} {124}},\ \bibinfo {eid} {190503} (\bibinfo {year} {2020})},\ \Eprint {https://arxiv.org/abs/1912.02449} {arXiv:1912.02449 [quant-ph]} \BibitemShut {NoStop}%
\bibitem [{\citenamefont {Giacomini}\ \emph {et~al.}(2016)\citenamefont {Giacomini}, \citenamefont {Castro-Ruiz},\ and\ \citenamefont {Časlav Brukner}}]{Giacomini_2016}%
  \BibitemOpen
  \bibfield  {author} {\bibinfo {author} {\bibfnamefont {F.}~\bibnamefont {Giacomini}}, \bibinfo {author} {\bibfnamefont {E.}~\bibnamefont {Castro-Ruiz}},\ and\ \bibinfo {author} {\bibnamefont {Časlav Brukner}},\ }\bibfield  {title} {\bibinfo {title} {Indefinite causal structures for continuous-variable systems},\ }\href {https://doi.org/10.1088/1367-2630/18/11/113026} {\bibfield  {journal} {\bibinfo  {journal} {New Journal of Physics}\ }\textbf {\bibinfo {volume} {18}},\ \bibinfo {pages} {113026} (\bibinfo {year} {2016})}\BibitemShut {NoStop}%
\bibitem [{\citenamefont {{Ban}}(2023{\natexlab{b}})}]{Ban2023Quantumness}%
  \BibitemOpen
  \bibfield  {author} {\bibinfo {author} {\bibfnamefont {M.}~\bibnamefont {{Ban}}},\ }\bibfield  {title} {\bibinfo {title} {{Quantumness of qubit states interacting with two structured reservoirs in indefinite causal order}},\ }\href {https://doi.org/10.1016/j.physleta.2023.128927} {\bibfield  {journal} {\bibinfo  {journal} {Physics Letters A}\ }\textbf {\bibinfo {volume} {479}},\ \bibinfo {eid} {128927} (\bibinfo {year} {2023}{\natexlab{b}})}\BibitemShut {NoStop}%
\bibitem [{\citenamefont {{Gao}}\ \emph {et~al.}(2023)\citenamefont {{Gao}}, \citenamefont {{Li}}, \citenamefont {{Mishra}}, \citenamefont {{Yan}}, \citenamefont {{Simonov}},\ and\ \citenamefont {{Chiribella}}}]{Gao2023MeasuringIncompatibility}%
  \BibitemOpen
  \bibfield  {author} {\bibinfo {author} {\bibfnamefont {N.}~\bibnamefont {{Gao}}}, \bibinfo {author} {\bibfnamefont {D.}~\bibnamefont {{Li}}}, \bibinfo {author} {\bibfnamefont {A.}~\bibnamefont {{Mishra}}}, \bibinfo {author} {\bibfnamefont {J.}~\bibnamefont {{Yan}}}, \bibinfo {author} {\bibfnamefont {K.}~\bibnamefont {{Simonov}}},\ and\ \bibinfo {author} {\bibfnamefont {G.}~\bibnamefont {{Chiribella}}},\ }\bibfield  {title} {\bibinfo {title} {{Measuring Incompatibility and Clustering Quantum Observables with a Quantum Switch}},\ }\href {https://doi.org/10.1103/PhysRevLett.130.170201} {\bibfield  {journal} {\bibinfo  {journal} {Phys. Rev. Lett.}\ }\textbf {\bibinfo {volume} {130}},\ \bibinfo {eid} {170201} (\bibinfo {year} {2023})},\ \Eprint {https://arxiv.org/abs/2208.06210} {arXiv:2208.06210 [quant-ph]} \BibitemShut {NoStop}%
\bibitem [{\citenamefont {{Pan}}(2023)}]{Pan2023LeggettGarg}%
  \BibitemOpen
  \bibfield  {author} {\bibinfo {author} {\bibfnamefont {A.~K.}\ \bibnamefont {{Pan}}},\ }\bibfield  {title} {\bibinfo {title} {{Leggett-Garg test of macrorealism using indefinite causal order of measurements}},\ }\href {https://doi.org/10.1016/j.physleta.2023.128898} {\bibfield  {journal} {\bibinfo  {journal} {Physics Letters A}\ }\textbf {\bibinfo {volume} {478}},\ \bibinfo {eid} {128898} (\bibinfo {year} {2023})}\BibitemShut {NoStop}%
\bibitem [{\citenamefont {{Krumm}}\ \emph {et~al.}(2022)\citenamefont {{Krumm}}, \citenamefont {{Allard Gu{\'e}rin}}, \citenamefont {{Zauner}},\ and\ \citenamefont {{Brukner}}}]{Krumm2022teleportation}%
  \BibitemOpen
  \bibfield  {author} {\bibinfo {author} {\bibfnamefont {M.}~\bibnamefont {{Krumm}}}, \bibinfo {author} {\bibfnamefont {P.}~\bibnamefont {{Allard Gu{\'e}rin}}}, \bibinfo {author} {\bibfnamefont {T.}~\bibnamefont {{Zauner}}},\ and\ \bibinfo {author} {\bibfnamefont {{\v{C}}.}~\bibnamefont {{Brukner}}},\ }\bibfield  {title} {\bibinfo {title} {{Quantum teleportation of quantum causal structures}},\ }\href {https://doi.org/10.48550/arXiv.2203.00433} {\bibfield  {journal} {\bibinfo  {journal} {arXiv e-prints}\ ,\ \bibinfo {eid} {arXiv:2203.00433}} (\bibinfo {year} {2022})},\ \Eprint {https://arxiv.org/abs/2203.00433} {arXiv:2203.00433 [quant-ph]} \BibitemShut {NoStop}%
\bibitem [{\citenamefont {{Bavaresco}}\ \emph {et~al.}(2022)\citenamefont {{Bavaresco}}, \citenamefont {{Murao}},\ and\ \citenamefont {{Quintino}}}]{bavaresco2022UnitaryChannelDiscrimination}%
  \BibitemOpen
  \bibfield  {author} {\bibinfo {author} {\bibfnamefont {J.}~\bibnamefont {{Bavaresco}}}, \bibinfo {author} {\bibfnamefont {M.}~\bibnamefont {{Murao}}},\ and\ \bibinfo {author} {\bibfnamefont {M.~T.}\ \bibnamefont {{Quintino}}},\ }\bibfield  {title} {\bibinfo {title} {{Unitary channel discrimination beyond group structures: Advantages of sequential and indefinite-causal-order strategies}},\ }\href {https://doi.org/10.1063/5.0075919} {\bibfield  {journal} {\bibinfo  {journal} {Journal of Mathematical Physics}\ }\textbf {\bibinfo {volume} {63}},\ \bibinfo {eid} {042203} (\bibinfo {year} {2022})},\ \Eprint {https://arxiv.org/abs/2105.13369} {arXiv:2105.13369 [quant-ph]} \BibitemShut {NoStop}%
\bibitem [{\citenamefont {{Quintino}}\ \emph {et~al.}(2019)\citenamefont {{Quintino}}, \citenamefont {{Dong}}, \citenamefont {{Shimbo}}, \citenamefont {{Soeda}},\ and\ \citenamefont {{Murao}}}]{quintino19Reversing}%
  \BibitemOpen
  \bibfield  {author} {\bibinfo {author} {\bibfnamefont {M.~T.}\ \bibnamefont {{Quintino}}}, \bibinfo {author} {\bibfnamefont {Q.}~\bibnamefont {{Dong}}}, \bibinfo {author} {\bibfnamefont {A.}~\bibnamefont {{Shimbo}}}, \bibinfo {author} {\bibfnamefont {A.}~\bibnamefont {{Soeda}}},\ and\ \bibinfo {author} {\bibfnamefont {M.}~\bibnamefont {{Murao}}},\ }\bibfield  {title} {\bibinfo {title} {{Reversing Unknown Quantum Transformations: Universal Quantum Circuit for Inverting General Unitary Operations}},\ }\href {https://doi.org/10.1103/PhysRevLett.123.210502} {\bibfield  {journal} {\bibinfo  {journal} {Phys. Rev. Lett.}\ }\textbf {\bibinfo {volume} {123}},\ \bibinfo {eid} {210502} (\bibinfo {year} {2019})},\ \Eprint {https://arxiv.org/abs/1810.06944} {arXiv:1810.06944 [quant-ph]} \BibitemShut {NoStop}%
\bibitem [{\citenamefont {Trillo}\ \emph {et~al.}(2023)\citenamefont {Trillo}, \citenamefont {Dive},\ and\ \citenamefont {Navascu\'es}}]{Trillo2023universal}%
  \BibitemOpen
  \bibfield  {author} {\bibinfo {author} {\bibfnamefont {D.}~\bibnamefont {Trillo}}, \bibinfo {author} {\bibfnamefont {B.}~\bibnamefont {Dive}},\ and\ \bibinfo {author} {\bibfnamefont {M.}~\bibnamefont {Navascu\'es}},\ }\bibfield  {title} {\bibinfo {title} {Universal quantum rewinding protocol with an arbitrarily high probability of success},\ }\href {https://doi.org/10.1103/PhysRevLett.130.110201} {\bibfield  {journal} {\bibinfo  {journal} {Phys. Rev. Lett.}\ }\textbf {\bibinfo {volume} {130}},\ \bibinfo {pages} {110201} (\bibinfo {year} {2023})}\BibitemShut {NoStop}%
\bibitem [{\citenamefont {Oreshkov}(2019)}]{oreshkov2019timeDelocalized}%
  \BibitemOpen
  \bibfield  {author} {\bibinfo {author} {\bibfnamefont {O.}~\bibnamefont {Oreshkov}},\ }\bibfield  {title} {\bibinfo {title} {Time-delocalized quantum subsystems and operations: on the existence of processes with indefinite causal structure in quantum mechanics},\ }\href {https://doi.org/10.22331/q-2019-12-02-206} {\bibfield  {journal} {\bibinfo  {journal} {{Quantum}}\ }\textbf {\bibinfo {volume} {3}},\ \bibinfo {pages} {206} (\bibinfo {year} {2019})},\ \Eprint {https://arxiv.org/abs/1801.07594} {arXiv:1801.07594 [quant-ph]} \BibitemShut {NoStop}%
\bibitem [{\citenamefont {{de la Hamette}}\ \emph {et~al.}(2022)\citenamefont {{de la Hamette}}, \citenamefont {{Kabel}}, \citenamefont {{Christodoulou}},\ and\ \citenamefont {{Brukner}}}]{DeLaHemette2022quantumDiffeomorphisms}%
  \BibitemOpen
  \bibfield  {author} {\bibinfo {author} {\bibfnamefont {A.-C.}\ \bibnamefont {{de la Hamette}}}, \bibinfo {author} {\bibfnamefont {V.}~\bibnamefont {{Kabel}}}, \bibinfo {author} {\bibfnamefont {M.}~\bibnamefont {{Christodoulou}}},\ and\ \bibinfo {author} {\bibfnamefont {{\v{C}}.}~\bibnamefont {{Brukner}}},\ }\bibfield  {title} {\bibinfo {title} {{Quantum diffeomorphisms cannot make indefinite causal order definite}},\ }\href@noop {} {\bibfield  {journal} {\bibinfo  {journal} {arXiv e-prints}\ } (\bibinfo {year} {2022})},\ \Eprint {https://arxiv.org/abs/2211.15685} {arXiv:2211.15685 [quant-ph]} \BibitemShut {NoStop}%
\bibitem [{\citenamefont {{Fellous-Asiani}}\ \emph {et~al.}(2023)\citenamefont {{Fellous-Asiani}}, \citenamefont {{Mothe}}, \citenamefont {{Bresque}}, \citenamefont {{Dourdent}}, \citenamefont {{Camati}}, \citenamefont {{Abbott}}, \citenamefont {{Auff{\`e}ves}},\ and\ \citenamefont {{Branciard}}}]{fellous2022comparing}%
  \BibitemOpen
  \bibfield  {author} {\bibinfo {author} {\bibfnamefont {M.}~\bibnamefont {{Fellous-Asiani}}}, \bibinfo {author} {\bibfnamefont {R.}~\bibnamefont {{Mothe}}}, \bibinfo {author} {\bibfnamefont {L.}~\bibnamefont {{Bresque}}}, \bibinfo {author} {\bibfnamefont {H.}~\bibnamefont {{Dourdent}}}, \bibinfo {author} {\bibfnamefont {P.~A.}\ \bibnamefont {{Camati}}}, \bibinfo {author} {\bibfnamefont {A.~A.}\ \bibnamefont {{Abbott}}}, \bibinfo {author} {\bibfnamefont {A.}~\bibnamefont {{Auff{\`e}ves}}},\ and\ \bibinfo {author} {\bibfnamefont {C.}~\bibnamefont {{Branciard}}},\ }\bibfield  {title} {\bibinfo {title} {{Comparing the quantum switch and its simulations with energetically constrained operations}},\ }\href@noop {} {\bibfield  {journal} {\bibinfo  {journal} {Physical Review Research}\ }\textbf {\bibinfo {volume} {5}} (\bibinfo {year} {2023})},\ \Eprint {https://arxiv.org/abs/2208.01952} {arXiv:2208.01952 [quant-ph]} \BibitemShut {NoStop}%
\bibitem [{\citenamefont {Paunkovi{\'{c}}}\ and\ \citenamefont {Vojinovi{\'{c}}}(2020)}]{paunkovic2020distinguishing}%
  \BibitemOpen
  \bibfield  {author} {\bibinfo {author} {\bibfnamefont {N.}~\bibnamefont {Paunkovi{\'{c}}}}\ and\ \bibinfo {author} {\bibfnamefont {M.}~\bibnamefont {Vojinovi{\'{c}}}},\ }\bibfield  {title} {\bibinfo {title} {Causal orders, quantum circuits and spacetime: distinguishing between definite and superposed causal orders},\ }\href {https://doi.org/10.22331/q-2020-05-28-275} {\bibfield  {journal} {\bibinfo  {journal} {{Quantum}}\ }\textbf {\bibinfo {volume} {4}},\ \bibinfo {pages} {275} (\bibinfo {year} {2020})},\ \Eprint {https://arxiv.org/abs/1905.09682} {arXiv:1905.09682 [quant-ph]} \BibitemShut {NoStop}%
\bibitem [{\citenamefont {{Ormrod}}\ \emph {et~al.}(2023)\citenamefont {{Ormrod}}, \citenamefont {{Vanrietvelde}},\ and\ \citenamefont {{Barrett}}}]{ormrod2022sectorialConstraints}%
  \BibitemOpen
  \bibfield  {author} {\bibinfo {author} {\bibfnamefont {N.}~\bibnamefont {{Ormrod}}}, \bibinfo {author} {\bibfnamefont {A.}~\bibnamefont {{Vanrietvelde}}},\ and\ \bibinfo {author} {\bibfnamefont {J.}~\bibnamefont {{Barrett}}},\ }\bibfield  {title} {\bibinfo {title} {{Causal structure in the presence of sectorial constraints, with application to the quantum switch}},\ }\href {https://doi.org/10.22331/q-2023-06-01-1028} {\bibfield  {journal} {\bibinfo  {journal} {Quantum}\ }\textbf {\bibinfo {volume} {7}},\ \bibinfo {pages} {1028} (\bibinfo {year} {2023})},\ \Eprint {https://arxiv.org/abs/2204.10273} {arXiv:2204.10273 [quant-ph]} \BibitemShut {NoStop}%
\bibitem [{\citenamefont {Vilasini}\ and\ \citenamefont {Renner}(2022)}]{vilasini2022embedding}%
  \BibitemOpen
  \bibfield  {author} {\bibinfo {author} {\bibfnamefont {V.}~\bibnamefont {Vilasini}}\ and\ \bibinfo {author} {\bibfnamefont {R.}~\bibnamefont {Renner}},\ }\bibfield  {title} {\bibinfo {title} {Embedding cyclic causal structures in acyclic spacetimes: no-go results for process matrices},\ }\href@noop {} {\bibfield  {journal} {\bibinfo  {journal} {arXiv preprint}\ } (\bibinfo {year} {2022})},\ \Eprint {https://arxiv.org/abs/2203.11245} {arXiv:2203.11245 [quant-ph]} \BibitemShut {NoStop}%
\bibitem [{\citenamefont {Saleh}\ and\ \citenamefont {Teich}(2007)}]{SalehFundamentals}%
  \BibitemOpen
  \bibfield  {author} {\bibinfo {author} {\bibfnamefont {B.~E.~A.}\ \bibnamefont {Saleh}}\ and\ \bibinfo {author} {\bibfnamefont {M.~C.}\ \bibnamefont {Teich}},\ }\href {https://cds.cern.ch/record/1084451} {\emph {\bibinfo {title} {{Fundamentals of photonics; 2nd ed.}}}},\ Wiley series in pure and applied optics\ (\bibinfo  {publisher} {Wiley},\ \bibinfo {address} {New York, NY},\ \bibinfo {year} {2007})\BibitemShut {NoStop}%
\bibitem [{\citenamefont {{Felce}}\ \emph {et~al.}(2022)\citenamefont {{Felce}}, \citenamefont {{Vidal}}, \citenamefont {{Vedral}},\ and\ \citenamefont {{Dias}}}]{Felce2022superpositionsInTime}%
  \BibitemOpen
  \bibfield  {author} {\bibinfo {author} {\bibfnamefont {D.}~\bibnamefont {{Felce}}}, \bibinfo {author} {\bibfnamefont {N.~T.}\ \bibnamefont {{Vidal}}}, \bibinfo {author} {\bibfnamefont {V.}~\bibnamefont {{Vedral}}},\ and\ \bibinfo {author} {\bibfnamefont {E.~O.}\ \bibnamefont {{Dias}}},\ }\bibfield  {title} {\bibinfo {title} {{Indefinite causal orders from superpositions in time}},\ }\href {https://doi.org/10.1103/PhysRevA.105.062216} {\bibfield  {journal} {\bibinfo  {journal} {Phys. Rev. A}\ }\textbf {\bibinfo {volume} {105}},\ \bibinfo {eid} {062216} (\bibinfo {year} {2022})},\ \Eprint {https://arxiv.org/abs/2107.08076} {arXiv:2107.08076 [quant-ph]} \BibitemShut {NoStop}%
\bibitem [{\citenamefont {Cavalcanti}\ \emph {et~al.}(2023)\citenamefont {Cavalcanti}, \citenamefont {Chaves}, \citenamefont {Giacomini},\ and\ \citenamefont {Liang}}]{cavalcanti2023fresh}%
  \BibitemOpen
  \bibfield  {author} {\bibinfo {author} {\bibfnamefont {E.~G.}\ \bibnamefont {Cavalcanti}}, \bibinfo {author} {\bibfnamefont {R.}~\bibnamefont {Chaves}}, \bibinfo {author} {\bibfnamefont {F.}~\bibnamefont {Giacomini}},\ and\ \bibinfo {author} {\bibfnamefont {Y.-C.}\ \bibnamefont {Liang}},\ }\bibfield  {title} {\bibinfo {title} {Fresh perspectives on the foundations of quantum physics},\ }\href@noop {} {\bibfield  {journal} {\bibinfo  {journal} {Nature Reviews Physics}\ ,\ \bibinfo {pages} {1}} (\bibinfo {year} {2023})}\BibitemShut {NoStop}%
\bibitem [{\citenamefont {{Hensen}}\ \emph {et~al.}(2015)\citenamefont {{Hensen}}, \citenamefont {{Bernien}}, \citenamefont {{Dr{\'e}au}}, \citenamefont {{Reiserer}}, \citenamefont {{Kalb}}, \citenamefont {{Blok}}, \citenamefont {{Ruitenberg}}, \citenamefont {{Vermeulen}}, \citenamefont {{Schouten}}, \citenamefont {{Abell{\'a}n}}, \citenamefont {{Amaya}}, \citenamefont {{Pruneri}}, \citenamefont {{Mitchell}}, \citenamefont {{Markham}}, \citenamefont {{Twitchen}}, \citenamefont {{Elkouss}}, \citenamefont {{Wehner}}, \citenamefont {{Taminiau}},\ and\ \citenamefont {{Hanson}}}]{Hensen2015Loophole}%
  \BibitemOpen
  \bibfield  {author} {\bibinfo {author} {\bibfnamefont {B.}~\bibnamefont {{Hensen}}}, \bibinfo {author} {\bibfnamefont {H.}~\bibnamefont {{Bernien}}}, \bibinfo {author} {\bibfnamefont {A.~E.}\ \bibnamefont {{Dr{\'e}au}}}, \bibinfo {author} {\bibfnamefont {A.}~\bibnamefont {{Reiserer}}}, \bibinfo {author} {\bibfnamefont {N.}~\bibnamefont {{Kalb}}}, \bibinfo {author} {\bibfnamefont {M.~S.}\ \bibnamefont {{Blok}}}, \bibinfo {author} {\bibfnamefont {J.}~\bibnamefont {{Ruitenberg}}}, \bibinfo {author} {\bibfnamefont {R.~F.~L.}\ \bibnamefont {{Vermeulen}}}, \bibinfo {author} {\bibfnamefont {R.~N.}\ \bibnamefont {{Schouten}}}, \bibinfo {author} {\bibfnamefont {C.}~\bibnamefont {{Abell{\'a}n}}}, \bibinfo {author} {\bibfnamefont {W.}~\bibnamefont {{Amaya}}}, \bibinfo {author} {\bibfnamefont {V.}~\bibnamefont {{Pruneri}}}, \bibinfo {author} {\bibfnamefont {M.~W.}\ \bibnamefont {{Mitchell}}}, \bibinfo {author} {\bibfnamefont {M.}~\bibnamefont {{Markham}}}, \bibinfo {author} {\bibfnamefont {D.~J.}\ \bibnamefont
  {{Twitchen}}}, \bibinfo {author} {\bibfnamefont {D.}~\bibnamefont {{Elkouss}}}, \bibinfo {author} {\bibfnamefont {S.}~\bibnamefont {{Wehner}}}, \bibinfo {author} {\bibfnamefont {T.~H.}\ \bibnamefont {{Taminiau}}},\ and\ \bibinfo {author} {\bibfnamefont {R.}~\bibnamefont {{Hanson}}},\ }\bibfield  {title} {\bibinfo {title} {{Loophole-free Bell inequality violation using electron spins separated by 1.3 kilometres}},\ }\href {https://doi.org/10.1038/nature15759} {\bibfield  {journal} {\bibinfo  {journal} {Nature}\ }\textbf {\bibinfo {volume} {526}},\ \bibinfo {pages} {682} (\bibinfo {year} {2015})},\ \Eprint {https://arxiv.org/abs/1508.05949} {arXiv:1508.05949 [quant-ph]} \BibitemShut {NoStop}%
\bibitem [{\citenamefont {{Giustina}}\ \emph {et~al.}(2015)\citenamefont {{Giustina}}, \citenamefont {{Versteegh}}, \citenamefont {{Wengerowsky}}, \citenamefont {{Handsteiner}}, \citenamefont {{Hochrainer}}, \citenamefont {{Phelan}}, \citenamefont {{Steinlechner}}, \citenamefont {{Kofler}}, \citenamefont {{Larsson}}, \citenamefont {{Abell{\'a}n}}, \citenamefont {{Amaya}}, \citenamefont {{Pruneri}}, \citenamefont {{Mitchell}}, \citenamefont {{Beyer}}, \citenamefont {{Gerrits}}, \citenamefont {{Lita}}, \citenamefont {{Shalm}}, \citenamefont {{Nam}}, \citenamefont {{Scheidl}}, \citenamefont {{Ursin}}, \citenamefont {{Wittmann}},\ and\ \citenamefont {{Zeilinger}}}]{Giustina2015Significant}%
  \BibitemOpen
  \bibfield  {author} {\bibinfo {author} {\bibfnamefont {M.}~\bibnamefont {{Giustina}}}, \bibinfo {author} {\bibfnamefont {M.~A.~M.}\ \bibnamefont {{Versteegh}}}, \bibinfo {author} {\bibfnamefont {S.}~\bibnamefont {{Wengerowsky}}}, \bibinfo {author} {\bibfnamefont {J.}~\bibnamefont {{Handsteiner}}}, \bibinfo {author} {\bibfnamefont {A.}~\bibnamefont {{Hochrainer}}}, \bibinfo {author} {\bibfnamefont {K.}~\bibnamefont {{Phelan}}}, \bibinfo {author} {\bibfnamefont {F.}~\bibnamefont {{Steinlechner}}}, \bibinfo {author} {\bibfnamefont {J.}~\bibnamefont {{Kofler}}}, \bibinfo {author} {\bibfnamefont {J.-{\r{A}}.}\ \bibnamefont {{Larsson}}}, \bibinfo {author} {\bibfnamefont {C.}~\bibnamefont {{Abell{\'a}n}}}, \bibinfo {author} {\bibfnamefont {W.}~\bibnamefont {{Amaya}}}, \bibinfo {author} {\bibfnamefont {V.}~\bibnamefont {{Pruneri}}}, \bibinfo {author} {\bibfnamefont {M.~W.}\ \bibnamefont {{Mitchell}}}, \bibinfo {author} {\bibfnamefont {J.}~\bibnamefont {{Beyer}}}, \bibinfo {author} {\bibfnamefont {T.}~\bibnamefont
  {{Gerrits}}}, \bibinfo {author} {\bibfnamefont {A.~E.}\ \bibnamefont {{Lita}}}, \bibinfo {author} {\bibfnamefont {L.~K.}\ \bibnamefont {{Shalm}}}, \bibinfo {author} {\bibfnamefont {S.~W.}\ \bibnamefont {{Nam}}}, \bibinfo {author} {\bibfnamefont {T.}~\bibnamefont {{Scheidl}}}, \bibinfo {author} {\bibfnamefont {R.}~\bibnamefont {{Ursin}}}, \bibinfo {author} {\bibfnamefont {B.}~\bibnamefont {{Wittmann}}},\ and\ \bibinfo {author} {\bibfnamefont {A.}~\bibnamefont {{Zeilinger}}},\ }\bibfield  {title} {\bibinfo {title} {{Significant-Loophole-Free Test of Bell's Theorem with Entangled Photons}},\ }\href {https://doi.org/10.1103/PhysRevLett.115.250401} {\bibfield  {journal} {\bibinfo  {journal} {Physical Review Letters}\ }\textbf {\bibinfo {volume} {115}},\ \bibinfo {eid} {250401} (\bibinfo {year} {2015})},\ \Eprint {https://arxiv.org/abs/1511.03190} {arXiv:1511.03190 [quant-ph]} \BibitemShut {NoStop}%
\bibitem [{\citenamefont {{Shalm}}\ \emph {et~al.}(2015)\citenamefont {{Shalm}}, \citenamefont {{Meyer-Scott}}, \citenamefont {{Christensen}}, \citenamefont {{Bierhorst}}, \citenamefont {{Wayne}}, \citenamefont {{Stevens}}, \citenamefont {{Gerrits}}, \citenamefont {{Glancy}}, \citenamefont {{Hamel}}, \citenamefont {{Allman}}, \citenamefont {{Coakley}}, \citenamefont {{Dyer}}, \citenamefont {{Hodge}}, \citenamefont {{Lita}}, \citenamefont {{Verma}}, \citenamefont {{Lambrocco}}, \citenamefont {{Tortorici}}, \citenamefont {{Migdall}}, \citenamefont {{Zhang}}, \citenamefont {{Kumor}}, \citenamefont {{Farr}}, \citenamefont {{Marsili}}, \citenamefont {{Shaw}}, \citenamefont {{Stern}}, \citenamefont {{Abell{\'a}n}}, \citenamefont {{Amaya}}, \citenamefont {{Pruneri}}, \citenamefont {{Jennewein}}, \citenamefont {{Mitchell}}, \citenamefont {{Kwiat}}, \citenamefont {{Bienfang}}, \citenamefont {{Mirin}}, \citenamefont {{Knill}},\ and\ \citenamefont {{Nam}}}]{Shalm2015strong}%
  \BibitemOpen
  \bibfield  {author} {\bibinfo {author} {\bibfnamefont {L.~K.}\ \bibnamefont {{Shalm}}}, \bibinfo {author} {\bibfnamefont {E.}~\bibnamefont {{Meyer-Scott}}}, \bibinfo {author} {\bibfnamefont {B.~G.}\ \bibnamefont {{Christensen}}}, \bibinfo {author} {\bibfnamefont {P.}~\bibnamefont {{Bierhorst}}}, \bibinfo {author} {\bibfnamefont {M.~A.}\ \bibnamefont {{Wayne}}}, \bibinfo {author} {\bibfnamefont {M.~J.}\ \bibnamefont {{Stevens}}}, \bibinfo {author} {\bibfnamefont {T.}~\bibnamefont {{Gerrits}}}, \bibinfo {author} {\bibfnamefont {S.}~\bibnamefont {{Glancy}}}, \bibinfo {author} {\bibfnamefont {D.~R.}\ \bibnamefont {{Hamel}}}, \bibinfo {author} {\bibfnamefont {M.~S.}\ \bibnamefont {{Allman}}}, \bibinfo {author} {\bibfnamefont {K.~J.}\ \bibnamefont {{Coakley}}}, \bibinfo {author} {\bibfnamefont {S.~D.}\ \bibnamefont {{Dyer}}}, \bibinfo {author} {\bibfnamefont {C.}~\bibnamefont {{Hodge}}}, \bibinfo {author} {\bibfnamefont {A.~E.}\ \bibnamefont {{Lita}}}, \bibinfo {author} {\bibfnamefont {V.~B.}\ \bibnamefont
  {{Verma}}}, \bibinfo {author} {\bibfnamefont {C.}~\bibnamefont {{Lambrocco}}}, \bibinfo {author} {\bibfnamefont {E.}~\bibnamefont {{Tortorici}}}, \bibinfo {author} {\bibfnamefont {A.~L.}\ \bibnamefont {{Migdall}}}, \bibinfo {author} {\bibfnamefont {Y.}~\bibnamefont {{Zhang}}}, \bibinfo {author} {\bibfnamefont {D.~R.}\ \bibnamefont {{Kumor}}}, \bibinfo {author} {\bibfnamefont {W.~H.}\ \bibnamefont {{Farr}}}, \bibinfo {author} {\bibfnamefont {F.}~\bibnamefont {{Marsili}}}, \bibinfo {author} {\bibfnamefont {M.~D.}\ \bibnamefont {{Shaw}}}, \bibinfo {author} {\bibfnamefont {J.~A.}\ \bibnamefont {{Stern}}}, \bibinfo {author} {\bibfnamefont {C.}~\bibnamefont {{Abell{\'a}n}}}, \bibinfo {author} {\bibfnamefont {W.}~\bibnamefont {{Amaya}}}, \bibinfo {author} {\bibfnamefont {V.}~\bibnamefont {{Pruneri}}}, \bibinfo {author} {\bibfnamefont {T.}~\bibnamefont {{Jennewein}}}, \bibinfo {author} {\bibfnamefont {M.~W.}\ \bibnamefont {{Mitchell}}}, \bibinfo {author} {\bibfnamefont {P.~G.}\ \bibnamefont {{Kwiat}}}, \bibinfo
  {author} {\bibfnamefont {J.~C.}\ \bibnamefont {{Bienfang}}}, \bibinfo {author} {\bibfnamefont {R.~P.}\ \bibnamefont {{Mirin}}}, \bibinfo {author} {\bibfnamefont {E.}~\bibnamefont {{Knill}}},\ and\ \bibinfo {author} {\bibfnamefont {S.~W.}\ \bibnamefont {{Nam}}},\ }\bibfield  {title} {\bibinfo {title} {{Strong Loophole-Free Test of Local Realism$^{*}$}},\ }\href {https://doi.org/10.1103/PhysRevLett.115.250402} {\bibfield  {journal} {\bibinfo  {journal} {Physical Review Letters}\ }\textbf {\bibinfo {volume} {115}},\ \bibinfo {eid} {250402} (\bibinfo {year} {2015})},\ \Eprint {https://arxiv.org/abs/1511.03189} {arXiv:1511.03189 [quant-ph]} \BibitemShut {NoStop}%
\bibitem [{\citenamefont {Jamio{\l}kowski}(1972)}]{jamiolkowski1972linear}%
  \BibitemOpen
  \bibfield  {author} {\bibinfo {author} {\bibfnamefont {A.}~\bibnamefont {Jamio{\l}kowski}},\ }\bibfield  {title} {\bibinfo {title} {Linear transformations which preserve trace and positive semidefiniteness of operators},\ }\href {https://doi.org/10.1016/0034-4877(72)90011-0} {\bibfield  {journal} {\bibinfo  {journal} {Reports on Mathematical Physics}\ }\textbf {\bibinfo {volume} {3}},\ \bibinfo {pages} {275} (\bibinfo {year} {1972})}\BibitemShut {NoStop}%
\bibitem [{\citenamefont {Choi}(1975)}]{choi1975completely}%
  \BibitemOpen
  \bibfield  {author} {\bibinfo {author} {\bibfnamefont {M.-D.}\ \bibnamefont {Choi}},\ }\bibfield  {title} {\bibinfo {title} {Completely positive linear maps on complex matrices},\ }\href {https://doi.org/10.1016/0024-3795(75)90075-0} {\bibfield  {journal} {\bibinfo  {journal} {Linear algebra and its applications}\ }\textbf {\bibinfo {volume} {10}},\ \bibinfo {pages} {285} (\bibinfo {year} {1975})}\BibitemShut {NoStop}%
\bibitem [{\citenamefont {Chiribella}\ \emph {et~al.}(2009)\citenamefont {Chiribella}, \citenamefont {D'Ariano},\ and\ \citenamefont {Perinotti}}]{chiribella2009theoretical}%
  \BibitemOpen
  \bibfield  {author} {\bibinfo {author} {\bibfnamefont {G.}~\bibnamefont {Chiribella}}, \bibinfo {author} {\bibfnamefont {G.~M.}\ \bibnamefont {D'Ariano}},\ and\ \bibinfo {author} {\bibfnamefont {P.}~\bibnamefont {Perinotti}},\ }\bibfield  {title} {\bibinfo {title} {Theoretical framework for quantum networks},\ }\href {https://doi.org/10.1103/PhysRevA.80.022339} {\bibfield  {journal} {\bibinfo  {journal} {Phys. Rev. A}\ }\textbf {\bibinfo {volume} {80}},\ \bibinfo {pages} {022339} (\bibinfo {year} {2009})}\BibitemShut {NoStop}%
\end{thebibliography}%

\clearpage
\section*{Appendix}

\subsection{Process Matrices}
\label{box:PM}
A valuable toolkit to describe processes exhibiting an indefinite causal order and to capture their temporal correlations is the process matrix formalism.
In this framework, quantum mechanics is assumed to be valid locally, without referring to the existence of a global causal structure \cite{oreshkov2012quantum}. The process matrix formalism aims to characterize probability distributions in diverse scenarios involving the application of local operations. Generally these operations, encompassing deterministic processes such as unitaries, quantum channels, or generalized measurements, are described by completely positive (CP) trace non-increasing maps {$\mathcal{M}^A: \mathcal{H}^{A_I}\rightarrow \mathcal{H}^{A_O}$}, where $\mathcal{H}^{A_I}$ and $\mathcal{H}^{A_O}$ denote the Hilbert spaces of input and output, respectively. The CP map $\mathcal{M}_a^A$ is often associated with an outcome $a$, and Its representation is facilitated by the Choi–Jamiołkowski (CJ) isomorphism \cite{jamiolkowski1972linear,choi1975completely}. The CJ matrix $M_a^{A_I A_O}$ corresponding to $\mathcal{M}_a^A$ is defined as
\begin{equation}
    M_a^{A_I A_O}:=\mathcal{I}\otimes \mathcal{M}_a^A(|\mathbbm{1}\rangle \rangle \langle\langle\mathbbm{1}|) \in \mathcal{H}^{A_I}\otimes \mathcal{H}^{A_O}
\end{equation}
where $\mathcal{I}$ is the identity {map} and {$|\mathbbm{1}\rangle\rangle=\sum_j |j\rangle   ^{A_I}\otimes |j\rangle^{A_I} \in \mathcal{H}^{A_I}\otimes \mathcal{H}^{A_I}$}. 
The collection $\mathcal{J}_A=\left\{\mathcal{M}_a^A\right\}$ associated with all outcomes must fulfill the condition that $\sum_a \mathcal{M}_a^A$ {is} CP and trace-preserving (CPTP). The selection of operations, termed quantum instruments, serves as inputs of the following statistics. The key concept of the formalism is the process $W$, treated as a resource that dictates the statistical probability distribution under CP maps $M_{a_1|x_1}^{A_I^1 A_O^1},\cdots M_{a_N|x_N}^{A_I^N A_O^N}$ for a choice of operations $x_1, x_2\cdots x_N$
\begin{equation}
    P(M_{a_1|x_1}^{A^1},\cdots M_{a_N|x_N}^{A^N})=\mathrm{tr}\left[\left(M_{a_1|x_1}^{A_I^1 A_O^1}\otimes \cdots \otimes M_{a_N|x_N}^{A_I^N A_O^N}\right) W \right]
    \label{eq:bornrule}
\end{equation}
with {$W \in \mathcal{H}^{A_I^1}\otimes \mathcal{H}^{A_O^1}\otimes\cdots \mathcal{H}^{A_I^N}\otimes \mathcal{H}^{A_O^N}$} being a Hermitian. This process matrix $W$ can be perceived as a generalization of a density matrix, and the equation above serves as a generalization of Born’s rule.
This framework extends the formalism of the `quantum comb' \cite{chiribella2009theoretical}. 

In the bipartite scenario involving parties A and B, a process matrix $W_{sep}$ is called causally separable if it can be expressed as a convex combination of causally ordered processes, i.e., $W_{sep}=p W^{A \preceq B}+(1-p) W^{B \preceq A}$, where $W^{A \preceq B}$ describes the process where party A causally precedes party B, and $0 \leq p \leq 1$. Concretely for the quantum switch in the bipartite case, with the control qubit initialized in $\frac{1}{\sqrt{2}}(|0\rangle+|1\rangle)^C$, the process matrix is represented by $W_{switch}=|w_{switch}\rangle\langle w_{switch}|$, with the process vector

\begin{eqnarray}
    |w_{switch}\rangle & = \frac{1}{\sqrt{2}}(|w^{A \preceq B}\rangle|0\rangle^C + |w^{B \preceq A}\rangle|1\rangle^C), \\
    |w^{A \preceq B}\rangle & =  |\mathbbm{1}\rangle\rangle^{\mathcal{H}^{in}A_I}|\mathbbm{1}\rangle\rangle^{A_O B_I}|\mathbbm{1}\rangle\rangle^{B_O\mathcal{H}^{out}}
    \label{eq:switch}
\end{eqnarray}

Here, $\mathcal{H}^{in}$ ($\mathcal{H}^{out}$) is associated with the Hilbert space of the target state from the input (output) space of quantum switch.


\begin{figure}[t]
\centering
\includegraphics[width=0.5\columnwidth]{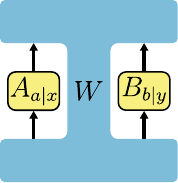}
    \caption{Bipartite process matrix with no global past or future system. Note, most formulations of the quantum switch process matrix have past and future systems.}
\end{figure}
\subsection{Counting Gate Uses}
\label{box:Counting}
For many applications the advantage of quantum switch is that they require fewer gate uses than traditional quantum circuits. This makes it important to define what is meant by a gate use, which is nontrivial when using essentially classical resources (such as a wave plate, laser or RF pulse) to coherently manipulate a quantum system.
For example, in a trapped-ion quantum computer a single laser pulse can simultaneously manipulate each ion in the trap.  Should this be considered as: one gate use since a single pulse was used; one gate use for each ion in the trap; or many many more gate uses, since not all of the light is absorbed by the ions, and in principle one could manipulate many more ions?
\\
\\
Using wave plates to manipulate photonic qubits, as is done in quantum switch experiments, is a very similar situation.
In principle, one could send a nearly infinite number of orthogonal optical modes through a given optical element.
To take a more concrete experimental perspective, one could ask how much energy flows through each optic.
The energy transmitted through a given area is given by the optical power integrated over the experimental time \cite{SalehFundamentals}. 
This power is proportional to the photon occupation number, and does not depend on the mode shape or even the number of modes.
Thus, when a \textit{single photon} is placed in a superposition of any number of modes (the spatial modes, spectral modes, orbital angular momentum modes, etc.), the energy traversing the gate is the same.
From this point of view, only one photon traverses the gate per experimental run.
\\
\\
Phrasing this argument in an operational quantum manner, we can imagine using an ancillary system to coherently count how many times the photon passes through the gate (Fig. \ref{fig:loopholes} a).
Since the energy is proportional to the photon flux, these two arguments lead to qualitatively similar results.
In particular, imagine a qudit counter system in state $\ket{0}$ that interacts with the switch photon via two ``controlled-plus'' gates. In this interaction if a particle traverses the control path, then the state of the qudit system evolves as $\ket{j}\rightarrow\ket{(j+1)~\mathrm{mod}~d}$, where $d$ is the dimension of the system. 
In these, the auxiliary system counts how many times the photon passes each gate.
A straightforward calculation shows that after the experiment the qudit is left in the state $\ket{1}$, indicating that the gate has only been used once.
One could also use a counting system in the unfolded switch (Fig. \ref{fig:loopholes} b), but then one cannot associate a single spatial region to either gate; i.e., the spatial regions, the (encircled yellow and green regions in Fig. \ref{fig:loopholes}b) must cross.
\\
\\
In another approach, Fellous-Asiani et al. theoretically analyzed a situation wherein an additional quantum system (an atom in a cavity) is used to implement a gate.
Since this requires the atom to transition between different energy states, this allowed them to associate an energy cost to each gate use.
They then computed the energy cost of implementing quantum gates both in the standard switch and in the unfolded switch. 
They showed that the unfolded configuration requires more energy than the standard switch \cite{fellous2022comparing}.

\end{document}